\documentclass[aps,prb,twocolumn,showpacs,floats]{revtex4}
\usepackage{latexsym}
\usepackage{amsmath}
\usepackage{amssymb}
\usepackage{bm}
\usepackage{wasysym}
\usepackage[dvips]{color}
\usepackage{graphicx}
\usepackage{graphicx}
\usepackage{subfigure}
\usepackage{epsfig}

\newcommand{\veps}{\varepsilon}
\newcommand{\bfk}{{\bm k}}

\newcommand{\ua}{\uparrow}
\newcommand{\da}{\downarrow}

\def\mbfE{\bm{E}}
\def\mbfe{\bm{e}}

\def\mbfk{\bm{k}}
\def\mbfq{\bm{q}}

\def\mbfd{\bm{d}}
\def\mbfa{\bm{a}}

\def\mbfr{\bm{r}}

\def\mbfn{\bm{n}}
\usepackage{amsmath,stmaryrd,graphicx}
\makeatletter
\newcommand{\fixed@sra}{$\vrule height 2\fontdimen22\textfont2 width 0pt\shortrightarrow$}
\newcommand{\shortarrow}[1]{%
  \mathrel{\text{\rotatebox[origin=c]{\numexpr#1*45}{\fixed@sra}}}
}
\makeatother

\begin{document}

\title{Spin-Orbit Coupling and Topological States in $F=\frac{3}{2}$ Cold
       Fermi Gas}

\author{
        Igor Kuzmenko$^1$, Tetyana Kuzmenko$^1$,
       Yshai Avishai$^{1,2,3}$ and Masatoshi Sato$^2$
         }
\affiliation{$^1$Department of Physics, Ben-Gurion University
of the Negev,
  Beer-Sheva, Israel \\
  $^2$Yukawa Institute for Theoretical Physics, Kyoto University,
  Kyoto 606-8502, Japan \\
  $^3$New-York University at Shanghai, Shanghai, China}

\date{\today}
\begin{abstract}
In this work we study the possible occurrence of topological
insulators for 2D fermions of high spin. They can be realized in
cold fermion systems with ground-state atomic spin
$F>\tfrac{1}{2}$, if the optical potential is properly designed,
and spin-orbit coupling is relevant. The latter is shown to be
induced by letting the fermions interact with a specially tuned
arrangement of polarized laser beams. When  the system is subject
to a perpendicular magnetic field, time reversal symmetry is
broken but the ensuing Hamiltonian is still endowed with a mirror
symmetry.

Topological insulators for fermions of higher spins are
fundamentally distinct from those pertaining to spin
$\frac{1}{2}$. The underlying physics reveals a plethora of
positive and negative mirror Chern numbers, respectively
corresponding to chiral and anti-chiral  edge states. Here, for
simplicity, we concentrate on the case $F=\tfrac{3}{2}$ (which is
suitable for $^{6}$Li or $^2$H atoms) but extension to higher
spins (such as $^{40}$K whose ground-state spin is
$F=\tfrac{9}{2}$), is straightforward.
\end{abstract}


\maketitle

%%%%%%%%%%%%%%%%%%%%%%%%%%%%%%
\section{Introduction}
  \label{Sec:Intro}
Spin-orbit coupling (SOC) in cold-atom systems are now realizable,
employing laser radiation impinging on a gas of cold atoms
\cite{Dalibard}. Furthermore, spin-orbit coupling with equal
Rashba and Dresselhaus strengths is synthetically induced by
applying Raman lasers on atomic gases with hyperfine spin degrees
of freedom \cite{Lin2011,Wu-Chin-Phys-Lett-2011}, whose scheme was discussed
theoretically in several earlier works
\cite{Higbie2002, Liu2009, Spielman2009}.
%whose practical scheme was first pointed out by Liu et al \cite{Liu2009}.
This scheme has recently been implemented using both cold boson
\cite{Lin2011,Zhang2012}
and fermionic degenerate gases
\cite{Wang2012, Cheuk2012}. In addition, schemes
for creating general Rashba and Dresselhaus SOC or
three-dimensional (3D) analogue to Rashba SOC have theoretically
been proposed \cite{Juzeliuas2010, Campbell2011, Anderson2012},
and experimental realization of 2D SOC has been demonstrated
\cite{Huang2016, Wu2016}.

In a manner similar to solid state physics, SOC in cold-atom
systems provides interesting non-trivial phenomena such as
anomalous quantum Hall and quantum spin Hall effects
\cite{SO-opt-latt-PRL04}. An $s$-wave superfluild of cold atoms
may host Majorana fermions in the presence of SOC and a Zeeman
magnetic field \cite{Sato-Takahashi-Fujimoto2009, SOC-1D-prl-11, %
Mizushima2013}.
%
%More than a decade ago, SOC has already been experimentally studied in 1D
%systems
% SOC and an effective magnetic field
%were generated in 1D fermionic atoms by employing an optical Raman transition
%enabling the detection of Majorana fermions\cite{SOC-1D-prl-11}.
Moreover, Rashba mechanism of SOC in 2D cold atomic system is
under intensive study \cite{SOC-atoms-nature-material-15, %
SOC-PRL10a,SOC-progress-15}.

In solid state physics, except for a few materials
\cite{Brydon2016, Venderbos2017,Yang2016,Ghorashi2017,Roy2017},
SOC pertaining to a degenerate Fermi gas is
analyzed for
electrons (or holes) whose intrinsic spin is $s=\frac{1}{2}$.
In cold atom systems, SOC is mostly studied for atoms with
atomic spin $F=\tfrac{1}{2}$. However, one of the advantages
of studying cold atom systems (as compared  with solid-state
systems) is the possibility to explore the physics of degenerate
Fermi gases in which the fermions have atomic  spin
$F>\tfrac{1}{2}$ \cite{WuHuZhang-PRL-2003,Wu-Mod-Phys-Lett-B-2006}.
For example, in such systems, the Kondo physics
is expected to be richer than it is for degenerate Fermi gases
with spin $F=\tfrac{1}{2}$, because it might lead to
over-screening and hence to a non-Fermi liquid ground-state
\cite{Kim-prb-96,KKAK-prb-15}.

The purpose of the present work is to employ this godsend of
controlling Fermi gases of higher spin fermions for exploring SOC
in such systems in general, and, in particular, systems exhibiting
topological properties such as topological insulators. To achieve
this goal, we consider a model wherein a degenerate Fermi gas of
spin $F=\tfrac{3}{2}$ atoms (for example, $^6$Li or $^2$H
atoms) occupies a 2D optical lattice and is subject to SOC.
The Bloch spectrum is composed of two bands, and the bulk system
can be described by an $8 \times 8$ Hamiltonian (two bands and
four spin states) with finite gap between them. This optical
potential can be generated by a specially designed pattern of
polarized laser fields. In the present work we concentrate on
the case where the system is also subject to a uniform
perpendicular magnetic field  of strength $B$, so that time
reversal symmetry (TRS) is broken. (Note that because the atoms
are neutral, the magnetic field acts only on the spin degrees
of freedom). In a future communication we will study the system
in the absence of magnetic field. The ensuing Hamiltonian contains
a real parameter $\Delta_0$ that, together with the external
magnetic field $B>0$, controls the shape of the gap. Crossing into
the region $B>\tfrac{2}{3} \Delta_0 \Theta(\Delta_0)$ in the half
plane ($B>0,\Delta_0$) drives the insulator to be topologically
non-trivial.

As it turns out, this Hamiltonian has additional symmetries beyond
those listed in the Altland-Zirnbauer (AZ) classification scheme
\cite{AZ}. Moreover, one of these symmetries is not broken when
the system is opened to have one edge (for example, when the 2D
system is constrained in a half plane). Hence it is susceptible
for an analysis within the theory of topological insulators with
additional symmetries (with respect to those listed by AZ),
developed by one of us \cite{Shizoaki}. As such, it is shown to
support ${\bm Z}\times{\bm Z}$ topological indices, and
the bulk-edge correspondence is beautifully demonstrated after
the mirror Chen numbers in the bulk are calculated together with
the edge state spectrum.

The main achievements reported here are as follows:
\vspace{-0.1 in}
\begin{enumerate}
\item{} The Hamiltonian
describing the topology of spin $F=\frac{2n+1}{2}$ fermions with SOC
for $n>0$  is derived and found to be
fundamentally distinct from that for $n=0$. In particular, the former
cannot be expressed solely in terms of spin operators that belong to
the  $2F+1$ irrep of SU(2) \cite{WuHuZhang-PRL-2003}.
\item{} The Hamiltonian possesses numerous symmetries that
endows it with rich topological structure. In particular, there is a mirror
symmetry that simplifies the calculations of the spectrum and the
topological numbers.
\item{} There are $2F+2$ domains in the half-plane
$(B,\Delta_0)$ (where $B>0$ is the strength of the magnetic field
and $\Delta_0$ is a gap parameter) such that in each domain
there is a specific pattern of mirror Chern numbers, as well as 1D winding numbers.
\item{} The bulk-edge  correspondence scenario is shown to be valid also for
fermions of higher spin, leading to the occurrence of
chiral and anti-chiral edge states that  {\it  propagate on the same edge in opposite directions}.
\item{} For $\Delta_0 <0$ and $B \to 0$ the pairs of chiral and anti-chiral edge states
tends smoothly to into $F+\tfrac{1}{2}$  Kramers pairs of helical states.
\end{enumerate}
For mere simplicity, in this work, the analysis and substantiation of these
achievements is carried out for fermions with
atomic spin $F=\tfrac{3}{2}$, for example $^6$Li or $^2$H atoms.
However, the formalism developed here can straightforwardly
be extended  for studying fermions with higher atomic spin, for example $^{40}$K,
that is an appropriate experimental candidate.

The paper is organized as follows:
In section \ref{Sec:Model} the model's bare $8 \times 8$
Hamiltonian $\hat{H}_\bfk$ is written down (within the long-wave approximation)
in spin$\otimes$band space. Then, employing a mirror symmetry
(that is a simple unitary transformation that mixes spin and band spaces),
enables us to transform it  into a couple of $4 \times 4$ block
matrices on its diagonal. Formally,
$\hat{H}_\bfk=\mbox{diag}(\hat {\cal H}_{\bfk,1},\hat {\cal H}_{\bfk,\bar{1}})$ with
$\bar{1}=-1$. The intriguing feature is that each $4 \time 4$
block cannot be written simply as a combination of operators
belonging to the  four dimensional (irreducible) representation
of the SU(2) group associated with spin $F=\tfrac{3}{2}$
\cite{WuHuZhang-PRL-2003}.
The spectra of the two 4$\times$4 Hamiltonians
$\hat {\cal H}_{\bfk, \eta}$ (with $\eta=\pm 1$) in the bulk are
displayed and shown to be qualitatively similar but quantitatively
distinct.

In section \ref{Sec:Symm} we elaborate on the symmetry and
classification of these 4$\times$4 operators beyond the standard
AZ scheme. In particular, it is shown that it has two additional
symmetries but only one of them is not broken when the system
contains an edge. This observation enables an analysis in terms
of the formalism developed for classification of topological insulators with
additional symmetries\cite{Shizoaki}.
Then, in section \ref{sec-edge} the edge states are studied
separately for each 4$\times$4 block.

A crucial question is whether the present model and its analysis
can be realized in cold atom gases. This question is
addressed in appendix \ref{Sec:Justification}. Thus, although
the analysis detailed in the main text starting from the long-wave
$8 \times 8$ Hamiltonian is self consistent, the reader who is
interested in justification and derivation of the Hamiltonian
will find it in Appendix \ref{Sec:Justification}.  Finally,
Appendix \ref{append-num-calc} is devoted
to the numerical solutions of Eqs.~(\ref{eq-for-kappa}) and
(\ref{eq-for-energy}) and in Appendix \ref{append-min-max}
we elaborate upon the main properties of the optical potential
defined through Eq.~(\ref{V-pot-res}).

\section{The Long-Wave Model Hamiltonian}
\label{Sec:Model}
Following the detailed analysis described in Appendix
\ref{Sec:Justification}, the model Hamiltonian of cold atoms
subject to SOC and an external magnetic field
${\mathbf{B}}=B\mbfe_z$ can be written in a matrix form as,
\begin{eqnarray}
  \hat{H} &=&
  \sum_{\mbfk}
  \hat\Psi_{\mbfk}^{\dag}~
  \hat{H}_{\mbfk}~
  \hat\Psi_{\mbfk},
  \label{H-matrix}
\end{eqnarray}
where
\begin{eqnarray}
  \hat{H}_{\mbfk}
  &=&
  \hat{H}_{\mbfk}^{(0)}+
  \hat{H}_{\mbfk}^{\mathrm{(SO)}}.
  \label{H-4x2D-def}
\end{eqnarray}
The operator $\hat\Psi_{\mbfk}$ is
\begin{eqnarray}
  \hat\Psi_{\mbfk} ~=~
  \left(
    \begin{array}{c}
      \hat\Psi_{{\mathrm{c}},\mbfk}
      \\
      \hat\Psi_{{\mathrm{v}},\mbfk}
    \end{array}
  \right),
  \ \ \ \ \
  \hat\Psi_{\xi,\mbfk} ~=~
  \left(
    \begin{array}{c}
      c_{\xi,\mbfk,\frac{3}{2}}
      \\
      c_{\xi,\mbfk,\frac{1}{2}}
      \\
      c_{\xi,\mbfk,\frac{\bar{1}}{2}}
      \\
      c_{\xi,\mbfk,\frac{\bar{3}}{2}}
    \end{array}
  \right),
  \label{basis}
\end{eqnarray}
where $\mbfk=k_x\mbfe_x+k_y\mbfe_y$  is the 2D wave vector of length
$k=|\mbfk|$.
Here $c_{\xi,\mbfk,f}$ and $c_{\xi,\mbfk,f}^{\dag}$ are annihilation
and creation operators for atom in the conduction ($\xi={\mathrm{c}}$)
or valence ($\xi={\mathrm{v}}$) band with wave vector
$\mbfk$
and magnetic quantum number $f=\pm\frac{1}{2}$, $\pm\frac{3}{2}$.
The notations $\bar{f}=-f$ is used throughout.

The first term on the RHS of eq. (\ref{H-4x2D-def})
describes the kinetic energy of the atoms that moves
in an external magnetic field
${\mathbf{B}}=B{\mathbf{e}}_{z}$. Explicitly
%%%%%%%%%%%%%%%%%%%%%%%Begin Footnote%%%%%%%%%%%%%%%%%%%%%
\footnote{It has been shown that spin $3/2$ systems can be described
by $4 \times 4$ SO(5) $\Gamma$ matrices
\cite{WuHuZhang-PRL-2003,Wu-Mod-Phys-Lett-B-2006}.
However, our Hamiltonian is a 8x8 matrix, so it has
an SO(7) structure rather than SO(5). In the mirror sub-sector, the Hamiltonian reduces to $4x4$ blocks, so it can be written in terms SO(5) 
$\Gamma$ matrices, but 
in that case, half of the degrees of freedom comes from the band index, so it is not directly related 
to the spin 3/2 SO(5) matrix.}
%%%%%%%%%%%%%%%%%End Footnote%%%%%%%%%%%%%%%%%%%%%%%%%%%
\begin{eqnarray}
  &&
  \hat{H}_{\mbfk}^{(0)} ~=~
  \Delta_{\mbfk}~
  \hat\tau^z
  \otimes
  \hat{F}^0+
  B~
  \hat\tau^0
  \otimes
  \hat{F}^z.
  \label{H0-4x2D-def}
\end{eqnarray}
The first term on the RHS of eq. (\ref{H0-4x2D-def})
describes atoms in the conduction or valence band with dispersion
$\pm{\Delta}_{\mbfk}$, where
\begin{eqnarray}
  \Delta_{\mbfk} ~=~
  \Delta_0+
  \frac{\hbar^2 k^2}{2 M_0}.
  \label{Mk-def}
\end{eqnarray}
The real parameter $-\infty < \Delta_0 < \infty$ determines the shape of the gap, and
plays a central r\^ole in driving the system through a topological transition.
The second term on the RHS of eq. (\ref{H0-4x2D-def})
is Zeeman interaction of the atoms with the external magnetic field.
It is assumed that $M_0>0$, and $B>0$.

The spin-orbit interaction is encoded by the
second term on the RHS of eq. (\ref{H-4x2D-def}),
\begin{eqnarray}
  \hat{H}_{\mbfk}^{\mathrm{(SO)}} ~=~
  2~
  \hat\tau^{x}
  \otimes
  \big(
      \mbfd_{\mbfk}
      \cdot
      \hat{\mathbf{F}}
  \big), \ \  \mbfd_{\mbfk} ~=~
  \hbar v \mbfk, \ (v>0),
  \label{HSO-4x2D-def}
\end{eqnarray}
where
$$
  \hat{\boldsymbol\tau}
  ~=~
  \big(
      \hat\tau^{x},~
      \hat\tau^{y},~
      \hat\tau^{z}
  \big)
$$
is a vector of Pauli matrices acting in the isospin space of
the conduction and valence bands, while
$$
  \hat{\mathbf{F}}
  ~=~
  \big(
      \hat{F}^{x},~
      \hat{F}^{y},~
      \hat{F}^{z}
  \big)
$$
is a vector of the spin $\frac{3}{2}$ tensors. Nontrivial
matrix elements of $\hat{F}^{\alpha}$ ($\alpha=x,y,z$)
are
\begin{eqnarray*}
  &&
  F^{x}_{f,f+1} ~=~
  F^{x}_{f+1,f} ~=~
  \frac{1}{2}~
  {\mathcal{L}}_{f},
  \\
  &&
  F^{y}_{f,f+1} ~=~
  -F^{y}_{f+1,f} ~=~
  \frac{i}{2}~
  {\mathcal{L}}_{f},
  \\
  &&
  F^{z}_{f,f} ~=~ f,
\end{eqnarray*}
where
$$
  {\mathcal{L}}_{f} ~=~
  \sqrt{(F-f)(F+1+f)}.
$$
$\hat\tau^0$ and
$\hat{F}^{0}$ are $2\times2$ and $4\times4$ identity matrices
acting in the isospin and spin spaces.
%whereas $\Delta_0$ can be either positive or negative.
\subsubsection{Mirror symmetry}
As an $8 \times 8$ matrix,
the Hamiltonian (\ref{H-4x2D-def}) is similar to a matrix that has
 two $4 \times 4$ matrices on its diagonal. Here we briefly
construct the similarity transformation matrix.
In the next section, the physical origin of this symmetry will be elaborated upon.
In brief, it is related to the mirror reflection symmetry of the Hamiltonian
with respect to the $x-y$-plane. Consequently,
the $8 \times 8$ matrix Hamiltonian commutes
with the mirror operator $\hat{M}_z$ defined as,
\begin{eqnarray}
\hat{M}_z=
-i\hat{\tau}^z\otimes\hat{R}^z,
\end{eqnarray}
where $\hat{R}^{z}$ is given by
\begin{eqnarray}
  \hat{R}^{z} =
  {\mathrm{diag}}
  \Big(
      1,~
      -1,~
      1,~
      -1
  \Big).
  \label{Rz-def}
\end{eqnarray}
Therefore, performing a unitary transformation $\hat{\mathcal U}$
\begin{eqnarray*}
  \hat{\mathcal{U}} &=&
  \frac{1}{2}~
  \Big\{
      \hat\tau^{0}
      \otimes
      \big[
          \hat{F}^{0}+
          \hat{R}^{z}
      \big]+
      \hat\tau^{x}
      \otimes
      \big[
          \hat{F}^{0}-
          \hat{R}^{z}
      \big]
  \Big\},
\end{eqnarray*}
which transforms $\hat{M}_z$ as,
$$
  \hat{\mathcal{U}}~
  \hat{M}_z~
  \hat{\mathcal{U}}^{\dag}
  ~=~
  -i~
  \hat\tau^z
  \otimes
  \hat{F}^0,
$$
one obtains a block diagonal form of the Hamiltonian
\begin{eqnarray}
  \hat{\mathcal{H}}_{\mbfk} &=&
  \hat{\mathcal{U}}~
  \hat{H}_{\mbfk}~
  \hat{\mathcal{U}}^{\dag},
\nonumber\\
&=&
\left(
    \begin{array}{cc}
      \hat{\mathcal{H}}_{\mbfk,1} &
      0
      \\
      0 &
      \hat{\mathcal{H}}_{\mbfk,\bar{1}}
    \end{array}
  \right),
  \label{H-4x2D-block}
\end{eqnarray}
with
\begin{eqnarray}
  \hat{\mathcal{H}}_{\mbfk,\eta} ~=~
  \eta
  \Delta_{\mbfk}
  \hat{R}^{z}+
  B
  \hat{F}^{z}+
  2
  \big(
      \mbfd_{\mbfk}
      \cdot
      \hat{\mathbf{F}}
  \big).
  \label{H-4D-H-4D}
\end{eqnarray}
Here $\eta=\pm 1$ denotes the eigenvalue of $i\hat{M}_z$.
Note that for high atomic spin (i.e., for $F\geq\frac{3}{2}$),
the operator $\hat{R}^{z}$ (\ref{Rz-def}) is not
a generator of the $2F+1$ dimensional irrep of the SU(2) group,
an important distinction from the model for
atoms with spin $\frac{1}{2}$. When the atomic spin is
$F=\frac{1}{2}$, the operator $\hat{R}^{z}=2\hat{F}^{z}$ is
a generator of the SU(2) group.

The $4\times 4$ matrix Hamilotonians
$\hat{\mathcal{H}}_{\mbfk,\eta}$ in the mirror
sub-sector can be analytically diagonalized.
The resultant energies depend on three
quantum numbers: The mirror quantum
number $\eta$=$\pm 1$, the band quantum
number $\xi$=$\pm 1$=c (conductance), v (valence),
and $s$=$\tfrac{1}{2},\tfrac{3}{2}$ (the positive
possible values of a pseudo spin).  Explicitly,
\begin{eqnarray}
  \veps_{\xi,s,\eta}(k) &=&
  \xi
  \sqrt{{\mathcal{A}}_{\eta}(k^2)+
        2(s-1)~{\mathcal{B}}_{\eta}(k^2)},
  \label{disp-bulk}
\end{eqnarray}
with
\begin{eqnarray}
  {\mathcal{A}}_{\eta}(k^2) &=&
  \Delta_{\mbfk}^{2}+
  \frac{5}{4}~
  B^2+
  5 d_{k}^{2}+
  B \eta \Delta_{\mbfk},
  \label{A-B-eta-k-def}
  \\
  {\mathcal{B}}_{\eta}(k^2) &=&
  \sqrt{
       \big(
           B^2+
           2 B \eta \Delta_{\mbfk}+
           4d_{k}^{2}
       \big)^{2}-
  24 B \eta \Delta_{\mbfk}
  d_{k}^{2}},
  \nonumber
\end{eqnarray}
where $d_k\equiv|\mbfd_{\mbfk}|=\hbar v k$ is the SOC
contribution defined in Eq.~(\ref{HSO-4x2D-def}).
Due to  rotation symmetry around the $z$-axis, $\hat{F}^z$
becomes a good quantum number at ${\mbfk}=0$ and
$s$ reduces to $s=|f|$ (Recall that $-F \le f \le F$ is
the bare magnetic quantum number defined after
Eq.~(\ref{basis})).
%---------------- Sp0 vs MF ---------------------------
\begin{figure}[htb]
%H=6.28, L=14.65
\centering
  \subfigure[]
  {\includegraphics[width=50 mm,angle=0]
   {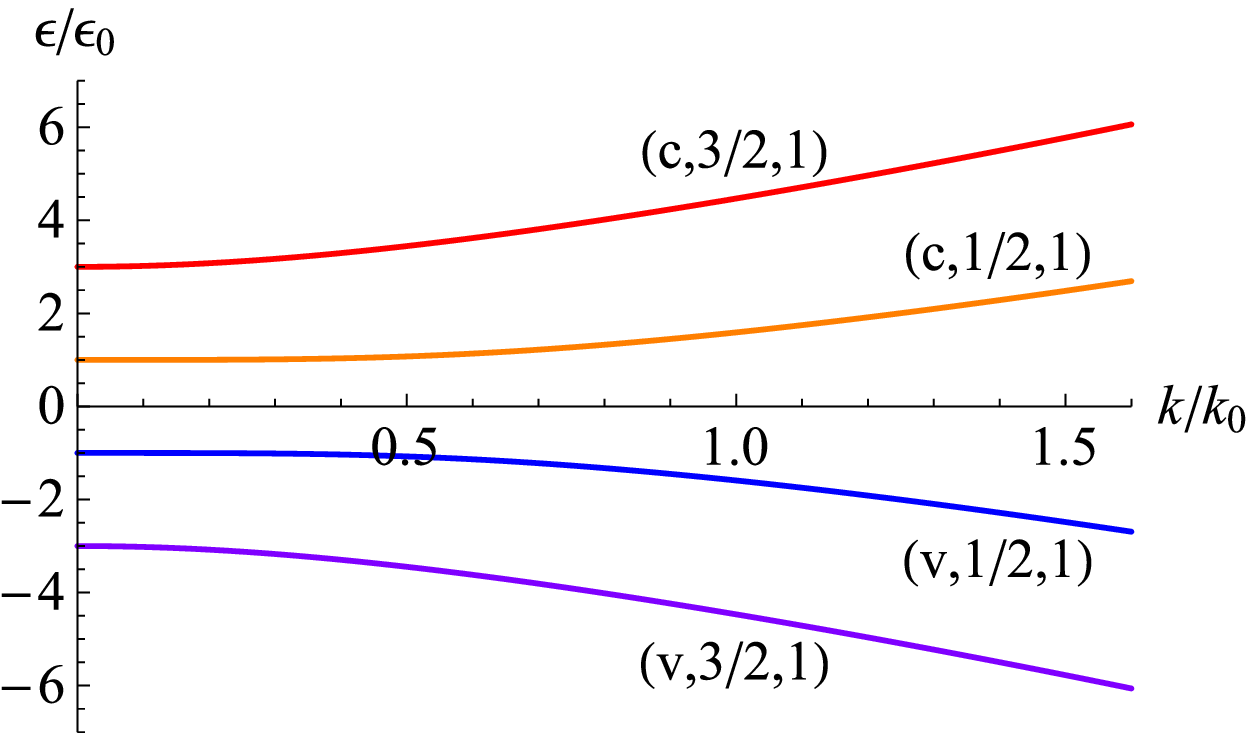}
   \label{Fig-Sp-bulk-p}}
  \subfigure[]
  {\includegraphics[width=50 mm,angle=0]
   {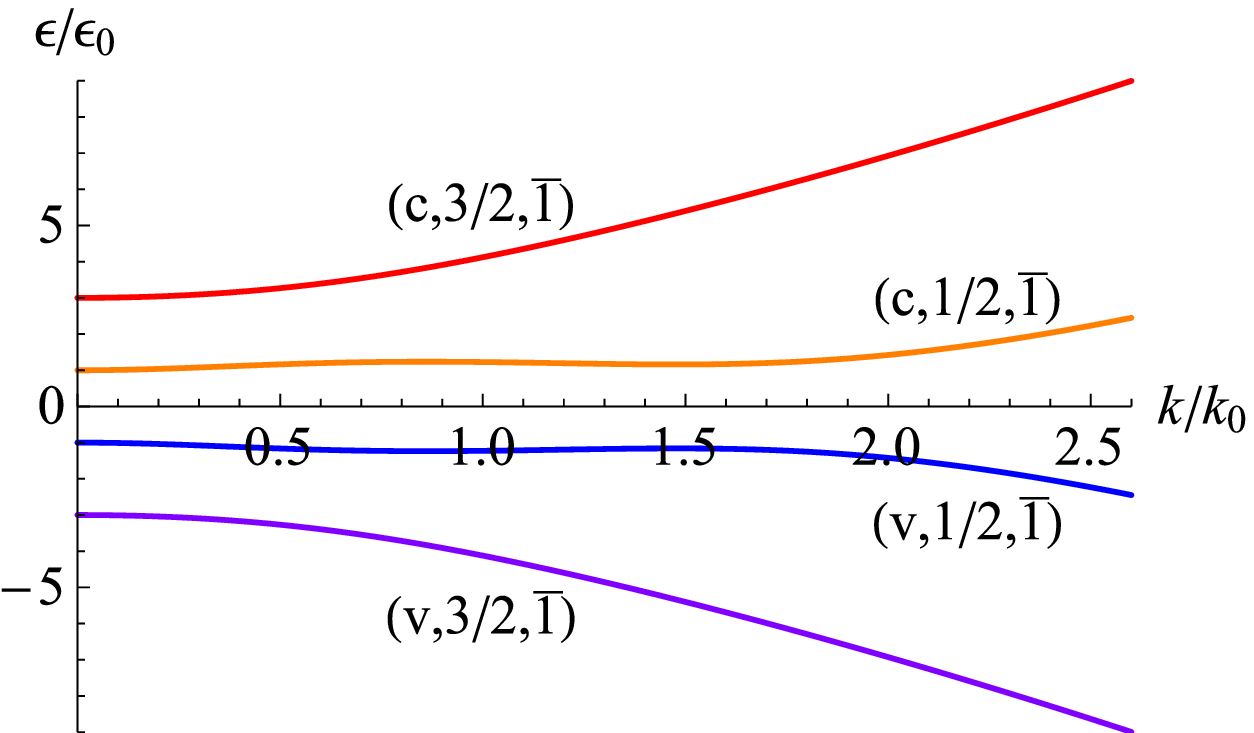}
   \label{Fig-Sp-bulk-m}}
 \caption{\footnotesize
   {\color{blue}(color online)}
   Energy dispersion (\ref{disp-bulk}) for $\eta=1$ [panel (a)]
   and $\eta=\bar{1}$ [panel (b)] for $\Delta_0=0$ and
   $B=2 \epsilon_0$.
   $\epsilon_0$ and $k_0$ are given by eq. (\ref{epsilon0-k0-def}).
  Note that for $\eta=\bar{1}$ [panel (b)], the gap has a shallow minimum
  implying a topologically non-trivial  insulator. Thus, the topological transition can be
  driven by both $\Delta_0$ and $B$.
  Here $(\xi,s,\eta)$ represent the band
  quantum number, pseudo spin, and
  the mirror quantum number for each
  band.}
 \label{Fig-Sp-bulk}
\end{figure}
The energy dispersion (\ref{disp-bulk}) of the bulk
system (without edges) is shown in Fig. \ref{Fig-Sp-bulk} for $\Delta_0=0$ and
$B=2\epsilon_0$.
The energy gap $D_{\eta}$ separating
the conduction and valence bands is,
\begin{eqnarray*}
  D_{\eta} &=&
  \veps_{{\mathrm{c}},\frac{1}{2},\eta}(0)-
  \veps_{{\mathrm{v}},\frac{1}{2},\eta}(0)
  ~=~
  2 \epsilon_0,
\end{eqnarray*}
with the notations
\begin{eqnarray}
  \epsilon_0 ~=~
  \hbar v k_0,
  \ \ \ \ \
  k_0 ~=~
  \frac{M_0 v}{\hbar}.
  \label{epsilon0-k0-def}
\end{eqnarray}

%%%%%%%%%%%%%%%Symmetries%%%%%%%%%%%%%%%

\section{Symmetry and Topology of $\hat {H}_\bfk$}
\label{Sec:Symm}
In this section we discuss some symmetries of the
$8 \times 8$ Hamiltonian $\hat {H}_\bfk$ defined through
Eqs.~(\ref{H-4x2D-def},\ref{H0-4x2D-def},\ref{HSO-4x2D-def}).
There are discrete as well as
 continuous symmetries that
endow the Hamiltonian with a rich topological
structure and enable its classification above that
implied within the Altland-Zirnbauer scheme.

For the sake of completeness we
very briefly discuss in subsection \ref{subsec-B=0}
also the case where the magnetic field is absent.
Then, in subsection \ref{subsec-Bne0} we introduce
the symmetries for the case $B \ne 0$.
In subsection \ref{subsec-Class} the topological
structure is exposed and the classification of
the Hamiltonian is identified. It is encoded by a set of 1D winding numbers
and a set of (mirror) Chern numbers. The
pattern of winding numbers is discussed in subsection
\ref{subsec-winding} and that of the mirror Chern numbers
is discussed in subsection \ref{subsec-Chern}.

%%%%%%%%%%%%%%Symmetries for B=0%%%%%%%%%%%%
\subsection{Symmetries for $B=0$}
\label{subsec-B=0}

Consider first symmetries of the system
in the absence of the external magnetic field,  that is, $B=0$.
In this case,
the system is (effectively) time-reversal invariant
\begin{eqnarray}
T
\hat{H}_{\mbfk}
T^{-1}=\hat{H}_{-{\mbfk}}, \quad
T=e^{i\pi\hat{F}^y}K,
\end{eqnarray}
where $K$ denotes the operation of complex conjugation.
In addition, there is point group symmetry $D_{2h}$ of the optical lattice,
which is generated by
inversion $P=\tau_z$.
\begin{eqnarray}
P \hat{H}_{\mbfk} P^{-1}=\hat{H}_{-{\mbfk}},
\end{eqnarray}
and two-fold rotations $C_{2i}=e^{i\hat{F}^i\pi}$ around the $i$-axis $(i=x, y, z)$,
\begin{eqnarray}
  C_{2x}
  \hat{H}_{\mbfk}
  C_{2x}^{-1}
  &=&
  \hat{H}_{(k_x, -k_y)},
  \nonumber
  \\
  C_{2y}
  \hat{H}_{\mbfk}
  C_{2y}^{-1}
  &=&
  \hat{H}_{(-k_x, k_y)},
  \label{rotation-symmetry}
  \\
  C_{2z}
  \hat{H}_{\mbfk}
  C_{2z}^{-1}
  &=&
  \hat{H}_{-\mbfk}.
  \nonumber
\end{eqnarray}
In the long wavelength approximation,
the two-fold rotations becomes the full rotation,
\begin{eqnarray}
U({\theta})
\hat{H}_{\mbfk}
U^{\dagger}({\theta})=
\hat{H}_{R({\theta}){\mbfk}},
\end{eqnarray}
where $U({\theta})=e^{i\theta\hat{F}^{z}}$ is the
rotation operator by $\theta$ and $R({\theta})$ is
the corresponding rotation matrix for $\mbfk$.
Consequently, the eigenvalues depend only on
$\mbfk^2$.

A somewhat less evident symmetry
 (referred to as chiral symmetry) reads,
\begin{eqnarray}
  \hat{\tau}^y
  \hat{H}_{\mbfk} (\hat{\tau}^{y})^{-1}=
  -\hat{H}_{\mbfk},
\end{eqnarray}
which constitutes a relation between the valence and
conduction bands.

\subsection{Symmetries for $B \ne 0$}
\label{subsec-Bne0}

Now consider the system under a finite (perpendicular) magnetic field $B$,
applied in the $z$-direction.
Inversion $P$ and two-fold rotation $C_{2z}$ around the $z$-axis are
still valid symmetries.
While the other symmetries are broken, some of their combinations
may survive:
Both time-reversal and two-fold-rotation around the $y$-axis flip
$B$, so their combination is preserved as a magnetic rotation symmetry,
\begin{eqnarray}
  &&
  [C_{2y} T]
  \hat{H}_{\mbfk}
  [C_{2y}T]^{-1}= \hat{H}_{(k_x, -k_y)},
  \label{eq:magnetic_rot}
\end{eqnarray}
which implies
\begin{eqnarray}
 \hat{ H}^*_{\mbfk}
= \hat{H}_{(k_x, -k_y)}.
\end{eqnarray}
Furthermore, the combination of chiral symmetry and $C_{2y}$ leads to the relation,
\begin{eqnarray}
[C_{2y}\hat{\tau}^y]
 \hat{H}_{\mbfk}
[C_{2y}\hat{\tau}^y]^{-1}
=-\hat{H}_{(-k_x, k_y)}.
\label{eq:chiral_rot}
\end{eqnarray}
In Eqs.(\ref{eq:magnetic_rot}) and (\ref{eq:chiral_rot}), two-fold
rotation around the $y$-axis can be
replaced with that around the $x$-axis.

By combining inversion $P$ and two-fold rotation $C_{2z}$ around the $z$-axis,
we have mirror reflection symmetry $M_z$ with respect to the $xy$-plane.
Since the Hamiltonian is independent of $k_z$, the mirror operator $M_{xy}$
commutes with the Hamiltonian.
We have already used the mirror reflection symmetry to block diagonalize the
Hamiltonian in Eq.(\ref{H-4x2D-block}).
For the resultant $4\times 4$ Hamiltonians $\hat{\cal H}_{{\mbfk},\eta}$,
the symmetries in Eqs.(\ref{eq:magnetic_rot}) and (\ref{eq:chiral_rot})
reduce to
\begin{eqnarray}
&& T'\hat {\cal H}_{k_x,k_y; \eta}(T')^{-1}=\hat {\cal H}_{k_x,-k_y; \eta},
\quad T'=K,
\nonumber\\
&&Q\hat {\cal H}_{k_x,k_y; \eta}Q^{-1}=-\hat {\cal H}_{-k_x,k_y; \eta},
\label{Transform}
\end{eqnarray}
where $Q$ is given by
\begin{eqnarray} \label{QC}
Q=\sigma_x \otimes \sigma_x=\begin{pmatrix}
0&0&0&1\\0&0&1&0\\0&1&0&0\\1&0&0&0 \end{pmatrix}.
\label{Q}
\end{eqnarray}

\subsection{Topological Structure}
\label{subsec-Class}

To check whether a symmetry transformation induces
a non-trivial topological structure, we need to confirm
that this symmetry is not broken when the system has
an edge. Note that we have to check the result of
the symmetry operation on the Hamiltonian
{\it when it is written in configuration space}, that is,
$\hat {\cal H}_\eta(x,y)$ (the Fourier transform of $\hat {\cal H}_{\bfk,\eta}$).
So far, the above symmetries have been considered in the bulk 2D system.
An important feature of the pertinent Hamiltonian is that
some symmetries are valid also when the system is opened.
In particular, mirror reflection symmetry
with respect to the $z$-axis is retained even in
the presence of an edge.
Therefore, we shall later on employ the $4\times 4$ Hamiltonians
$\hat{\cal H}_{{\mbfk},\eta}$ in the mirror subsector for
for the analysis of edge states. For definiteness, it is
assumed that the edge is normal
to the $y$-axis, and the open system is defined on
the half plane $y \geq 0$. In that case, the symmetries defined in
Eq. (\ref{Transform}) are not broken.
Moreover, using rotation symmetry around the $z$-axis,
we also have similar symmetries consistent with
a boundary normal to the $x$-axis.

Now we specify possible topological phases in
$\hat{\cal H}_{{\mbfk},\eta}$ by employing some techniques borrowed from $K$-theory.
The idea is to deform the Hamiltonian $\hat{\cal H}_{{\mbfk},\eta}$
into a Dirac form Hamiltonian, while
keeping the symmetries in Eq. (\ref{Transform}) intact and
assuring that the system remains gapped throughout the deformation.
Explicitly, for a given $\eta$ we consider the deformation,
\begin{eqnarray}
  &&
\hat{\cal H}_{{\mbfk},\eta} \to  \hat{\cal H}_{{\mbfk}}^{{\rm Dirac}} =
  k_x\gamma_x+
  k_y\gamma_y+
  m\gamma_z,
  \nonumber
  \\
  &&
  \big\{
      \gamma_i,
      \gamma_j
  \big\}
  ~=~
  2\delta_{i,j},
  \quad (i,j=x,y,z).
  \label{HDirac}
\end{eqnarray}
Since $\hat{\cal H}^{\rm Dirac}_{\mbfk}$ is adiabatically
connected to the original Hamiltonian $\hat{\cal H}_{{\mbfk},\eta}$,
we can identify candidate topological phases in
$\hat{\cal H}_{{\mbfk},\eta}$ by examining them in
$\hat{\cal H}^{\rm Dirac}_{\mbfk}$.
A possible topological phase in the latter
%$\hat{\cal H}^{\rm Dirac}_{\mbfk}$
can be specified when it contains a mass term.
% in $\hat{\cal H}^{\rm Dirac}_{\mbfk}$.
To identify such mass term, we examine algebraic
structures required by symmetry.

The symmetries specified in
Eq. (\ref{Transform}) impose the following relations on
the gamma matrices $\{ \gamma_i \}$,
\begin{eqnarray}
&&[T', \gamma_x]=[T', \gamma_z]=[Q, \gamma_x]=[T', Q]=0,
\nonumber\\
&&\{T', \gamma_y\}=\{Q, \gamma_y\}=\{Q, \gamma_z\}=0,
\nonumber\\
&&T'{}^2=Q^2=1,
\quad
\{T', i\}=0,
\quad
[Q, i]=0,
\label{eq3:relation}
\end{eqnarray}
where the last two equations are due to the anti-unitarity of $T'$ and the unitarity of $Q$.

Let $Cl_{p, q}$ denote the Clifford algebra $\{e_1, \dots. e_p, e_{p+1}, \dots, e_{p+q}\}$
obeying
$\{e_i, e_j\}=0$ ($i\neq j$), $e^2_i=-1$ ($i=1, \dots, p$),
and $e^2_{p+i}=1$ ($i=1, \dots, q$).
With the following identification,
\begin{eqnarray}
&& e_1=i\gamma_x,
\quad
e_2=i\gamma_z,
\quad
e_3=iT',
\nonumber \\
&& e_4=T',
\quad
e_5=\gamma_y,
\quad
S=\gamma_x Q.
\label{Clifford2}
\end{eqnarray}
Eq. (\ref{eq3:relation}) is commensurate with the Clifford algebra $Cl_{2,3}$ $\{e_1, \dots, e_5\}$ with
the commutating operator $S$
\begin{eqnarray}
[e_i, S]=0 \quad (i=1,\dots, 5), \quad S^2=1.
\label{eq3:S}
\end{eqnarray}
On the other hand, if the mass term
$m\gamma_z$ is absent, we encounter the $Cl_{1,3}$
algebra with $S$, since $e_2$ is missing.
Therefore, specifying a possible mass term of
$\hat{\cal H}^{\rm Dirac}_{\mbfk}$
is identical to specifying a possible extension of
the Clifford algebra from $Cl_{1,3}$ to $Cl_{2,3}$
with $S$. From the $K$-theory, the latter extension
is found to define the classification space
\cite{Shizoaki,Morimoto} $R_0\times R_0$,
which is endowed with topologically distinct subspaces characterized
by $\pi_0(R_0\times R_0)={\bm Z}\times {\bm Z}$.
Therefore, we have topologically distinct mass
terms in $\hat{\cal H}^{\rm Dirac}_{\mbfk}$, and
correspondingly, topologically different phases
with a ${\bm Z}\times {\bm Z}$ number.
For the Hamiltonian $\hat{\cal H}_{{\mbfk},\eta}$,
the ${\bm Z}\times {\bm Z}$ number is given by
the pair $(C_{\eta}, w_{\eta})$ composed of
 a mirror Chern
number $C_{\eta}$ and a one-dimensional
mirror winding number $w_{\eta}$. These two numbers are defined
below, where it is also shown that the parities of $C_{\eta}$ and $w_{\eta}$ coincide, that is,
$(-1)^{C_{\eta}}=(-1)^{w_{\eta}}$.

%%%%%%%%%%%%%%%%%%%%
\subsection{Mirror Chern Numbers}
  \label{subsec-Chern}

The mirror Chern numbers $\{ C_{\eta} \}$ are  defined (in the standard  way)
in terms of the eigenfunctions $\{ \psi_{\xi,s,\eta}(\bfk) \} $ that obey the Schr\"odinger equation {\it in the bulk system}\cite{TI-Chern-RepProgPhys-14},
\begin{equation} \label{SEbulk}
\hat{\cal H}_{{\bf \bfk},\eta} \psi_{\xi,s,\eta}(\bfk)=\veps_{\xi,s,\eta}(k) \psi_{\xi,s,\eta}(k).
\end{equation}
Recall that the "isospin"  quantum number $\xi=$c,(v) refers
to the conduction (valence) band, $s=\tfrac{1}{2}, \tfrac{3}{2}$ is a pseudo-spin quantum number and $\eta$ is the block number. Explicitly,
\begin{eqnarray}
  C_{\eta} =
  \frac{1}{2\pi}
  \sum_{s}
  \iint
  F_{{\mathrm{v}},s,\eta}(\mbfk)~
  d^2\mbfk.
  \label{Chern-number-def}
\end{eqnarray}
Here the Berry curvature is
\begin{eqnarray}
  F_{{\mathrm{v}},s,\eta}(\mbfk) =
  i~
  \big(
      {\mathbf{e}}_{z}
      \cdot
      \boldsymbol\Omega_{{\mathrm{v}},s,\eta}
  \big),
  \label{Berry-curv-def}
\end{eqnarray}
where
\begin{eqnarray}
  &&
  \boldsymbol\Omega_{{\mathrm{v}},s,\eta}
  =
  \nabla_{\mbfk}
  \times
  \big\langle
      \psi_{{\mathrm{v}},s,\eta}(\mbfk)
  \big|
      \vec\psi_{{\mathrm{v}},s,\eta}(\mbfk)
  \big\rangle,
  \label{Omega-def}
  \\
  &&
  \big|
      \vec\psi_{{\mathrm{v}},s,\eta}(\mbfk)
  \big\rangle
  =
  \nabla_{\mbfk}
  \big|
      \psi_{{\mathrm{v}},s,\eta}(\mbfk)
  \big\rangle.
  \label{vec-psi-def}
\end{eqnarray}

\begin{figure}[htb]
%H=6.28, L=14.65
\centering
  \includegraphics[width=60 mm,angle=0]
   {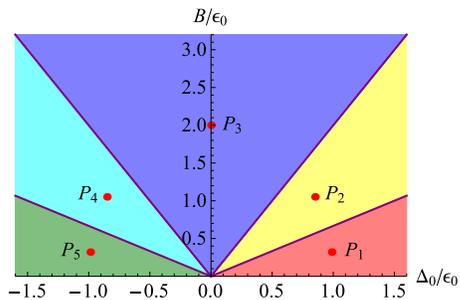}
 \caption{\footnotesize
   {\color{blue}(color online)}
   Relevant domains in the half-plane
   $\Delta_0$-$B$ with $B>0$, see details in the text.
   The purple lines are $B=\pm2\Delta_0$ and
   $B=\pm\frac{2}{3}\Delta_0$.
   $\epsilon_0$ is given by eq. (\ref{epsilon0-k0-def}).
   We illustrate formation of the edge states for
   the points $P_1$, $P_2$, $P_3$, $P_4$ and $P_5$.}
 \label{Fig-intervals-MF}
\end{figure}

As shown
in Appendix \ref{append-Chern}, there are five relevant domains
in the half-plane $\Delta_0$-$B$ with $B>0$, illustrated in
Fig. \ref{Fig-intervals-MF}. We derive the mirror Chern numbers
for each of the domains.
%---------------- domains Delta0-B ----------------
Explicitly, the mirror Chern numbers $C_{\eta}$ for
the domains (1) -- (5) are:
\begin{itemize}
\item[(1)] $C_{1}=C_{\bar{1}}=0$.

\item[(2)] $C_{1}=0$ and $C_{\bar{1}}=3$.

\item[(3)] $C_{1}=1$ and $C_{\bar{1}}=3$.

\item[(4)] $C_{1}=1$ and $C_{\bar{1}}=2$.

\item[(5)] $C_{1}=-2$ and $C_{\bar{1}}=2$.
\end{itemize}
These results indicate that the topological
phases are driven by both $\Delta_0$ and $B$.
For $B<\tfrac{2}{3} \Delta_0$ the mirror Chern numbers
occur only for $\Delta_0<0$, but for $B > \tfrac{2}{3} \Delta_0$
there are non-zero mirror Chern numbers also for $\Delta_0>0$.
From the bulk-boundary correspondence,
when $C_{\eta}$ is nonzero,
$\hat{\cal H}_{k_x, k_y;\eta}$
should host $|C_{\eta}|$ chiral edge modes on a boundary normal to
the $y$-axis. This result will be confirmed in the next section.

%%%%%%%%%%%%%%%%%%%%
\subsection{Mirror Winding Numbers}
  \label{subsec-winding}

The one-dimensional mirror winding number $w_{\eta}$ is defined by
\begin{eqnarray}
w_{\eta}=\frac{i}{4\pi}\int dk_y {\rm tr}
\left[Q \hat{\cal H}_{0,k_y;\eta}^{-1}
\partial_{k_y} \hat{\cal H}_{0,k_y;\eta}\right].
\end{eqnarray}
This number can be calculated analytically.
Like the mirror Chern numbers, their values
depend on the pertinent regions in the half plane
$\Delta_0$-$B$
with $B>0$ shown in Fig. \ref{Fig-intervals-MF}.
Thus, the mirror winding numbers for the domains (1) -- (5) are:
\begin{itemize}
\item[(1)] $w_1=w_{\bar{1}}=0$.

\item[(2)] $w_1=0$ and $w_{\bar{1}}=1$.

\item[(3)] $w_1=w_{\bar{1}}=1$.

\item[(4)] $w_1=1$ and $w_{\bar{1}}=0$.

\item[(5)] $w_1=w_{\bar{1}}=0$.
\end{itemize}
The generalized index theorem implies that when
$w_\eta$ is nonzero, there exist $|w_\eta|$
zero modes with $k_x=0$ on a boundary normal to
the $y$-axis.\cite{Sato11}  As will be confirmed below,
these zero modes are realized as the $k_x=0$ part of
chiral edge states.
%%%%%%%%%%%%%%%%%%%%%%%%%%%%%%
\section{Edge States}
  \label{sec-edge}
Having defined and exposed the mirror Chern and winding numbers in the bulk,
and their values in the five domains in the $(B, \Delta_0)$
plane, the next task is to analyze the corresponding
pattern of edge states, whose existence is a consequence
of the bulk-edge correspondence scenario. For this purpose, it is convenient to consider a system with single
edge, such as a 2D half plane $-\infty < x < \infty, \ \ 0 \le y < \infty$. After deriving and
solving the equations for the edge state wave functions
in subsection \ref{subsec-Edge-Equations}, we expose
their dispersion curves in the five domains specified in
Fig.\ref{Fig-intervals-MF}.

\subsection{Equations for the Edge States}
\label{subsec-Edge-Equations}

Let us then consider an optical lattice occupying the half plane $y>0$.
In this case, $k_x$ is still a good quantum number, but $k_y$ is
not. Rather, it becomes an operator $k_y \to -i \partial_y$.
The edge states wave functions
are the solutions of the Schr\"odinger equation,
\begin{eqnarray}
  \hat{\mathcal{H}}_{(k_x,-i \partial_y),\eta}~
      \Psi_{k_x,\eta}(y)
  &=&
  \veps_{\eta}({k_x})~
      \Psi_{k_x,\eta}(y),
  \label{eq-Schrodinger-edge}
\end{eqnarray}
that satisfy
the boundary conditions,
\begin{subequations}
\begin{eqnarray}
  &&
      \Psi_{k_x,\eta}(0)
  ~=~ 0,
  \label{bound-cond-0}
  \\
  &&
  \lim_{y\to\infty}
      \Psi_{k_x,\eta}(y)
  ~=~ 0, \ \ \mbox{(exponential decay)}.
  \label{bound-cond-infty}
\end{eqnarray}
  \label{subeqs-bound-cond}
\end{subequations}
The general solution of eq. (\ref{eq-Schrodinger-edge})
that satisfies the boundary condition at $y \to \infty$
(but {\it not} the boundary condition at $y=0$) is
a four component (pseudo) spinor,
\begin{eqnarray}
      \Psi_{k_x,\eta}(y)
  &=&
  \left(
    \begin{array}{c}
      \chi_{\frac{3}{2}} \\
      \chi_{\frac{1}{2}} \\
      \chi_{\frac{\bar{1}}{2}} \\
      \chi_{\frac{\bar{3}}{2}}
    \end{array}
  \right)~
  e^{-\kappa y},
  \label{sol-gen-Schrodinger-edge}
\end{eqnarray}
where the components $\{ \chi_f \}$ and the exponent $\kappa$ are (generically complex) constants.
The boundary condition (\ref{bound-cond-infty}) requires
${\mathrm{Re}}(\kappa)>0$. Substituting eq.
(\ref{sol-gen-Schrodinger-edge}) into eq.
(\ref{eq-Schrodinger-edge}), we get the following set of
equations,
\begin{subequations}
\begin{eqnarray}
  \Big(
      \veps_{\frac{3}{2},\eta}^{(0)}(k_x,i\kappa)-
      \veps_{\eta}(k_x)
  \Big)~
  \chi_{\frac{3}{2}}+
  \ \ \ \ \
  \nonumber \\ +~
  h_{\frac{3}{2},\frac{1}{2}}(k_x,i\kappa)~
  \chi_{\frac{1}{2}}
  ~=~ 0,
  \label{eq1-edge-states}
  %%%%%%%%%%
  \\
  \nonumber
  \\
  \Big(
      \veps_{\frac{1}{2},\eta}^{(0)}(k_x,i\kappa)-
      \veps_{\eta}(k_x)
  \Big)~
  \chi_{\frac{1}{2}}+
  \ \ \ \ \
  \nonumber \\ +~
  h_{\frac{1}{2},\frac{3}{2}}(k_x,i\kappa)~
  \chi_{\frac{3}{2}}+
  \ \ \ \ \
  \nonumber \\ +~
  h_{\frac{1}{2},\frac{\bar{1}}{2}}(k_x,i\kappa)~
  \chi_{\frac{\bar{1}}{2}} ~=~ 0,
  \label{eq2-edge-states}
  %%%%%%%%%%
\end{eqnarray}
\begin{eqnarray}
  \Big(
      \veps_{\frac{\bar{1}}{2},\eta}^{(0)}(k_x,i\kappa)-
      \veps_{\eta}(k_x)
  \Big)~
  \chi_{\frac{\bar{1}}{2}}+
  \ \ \ \ \
  \nonumber \\ +~
  h_{\frac{\bar{1}}{2},\frac{1}{2}}(k_x,i\kappa)~
  \chi_{\frac{1}{2}}+
  \ \ \ \ \
  \nonumber \\ +~
  h_{\frac{\bar{1}}{2},\frac{\bar{3}}{2}}(k_x,i\kappa)~
  \chi_{\frac{\bar{3}}{2}} ~=~ 0,
  \label{eq3-edge-states}
  %%%%%%%%%%
  \\
  \nonumber
  \\
  \Big(
      \veps_{\frac{\bar{3}}{2},\eta}^{(0)}(k_x,i\kappa)-
      \veps_{\eta}(k_x)
  \Big)~
  \chi_{\frac{\bar{3}}{2}}+
  \ \ \ \ \
  \nonumber \\ +~
  h_{\frac{\bar{3}}{2},\frac{\bar{1}}{2}}(k_x,i\kappa)~
  \chi_{\frac{\bar{1}}{2}}
  ~=~ 0.
  \label{eq4-edge-states}
\end{eqnarray}
  \label{subeqs-edge-states}
\end{subequations}
Here $\veps_{f,\eta}^{(0)}(k_x,i\kappa)$ are,
\begin{eqnarray}
  \veps_{f,\eta} &=&
  f B+
  \big(
      -1
  \big)^{F-f}~
  \eta \Delta_{k_x,i\kappa},
  \label{e0-def}
\end{eqnarray}
where $F=\frac{3}{2}$,
$f=-\tfrac{3}{2},-\tfrac{1}{2},\tfrac{1}{2},\tfrac{3}{2}$, and
\begin{eqnarray}
  &&
  \Delta_{k_x,i\kappa} =
  M_0 v^2+
  \frac{\hbar^2 k_{x}^{2}}{2 M_0}-
  \frac{\hbar^2 \kappa^2}{2 M_0}.
  \label{Delta-kx-ikappa}
\end{eqnarray}
The nontrivial matrix elements $h_{f,f'}(k_x,i\kappa)$ are
\begin{eqnarray}
  &&
  h_{\frac{3}{2},\frac{1}{2}}(k_x,i\kappa)
  ~=~
  h_{\frac{\bar{1}}{2},\frac{\bar{3}}{2}}(k_x,i\kappa)
  ~=~
  \sqrt{3}~
  A~
  k^{-},
  \nonumber
  \\
  &&
  h_{\frac{1}{2},\frac{3}{2}}(k_x,i\kappa)
  ~=~
  h_{\frac{\bar{3}}{2},\frac{\bar{1}}{2}}(k_x,i\kappa)
  ~=~
  \sqrt{3}~
  A~
  k^{+},
  \label{H-nondiag}
  \\
  &&
  h_{\frac{1}{2},\frac{\bar{1}}{2}}(k_x,i\kappa)
  ~=~
  2 A k^{-},
  \ \ \
  h_{\frac{\bar{1}}{2},\frac{1}{2}}(k_x,i\kappa)
  ~=~
  2 A k^{+},
  \nonumber
\end{eqnarray}
where
\begin{eqnarray}
  k^{\pm} ~=~
  k_x
  \mp
  \kappa.
  \label{d-pm-def}
\end{eqnarray}
The set of equations (\ref{subeqs-edge-states}) has
nontrivial solutions when the corresponding determinant vanishes, that is,
\begin{eqnarray*}
  \det
  \big(
      \hat{\mathcal{H}}_{{k_x,i\kappa},\eta}-
      \veps_{\eta}(k_x)
      \hat{F}^{0}
  \big)
  &=& 0,
\end{eqnarray*}
where $\hat{F}^{0}$ is the $4\times4$ unit matrix. This equality yields the following equation for $\kappa$,
\begin{eqnarray}
  \Big(
      \veps^{2}_{\eta}(k_x)-
      {\mathcal{A}}_{\eta}\big(k_{x}^{2}-\kappa^2\big)
  \Big)^{2}-
  {\mathcal{B}}_{\eta}^{2}\big(k_{x}^{2}-\kappa^2\big)
  = 0,
  \label{eq-for-kappa}
\end{eqnarray}
where ${\mathcal{A}}_{\eta}(k^2)$ and ${\mathcal{B}}_{\eta}(k^2)$
are given by eq. (\ref{A-B-eta-k-def}).
Taking into account that both
${\mathcal{A}}_{\eta}(k_{x}^{2}-\kappa^2)$ and
${\mathcal{B}}_{\eta}(k_{x}^{2}-\kappa^2)$ are second
order polynomials of $\kappa^2$, implies that
eq. (\ref{eq-for-kappa}) is a quartic  equation for
$\kappa^2$. Consequently, it yields four solutions
$\{ \kappa_n, n=1,2,3,4 \}$ with
${\mathrm{Re}}(\kappa_n)>0$. These solutions are retained, as they
satisfy the boundary conditions
(\ref{bound-cond-infty}). The other four solutions
with ${\mathrm{Re}}(\kappa)<0$ diverge as $y \to \infty$.

The remaining task is to form combinations of
these basic solutions that satisfy also the boundary
condition at $y=0$. Knowing the four exponent
$\{ \kappa_n\}$ [the four solutions of eq. (\ref{eq-for-kappa})],
we can express the constants $\{ \chi_{f}^{(n)}\}$ in terms
of normalization constants $N_n$,
\begin{eqnarray}
  \chi_{f}^{(n)} &=&
  \big(-1\big)^{F-f}~
  {\mathcal{D}}_{f}^{(n)}~
  N_n,
  \label{Csn-vs-Nn}
\end{eqnarray}
where $F=\frac{3}{2}$ and,
\begin{eqnarray}
  {\mathcal{D}}_{f}^{(n)}
  &=&
  \det
  \Big(
      \hat{\mathcal{M}}_{f,\eta}(k_x,i\kappa_n)
  \Big).
  \label{D-s-n=det}
\end{eqnarray}
The $3\times3$ matrices
$\hat{\mathcal{M}}_{f,\eta}(k_x,i\kappa_n)$ are,
\begin{widetext}
\begin{subequations}
\begin{eqnarray}
  \hat{\mathcal{M}}_{\frac{3}{2},\eta}(k_x,i\kappa_n)
  &=&
  \left(
    \begin{array}{ccc}
      h_{\frac{3}{2},\frac{1}{2}}(k_x,i\kappa_n)
      &
      0
      &
      0
      \\
      \veps_{\frac{1}{2},\eta}(k_x,i\kappa_n)-
      \veps_{\eta}(k_x)
      &
      h_{\frac{1}{2},\frac{\bar{1}}{2}}(k_x,i\kappa_n)
      &
      0
      \\
      h_{\frac{\bar{1}}{2},\frac{1}{2}}(k_x,i\kappa_n)
      &
      \veps_{\frac{\bar{1}}{2},\eta}(k_x,i\kappa_n)-
      \veps_{\eta}(k_x)
      &
      h_{\frac{\bar{\bar{1}}}{2},\frac{\bar{3}}{2}}(k_x,i\kappa_n)
    \end{array}
  \right),
  \label{M1-3x3}
  \\
  %%%%%%%%%%
  \hat{\mathcal{M}}_{\frac{1}{2},\eta}(k_x,i\kappa_n)
  &=&
  \left(
    \begin{array}{ccc}
      \veps_{\frac{3}{2},\eta}(k_x,i\kappa_n)-
      \veps_{\eta}(k_x)
      &
      0
      &
      0
      \\
      h_{\frac{1}{2},\frac{3}{2}}(k_x,i\kappa_n)
      &
      h_{\frac{1}{2},\frac{\bar{1}}{2}}(k_x,i\kappa_n)
      &
      0
      \\
      0
      &
      \veps_{\frac{\bar{1}}{2},\eta}(k_x,i\kappa_n)-
      \veps_{\eta}(k_x)
      &
      h_{\frac{\bar{\bar{1}}}{2},\frac{\bar{3}}{2}}(k_x,i\kappa_n)
    \end{array}
  \right),
  \label{M2-3x3}
  \\
  %%%%%%%%%%
  \hat{\mathcal{M}}_{\frac{\bar{1}}{2},\eta}(k_x,i\kappa_n)
  &=&
  \left(
    \begin{array}{ccc}
      h_{\frac{3}{2},\frac{1}{2}}(k_x,i\kappa_n)
      &
      \veps_{\frac{3}{2},\eta}(k_x,i\kappa_n)-
      \veps_{\eta}(k_x)
      &
      0
      \\
      \veps_{\frac{1}{2},\eta}(k_x,i\kappa_n)-
      \veps_{\eta}(k_x)
      &
      h_{\frac{1}{2},\frac{3}{2}}(k_x,i\kappa_n)
      &
      0
      \\
      h_{\frac{\bar{1}}{2},\frac{1}{2}}(k_x,i\kappa_n)
      &
      0
      &
      h_{\frac{\bar{\bar{1}}}{2},\frac{\bar{3}}{2}}(k_x,i\kappa_n)
    \end{array}
  \right),
  \label{M3-3x3}
  \\
  %%%%%%%%%%
  \hat{\mathcal{M}}_{\frac{\bar{3}}{2},\eta}(k_x,i\kappa_n)
  &=&
  \left(
    \begin{array}{ccc}
      h_{\frac{3}{2},\frac{1}{2}}(k_x,i\kappa_n)
      &
      0
      &
      \veps_{\frac{3}{2},\eta}(k_x,i\kappa_n)-
      \veps_{\eta}(k_x)
      \\
      \veps_{\frac{1}{2},\eta}(k_x,i\kappa_n)-
      \veps_{\eta}(k_x)
      &
      h_{\frac{1}{2},\frac{\bar{1}}{2}}(k_x,i\kappa_n)
      &
      h_{\frac{1}{2},\frac{3}{2}}(k_x,i\kappa_n)
      \\
      h_{\frac{\bar{1}}{2},\frac{1}{2}}(k_x,i\kappa_n)
      &
      \veps_{\frac{\bar{1}}{2},\eta}(k_x,i\kappa_n)-
      \veps_{\eta}(k_x)
      &
      0
    \end{array}
  \right).
  \label{M4-3x3}
\end{eqnarray}
  \label{subeqs-M-3x3}
\end{subequations}
\end{widetext}
The wave function describing the edge state is a combination,
\begin{eqnarray}
  \Psi_{k_x,\eta}(y)
  &=&
  \sum_{n=1}^{4}
  {\mathcal{N}}_{n}~
  \hat\chi_n~
  e^{-\kappa_n y},
  \label{psi-edge-res}
\end{eqnarray}
where
$$
  \hat\chi_n ~=~
  \left(
      \chi_{\frac{3}{2}}^{(n)},
      \chi_{\frac{1}{2}}^{(n)},
      \chi_{\frac{\bar{1}}{2}}^{(n)},
      \chi_{\frac{\bar{3}}{2}}^{(n)}
  \right)^T,
$$
while $\chi_{f}^{(n)}$ is given by eq. (\ref{Csn-vs-Nn}).
The boundary condition (\ref{bound-cond-0}) implies,
\begin{eqnarray}
  \sum_{n=1}^{4}
  \chi_{f}^{(n)}~
  {\mathcal{N}}_n
  &=& 0.
  \label{eqs-set-for-energy-y=0}
\end{eqnarray}
This is a system of linear homogeneous
equations for ${\mathcal{N}}_n$.
[Recall that $f=\pm\frac{1}{2}$,
$\pm\frac{3}{2}$, and $n=1,2,3,4$,
so that the matrix $\|\chi_{f}^{(n)}\|$
is a square $4\times4$ matrix.]
A nontrivial solution of
Eq. (\ref{eqs-set-for-energy-y=0}) obtains
when
\begin{eqnarray}
  {\mathcal{F}}_{\eta}
  \big(\veps_{\eta}(k_x)\big)
  &=&
  \det
  \Big\|
      \chi_{f}^{(n)}
  \Big\| =0.
  \label{eq-for-energy}
\end{eqnarray}
Since the matrix elements $\chi_{f}^{(n)}$ are functions of
the energy $\veps_{\eta}(k_x)$, eq.
(\ref{eq-for-energy}) yields the energies of the (topological)
edge states (provided they exist). As we shall see below,
there may be several solutions, denoted as $\veps_{\eta, a}(k_x)$.

In order to compute the energies for which solutions of eq. (\ref{eq-for-energy}) exist,
we apply
a special numerical technique that is
developed and explained in
Appendix \ref{append-num-calc}.
The edge state spectrum will be displayed
below for $\eta=1$ and $\eta=-1$ at the five
points $P_i, \ i=1,2,3,4,5$ in the ($B>0$,
$\Delta_0$) half plane, representing the five
domains  as marked in Fig.\ref{Fig-intervals-MF}.

Since the atoms are neutral, the magnetic field acts only on the spin, so, unlike the electronic
version, the direction  of propagation of the edge states is not determined by the Lorenz force,
that has a classical origin. In other words, when TRS is broken solely through the Zeeman effect,
the chirality cannot be predicted a-priory. In fact, we shall see that there is a scenario where
edge states propagate in {\it both} directions. To distinguish between them we refer to the states
with negative (averaged) group velocity as {\it chiral} whereas states with
positive (averaged) group velocity are referred to as
{\it anti-chiral}. The bulk-edge correspondence implies that the positive
mirror Chern numbers are equal to the number of chiral states and the negative
mirror Chern numbers are equal to the
number of anti-chiral states.

%%%%%%%%%%%%%%%%%%%%
\subsection{Domain (1): $(C_1,w_1)=(C_{\bar{1}},w_{\bar{1}})=(0,0)$}
\label{domain-1}

%---------------- Sp 1st ---------------------------
\begin{figure}[htb]
%H=6.28, L=14.65
\centering
  \subfigure[]
  {\includegraphics[width=40 mm,angle=0]
   {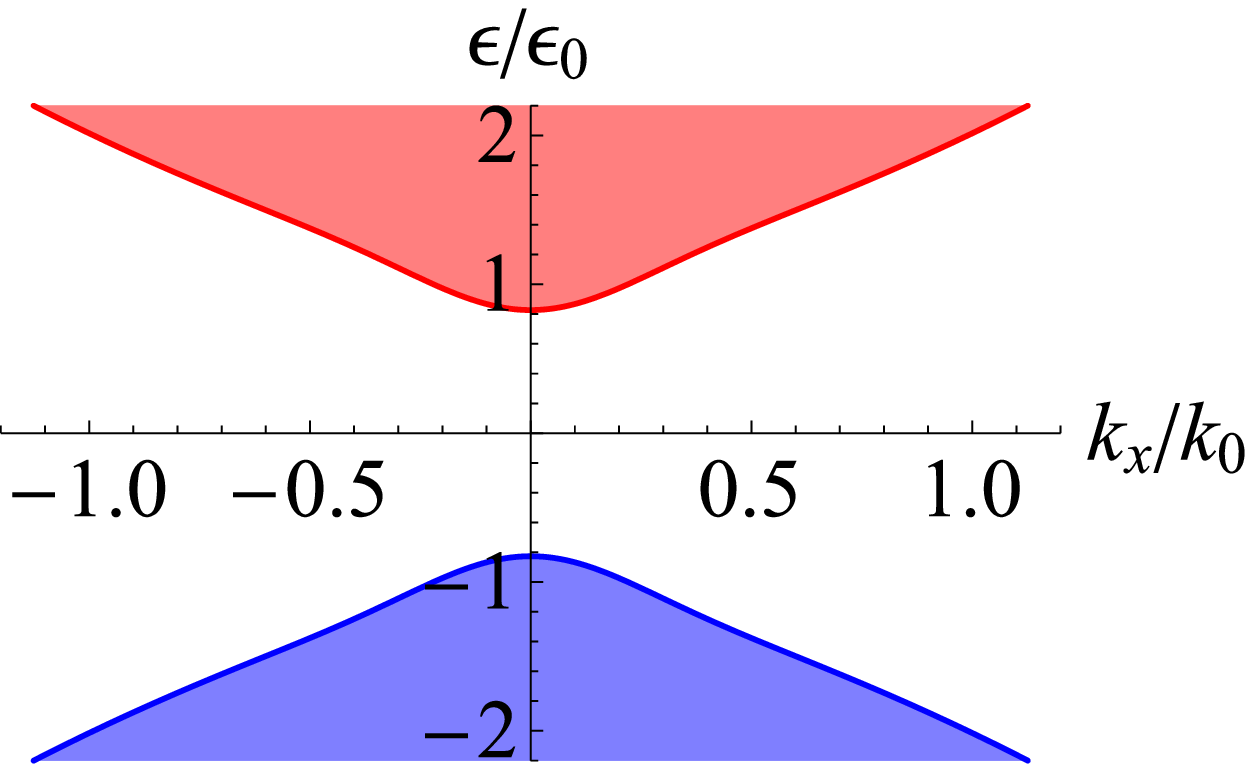}
   \label{Fig-sp-p-1st}}
  \subfigure[]
  {\includegraphics[width=40 mm,angle=0]
   {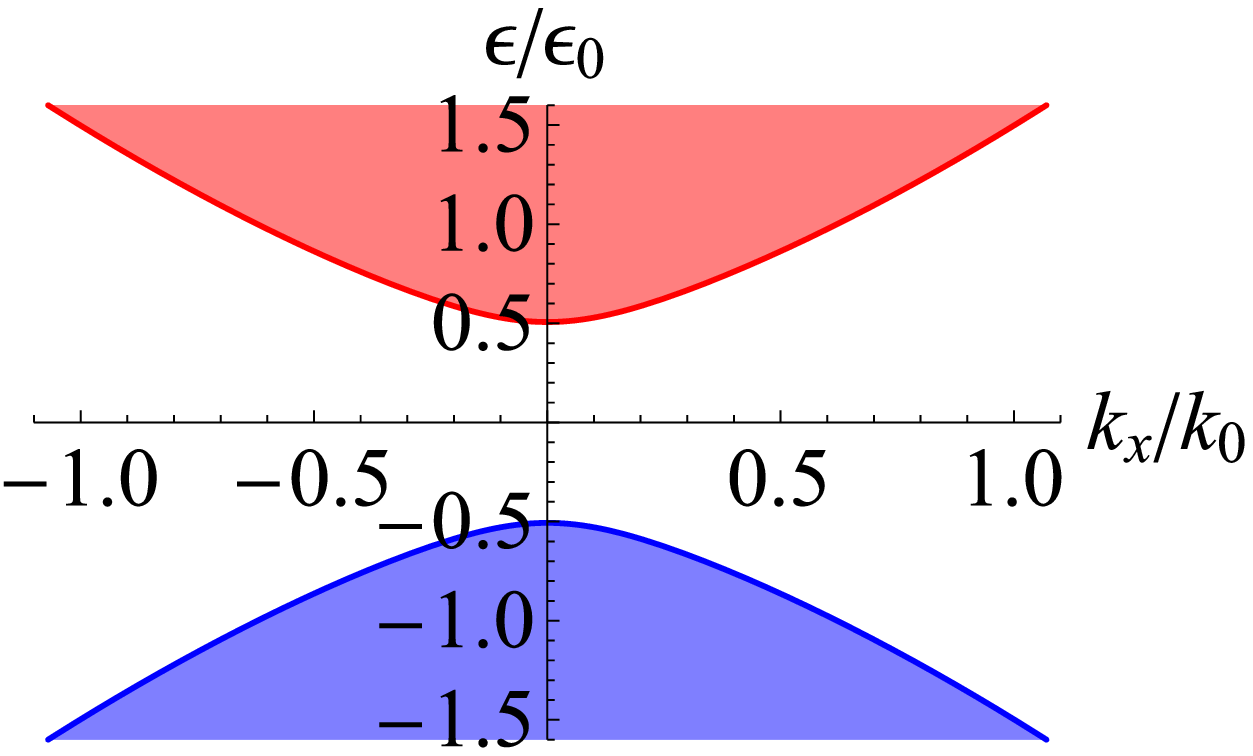}
   \label{Fig-sp-m-1st}}
 \caption{\footnotesize
   {\color{blue}(color online)}
   Energy dispersion for $\eta=1$ [panel (a)]
   and $\eta=\bar{1}$ [panel (b)]. For both panels:
   $\Delta_0=0.987087\epsilon_0$
   and $B=0.320364\epsilon_0$,
   corresponding to point $P_1$ in Fig. \ref{Fig-intervals-MF}.
   The red and blue areas denote the conduction and
   valence bands, eq. (\ref{disp-bulk}).
   $\epsilon_0$ and $k_0$ are given by eq. (\ref{epsilon0-k0-def}).
   }
 \label{Fig-sp-1st}
\end{figure}
%\end{document}
The energy dispersion for domain (1) is displayed in Fig. \ref{Fig-sp-1st}
for $\Delta_0=0.987087\epsilon_0$ and $B=0.320364\epsilon_0$
[point $P_1$ in Fig. \ref{Fig-intervals-MF}].
For this domain, the mirror Chern and winding numbers are zero
and there are no (gap closing) edge states.
The reason is clear: For $\Delta_0>0$ the topological
phase driven solely by the magnetic field, but in
the first domain, the magnetic field is too weak.

%%%%%%%%%%%%%%%%%%%%
\subsection{Domain (2): $(C_1,w_1)=(0,0), (C_{\bar{1}},w_{\bar{1}})=(3,1)$ }
\label{domain-2}

%---------------- Sp 2nd ---------------------------
\begin{figure}[htb]
%H=6.28, L=14.65
\centering
  \subfigure[]
  {\includegraphics[width=40 mm,angle=0]
   {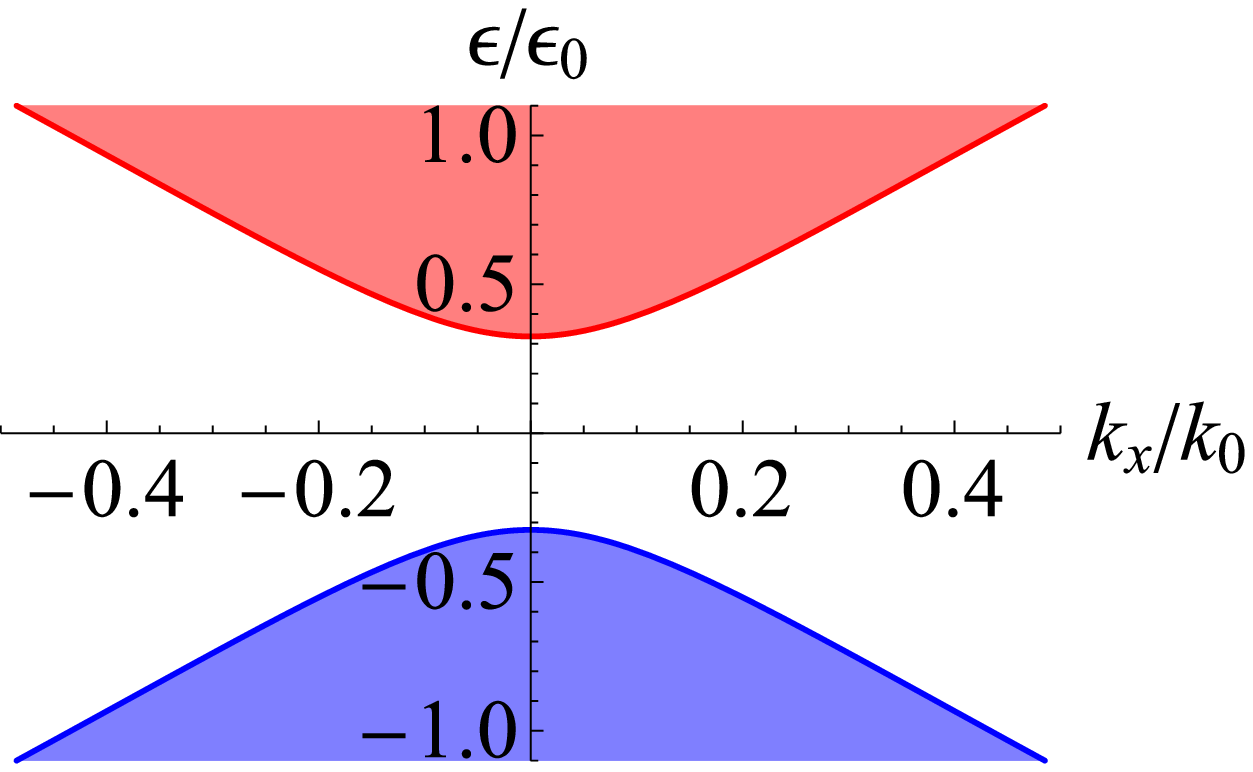}
   \label{Fig-sp-p-2nd}}
  \subfigure[]
  {\includegraphics[width=40 mm,angle=0]
   {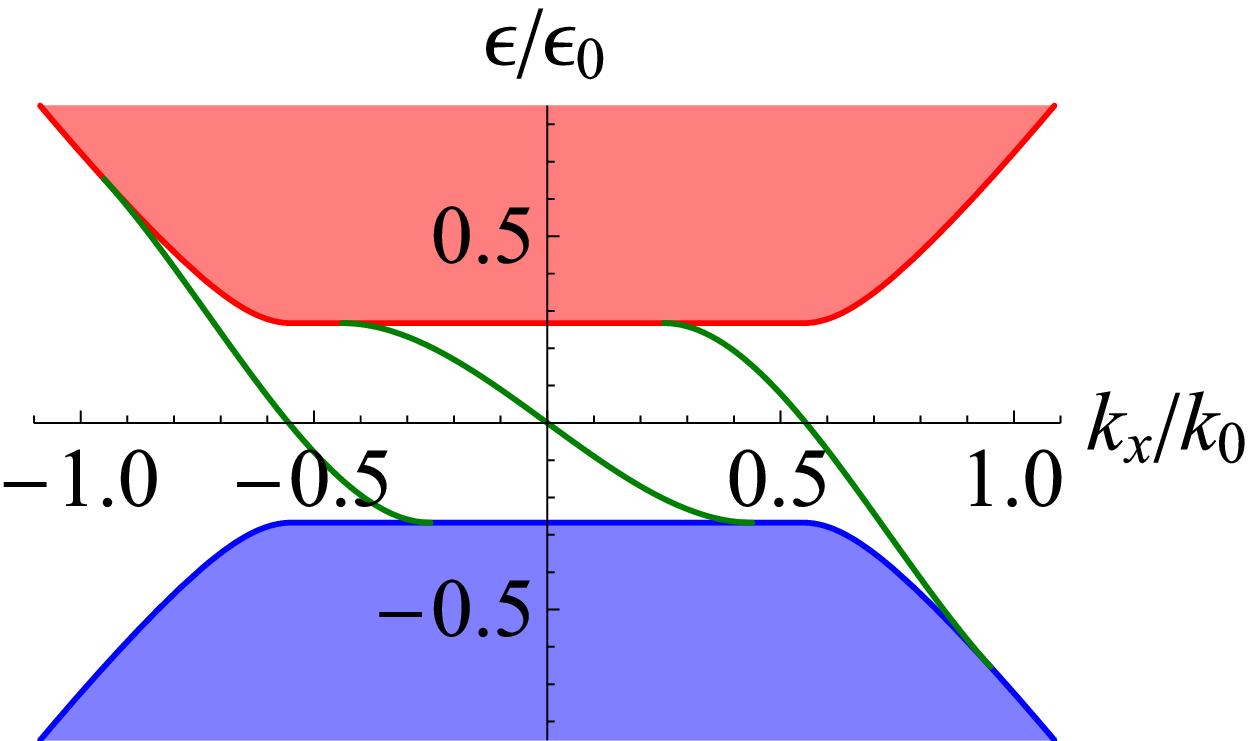}
   \label{Fig-sp-m-2nd}}
 \caption{\footnotesize
   {\color{blue}(color online)}
   Same as in Fig. \ref{Fig-sp-1st} albeit with
   $\Delta_0=0.850651\epsilon_0$
   and $B=1.05146\epsilon_0$,
   corresponding to point $P_2$ in Fig. \ref{Fig-intervals-MF}.}
 \label{Fig-sp-2nd}
\end{figure}
The energy dispersion for domain (2) is displayed in Fig. \ref{Fig-sp-2nd}
for $\Delta_0=0.850651\epsilon_0$ and $B=1.05146\epsilon_0$
[point $P_2$ in Fig. \ref{Fig-intervals-MF}].
For this domain, the mirror Chern numbers are $C_1=0$, and $C_{\bar{1}}=3$.
Accordingly, there are no anti-chiral states, whereas
there are three chiral edge-states, denoted as
$c,a,b$ [green curves in Fig. \ref{Fig-sp-m-2nd} ordered
from left to right]. The dispersion  curve for mode $a$
satisfies the equality $\veps_{\bar{1},a}(0)=0$, namely,
it crosses the middle of the gap at $k_x=0$.
The two other modes, $b$ and $c$ cross the middle of the gap at $\pm k_x \ne 0$.
 Explicitly, mode $b$ enters the gap for $k_1<k_x<k_2$,
(here $k_1\approx0.25k_0$ and $k_2\approx0.95k_0$), while mode
$c$ is determined from mode $b$ by
the symmetry (\ref{Transform}), so that,
\begin{eqnarray}
  &&
  \veps_{\bar{1},b}(k_x) ~=~
  -\veps_{\bar{1},c}(-k_x),
  \ \ \ \ \
  k_1 < k_x < k_2,
  \nonumber \\
  &&
  \veps_{\bar{1},a}(-k_x) ~=~
  -\veps_{\bar{1},a}(k_x).
  \label{sym-edges}
\end{eqnarray}

%%%%%%%%%%%%%%%%%%%%
\subsection{Domain (3): $(C_1,w_1)=(1,1), (C_{\bar{1}},w_{\bar{1}})=(3,1)$}
\label{domain-3}

%---------------- Sp 3rd ---------------------------
\begin{figure}[htb]
%H=6.28, L=14.65
\centering
  \subfigure[]
  {\includegraphics[width=40 mm,angle=0]
   {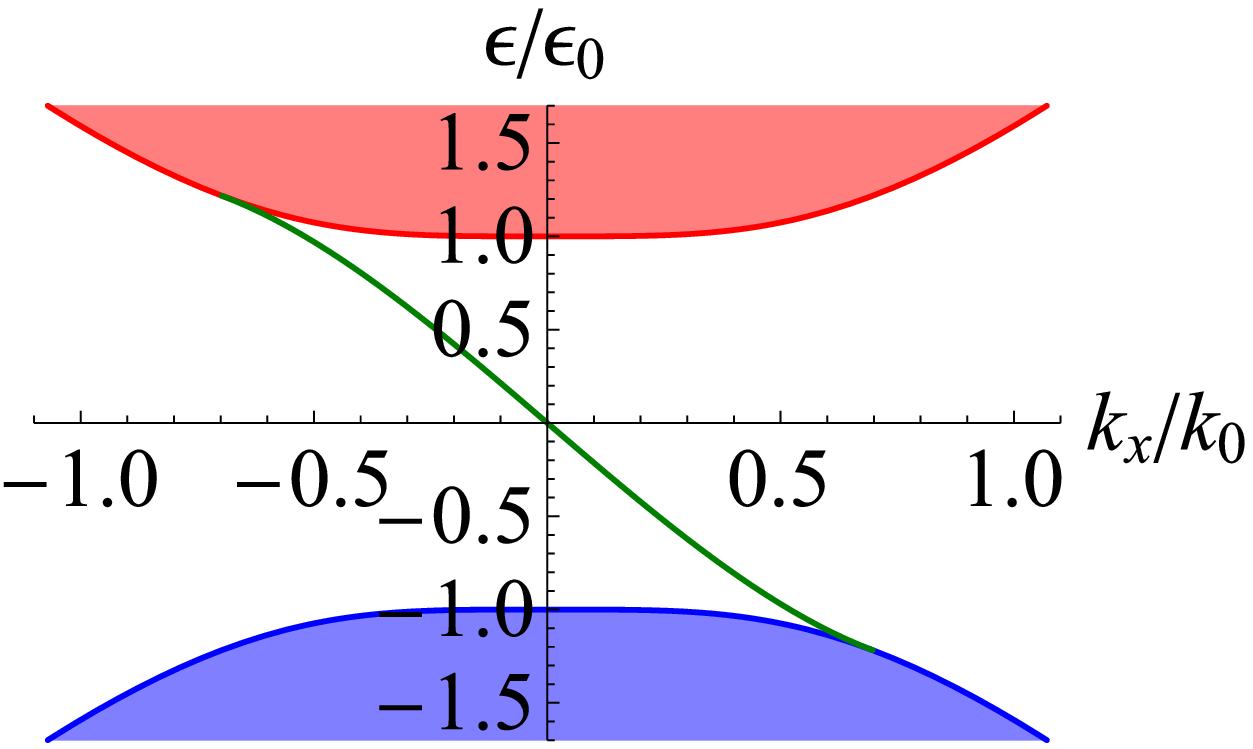}
   \label{Fig-sp-p-3rd}}
  \subfigure[]
  {\includegraphics[width=40 mm,angle=0]
   {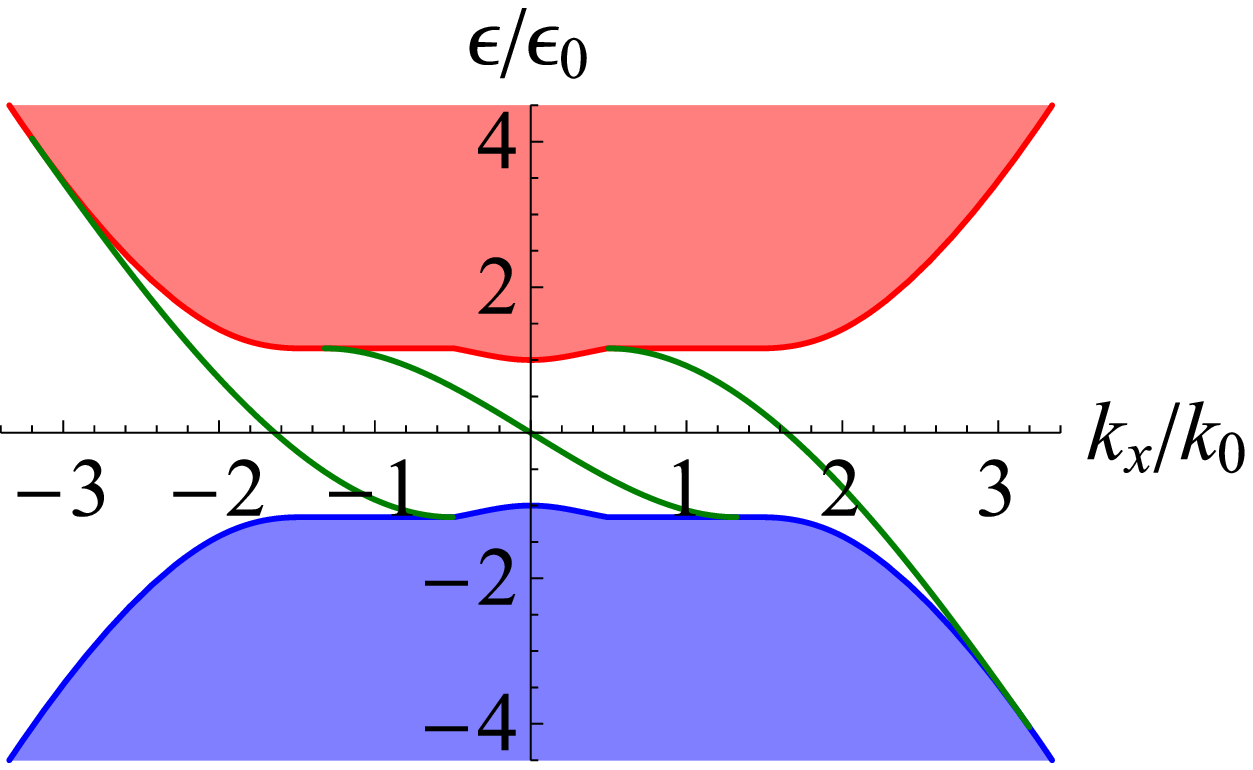}
   \label{Fig-sp-m-3rd}}
 \caption{\footnotesize
   {\color{blue}(color online)}
   Same as in Fig. \ref{Fig-sp-1st} albeit with
   $\Delta_0=0$
   and $B=2\epsilon_0$, corresponding for
   point $P_3$ in Fig. \ref{Fig-intervals-MF}.
   }
 \label{Fig-sp-3rd}
\end{figure}

The energy dispersion for domain (3) is displayed in Fig. \ref{Fig-sp-3rd}
for $\Delta_0=0$ and $B=2\epsilon_0$
[point $P_3$ in Fig. \ref{Fig-intervals-MF}].
For this domain, the mirror Chern numbers are $C_1=1$, and $C_{\bar{1}}=3$.
Thus, there is a single chiral mode for $\eta=1$
[green curve in Fig. \ref{Fig-sp-p-3rd}], and
three chiral modes associated with $\eta=\bar{1}$
[green curves in Fig. \ref{Fig-sp-m-3rd}].
The properties of the latter three modes are
the same as discussed in Fig.~\ref{Fig-sp-m-2nd}, except that here
$k_1\approx0.5k_0$ and $k_2\approx3.2k_0$.
In addition to the symmetries
exposed in Eq.~(\ref{sym-edges})
(relevant to $\bar{\eta}=-1$), the symmetry (\ref{Transform}) also implies
\begin{eqnarray}
  &&
  \veps_{1}(k_x) ~=~
  -\veps_{1}(-k_x).
  \label{symmetry-p}
\end{eqnarray}

%%%%%%%%%%%%%%%%%%%%
\subsection{Domain (4): $(C_1, w_1)=(1,1), (C_{\bar{1}}, w_{\bar{1}})=(2,0)$}
\label{domain-4}

%---------------- Sp 4th ---------------------------
\begin{figure}[htb]
%H=6.28, L=14.65
\centering
  \subfigure[]
  {\includegraphics[width=40 mm,angle=0]
   {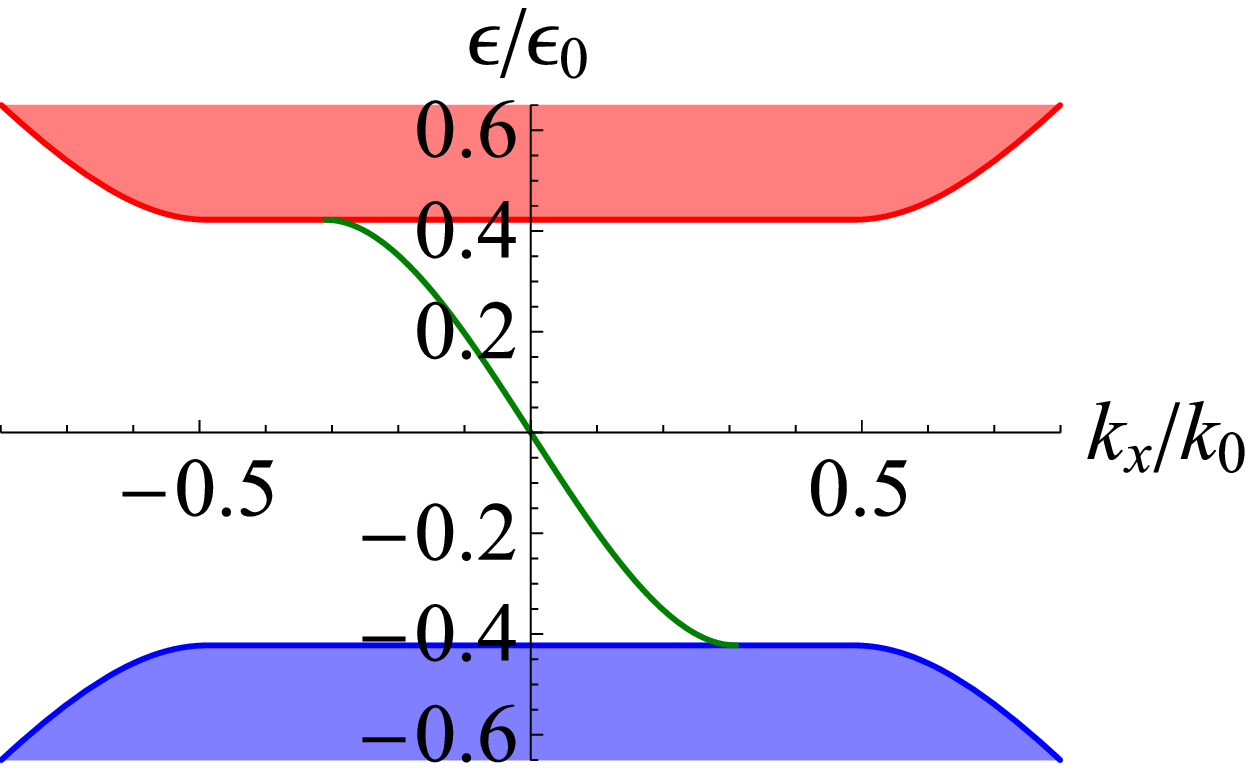}
   \label{Fig-sp-p-4th}}
  \subfigure[]
  {\includegraphics[width=40 mm,angle=0]
   {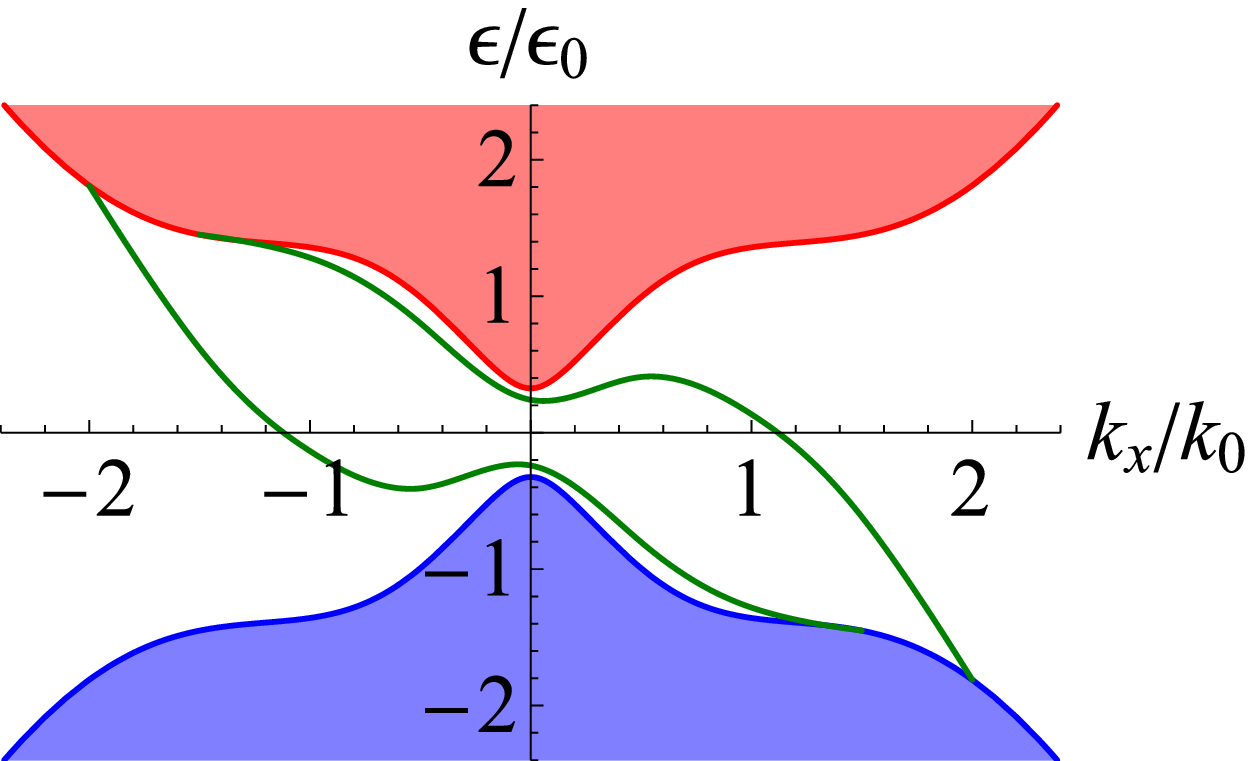}
   \label{Fig-sp-m-4th}}
 \caption{\footnotesize
   {\color{blue}(color online)}
   Same as in Fig. \ref{Fig-sp-1st} albeit with
   $\Delta_0=-0.850651\epsilon_0$
   and $B=1.05146\epsilon_0$, corresponding to
   point $P_4$ in Fig. \ref{Fig-intervals-MF}.
   }
 \label{Fig-sp-4th}
\end{figure}
The energy dispersion for domain (4) is displayed in Fig. \ref{Fig-sp-4th}
for $\Delta_0=-0.850651\epsilon_0$ and $B=1.05146\epsilon_0$
[point $P_4$ in Fig. \ref{Fig-intervals-MF}].
For this domain, the mirror Chern numbers are $C_1=1$, and $C_{\bar{1}}=2$.
There is a single chiral mode for the block $\eta=1$
[green curve in Fig. \ref{Fig-sp-p-4th}], and
two chiral modes for $\eta=\bar{1}$
[green curves in Fig. \ref{Fig-sp-m-4th}].
As noted for states $b,c$ in the analysis of
Figs. \ref{Fig-sp-m-2nd} and \ref{Fig-sp-m-3rd},
the energies of the two edge states for $\bar{\eta}=-1$, displayed
in Fig. \ref{Fig-sp-m-4th} are not zero at $k_x=0$.
Moreover, the corresponding energies $\veps_{\eta,b}(k_x)$ and
$\veps_{\eta,c}(k_x)$
are not monotonic as function of $k_x$.
This result does not contradict the symmetry
(\ref{Transform}), according to which
\begin{eqnarray*}
  &&
  \veps_{\eta,c}(-k_x) ~=~
  -\veps_{\eta,b}(k_x).
\end{eqnarray*}
However, this pattern of non-monotonicity
implies that the group velocity changes sign twice inside the gap.
The symmetry (\ref{Transform})  implies property (\ref{symmetry-p})
of $\veps_{1}(k_x)$.

%%%%%%%%%%%%%%%%%%%%
\subsection{Domain (5): $(C_1,w_1)=(-2,0), (C_{\bar{1}},w_{\bar{1}})=(2,0)$}
\label{domain-5}

%---------------- Sp 5th ---------------------------
\begin{figure}[htb]
%H=6.28, L=14.65
\centering
  \subfigure[]
  {\includegraphics[width=40 mm,angle=0]
   {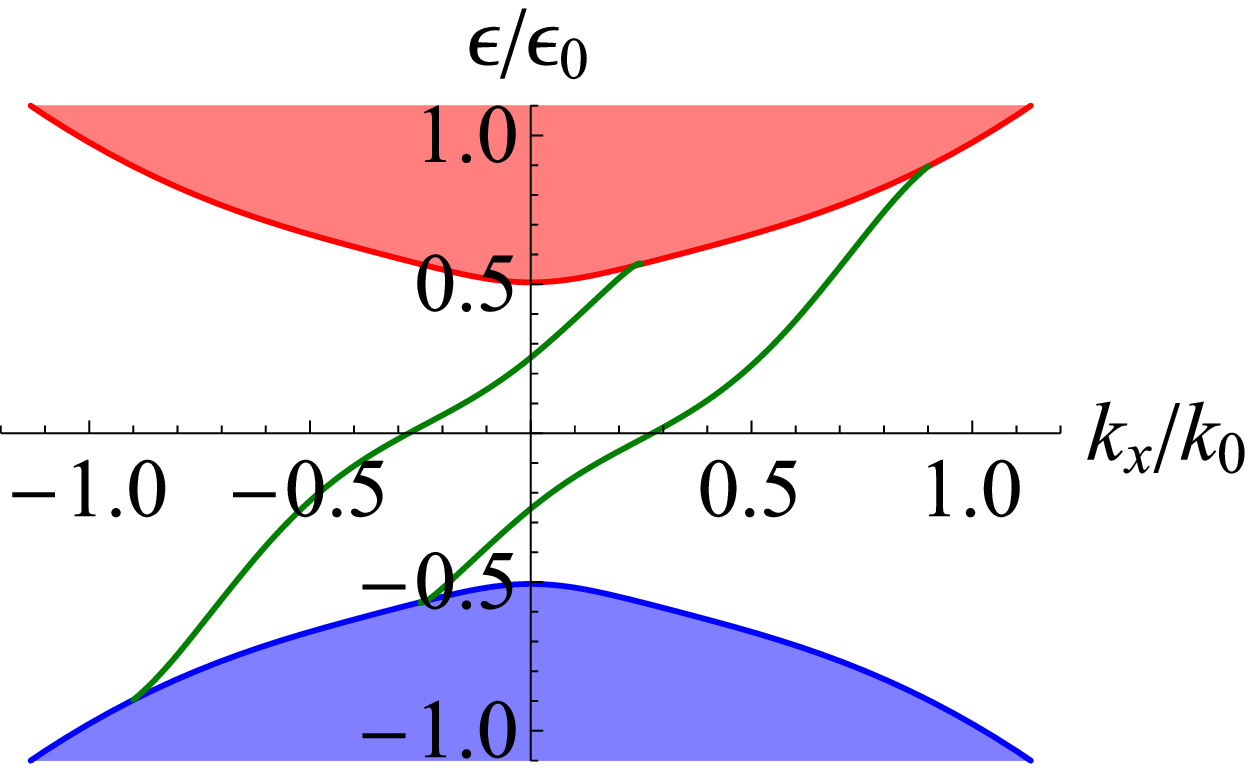}
   \label{Fig-sp-p-5th}}
  \subfigure[]
  {\includegraphics[width=40 mm,angle=0]
   {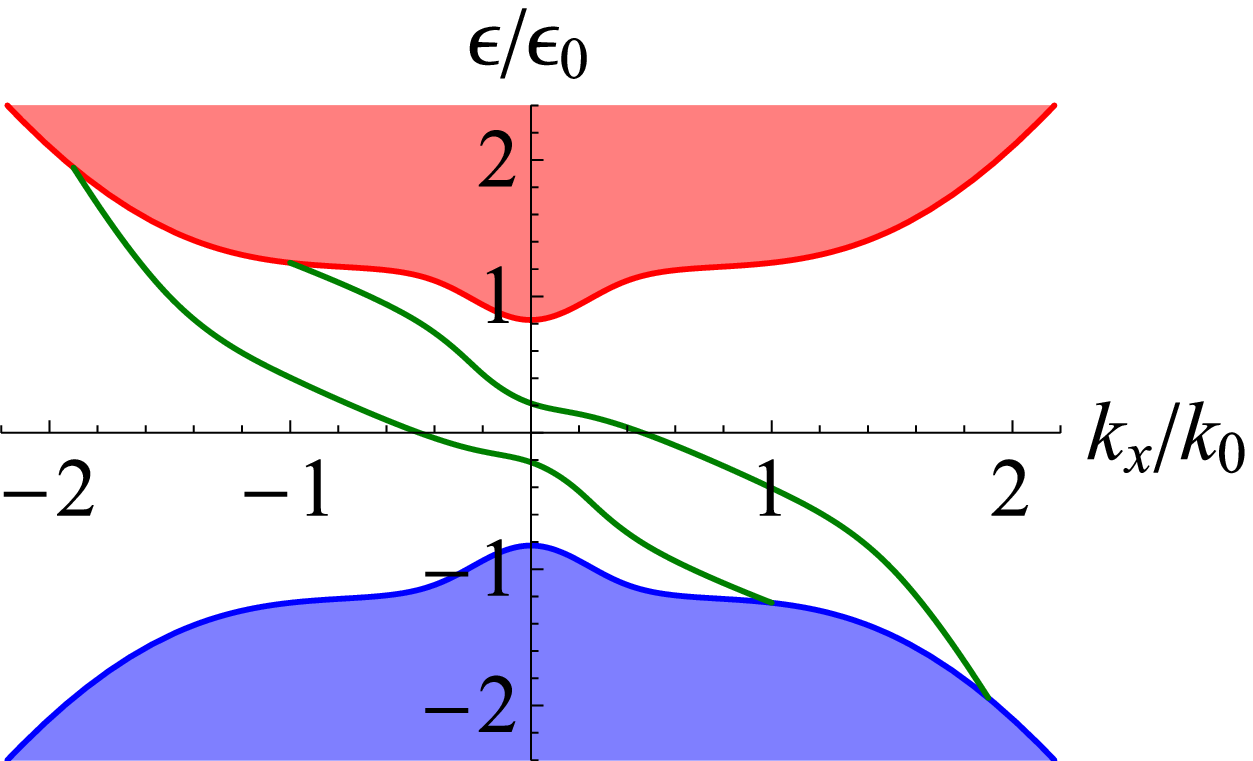}
   \label{Fig-sp-m-5th}}
 \caption{\footnotesize
   {\color{blue}(color online)}
   Same as in Fig. \ref{Fig-sp-1st} albeit with
   $\Delta_0=-0.987087\epsilon_0$
   and $B=0.320364\epsilon_0$, corresponding to
   point $P_5$ in Fig. \ref{Fig-intervals-MF}.
   Note the opposite directions of the group
   velocities for $\eta=-1$ and $\eta=1$.
   In other words, there are two chiral states and  two anti-chiral states.  }
 \label{Fig-sp-5th}
\end{figure}

Energy dispersion for the domain (5) is displayed in Fig. \ref{Fig-sp-5th}
for $\Delta_0=-0.987087\epsilon_0$ and $B=0.320364\epsilon_0$
[point $P_5$ in Fig. \ref{Fig-intervals-MF}].
For this domain, the mirror Chern numbers are $C_1=-2$, and $C_{\bar{1}}=2$.
Thus, there are two edge states denoted as $b,c$ for each value of $\eta=\pm1$
[green curves in Fig. \ref{Fig-sp-p-5th} and
 Fig. \ref{Fig-sp-m-5th} respectively].
Here, as in the dispersion curves
displayed in Fig. \ref{Fig-sp-m-4th}, the energies of the edge states
do not vanish at $k_x=0$. Indeed, following Eq. (\ref{Transform}),
these energies satisfy the relation,
\begin{eqnarray*}
  &&
  \veps_{\eta,b}(-k_x) ~=~
  -\veps_{\eta,c}(k_x),
\end{eqnarray*}
which does not constrain them to vanish at $k_x=0$.

The occurrence of two chiral and two anti-chiral edge states
in domain 5 is somewhat expected. In this domain, $\Delta_0<0$ and $B$ is rather small.
The obvious question is what happens when $B \to 0$. Since $\Delta_0<0$
guarantees the occurrence of a topological phase also for $B=0$ we
have a TRS conserving topological insulator where the fermions have spin
$F=\tfrac{3}{2}$. As will be shown in a future publication, in this case for fermion of spin $F$,
there are $\frac{n_F}{2}=F+\tfrac{1}{2}$ Kramers
pairs of {\it helical} edge states. The two helical states
in each pair carry a pseudo-spin magnetic quantum numbers $\pm s$, and different pairs
have the quantum numbers
$\pm s=\pm \tfrac{1}{2},\pm \tfrac{3}{2}...,\pm F$.
All states with positive pseudo-spin magnetic quantum numbers have the same
Kramers parity (they propagate along the same direction along the edge),
but still they cannot scatter each other since they
are protected by the quantum number $s$. Moreover, the energies of all
helical states are antisymmetric functions of $k_x$ (so that they vanish at $k_x=0$) and obey the symmetry
$$ \veps_{\eta s}(k_x)=\veps_{\bar {\eta} \bar {s}}(-k_x).$$
Moreover, near $k_x=0$ the dispersion relations are proportional to odd powers of $k_x$, that is,
$\veps_{\eta s}(k_x) \propto k_x^{2|s|}$. For $F=\tfrac{3}{2}$ the number of Kramers pairs is $N=2$.
As $B \to 0$, the chiral and anti chiral edge states shown in Fig.~\ref{Fig-sp-5th}
of the TRS breaking system
in region 5 of Fig. \ref{Fig-intervals-MF}, fuse smoothly into the two Kramers pairs of the TRS
conserving system discussed above.

\subsection{Summary of section \ref{sec-edge}}
  \label{subsec-sum-edge}

The results presented in this section underline
the profound distinction between the nature of
topological insulators wherein the fermions have
spin $S=\frac{1}{2}$ (e.g electrons or holes
encountered in solid state physics), and
topological insulators wherein the fermion
have spin $S>\frac{1}{2}$ (e.g cold atoms with
spin $F=\frac{3}{2}$). On the fundamental level,
it should be mentioned that in the former case,
the SO coupling is a direct consequence of
the Pauli equation that is derived from the Dirac
theory of the electron. In the latter case, in contrast,
SO coupling is not derived from a relativistic equation.
Rather, it is due to an interaction between a polarized
laser field and the atomic spin.

Once this interaction is justified and formulated,
it is evidently clear  that the pattern of edge states
exposed in this section is much richer than that
encountered in solid state physics. In order to
highlight the main features of this pattern, it is
crucial to stress that it pertains to edge states
along a {\it single edge} of the {\it original
Hamiltonian} (\ref{H-4x2D-def}). Recasting
this $8 \times 8$ Hamiltonian as a two block
$4 \times 4$ operators, Eq.~(\ref{H-4x2D-block}),
reflects an important symmetry of
$\hat{\mathcal{H}}_{\mbfk}$
but the separate Hamiltonians
$\hat{\mathcal{H}}_{\mbfk,\eta}$ by themselves
do not correspond to a physically realisable system.
The unique features of the edge states pattern for
the present system are now listed below.
\begin{enumerate}
\item{} Depending on the domain of parameters in
        the $(B, \Delta_0)$ ($B>0$) half plane, the number of edge states
        varies between $0$ and $4$.

\item{} Edge states may cross the middle of the gap
        at $k_x \ne 0$. The only restriction is imposed
        by the symmetry (\ref{Transform}).

\item{} As function of $k_x$, the edge state dispersion
        curves need not be  monotonic. Physically, it
        means that the group velocity of the edge states
        might change sign at certain values of $k_x$.

\item{} For the same value of the magnetic field, it is
        possible to have chiral and anti-chiral edge states
        {\it propagating on the same edge}.

\item{} The total number of edge states is equal to
        $C=|C_1|+|C_{\bar{1}}|$
        and the number of anti-chiral edge states appears with
        negative sign.

\item{} Properties 2, 4 and 5 are intimately related to
        the question whether the edge states are even or odd
        eigenstates of the parity operator $Q$ defined in
        Eq.~(\ref{Q}).
\item{} For $\Delta_0<0$ and $B \to 0$ the chiral
     and anti-chiral edge states fuse smoothly into
     Kramers pairs of helical states. The number of
     Kramers pairs is $F+\tfrac{1}{2}$. Two different
     helical states of the same Kramers parity are
     protected by the pseudo-spin magnetic quantum
     number $s$. The dispersion relation of helical
     edge states near $k_x=0$ is proportional to
     an odd power $k_x^{|2s|}$.
\end{enumerate}

%%%%%%%%%%%%Conclusion%%%%%%%%%%%%%%%%%%
\section{Conclusions}
  \label{sec-comclud}

We have developed an {\it ab-initio} theoretical
framework for studying the physics of topological
insulators in 2D  gas of fermionic atoms in which
the ground-state spin is $F>\tfrac{1}{2}$.
These systems can be realized if SOC is relevant
and if the pertinent optical potential is properly
tuned to have two bands (conduction and valence),
and its gap dispersion has a Mexican hat shape.
It is suggested that SOC can be induced by
irradiating the gas of trapped cold fermionic
atoms with a specially designed arrangement
of four polarized laser beams. Within this
construction the SOC term has a Dresselhaus
form displayed in Eq.~(\ref{HSO-4x2D-def}).
In the long wavelength approximation
the bare Hamiltonian, Eq.~(\ref{H-matrix}) is
a matrix of dimension $(2n_F) \times (2n_F)$
where $n_F=2 F+1$. The occurrence of mirror
symmetry enables its two-block decomposition
as in Eq.~(\ref{H-4D-H-4D}).  For $F>\frac{1}{2}$,
the SOC couples conduction and valence bands
in a peculiar way. Thus, although each
$n_F \times n_F$ block Hamiltonian describes
a quasiparticle with (pseudo) spin $F$, it is not
possible to express it in terms of generators of
the $n_F$ dimensional irrep of SU(2).
This implies that the underlying physics is
qualitatively distinct from that of spin
$\tfrac{1}{2}$ topological insulators.

In this work we concentrated on the case
$F=\tfrac{3}{2}$ and assumed that the 2D fermion
gas is subject to an external magnetic field.
Analysis of the topological properties revealed
a rich pattern of (mirror) Chern numbers that,
depending on the strength of the magnetic
field and the gap parameter $\Delta_0$,
can be either zero, positive or negative
integers. The corresponding edge states
are either chiral or anti-chiral, as explained
in more details in section \ref{sec-edge}.
In particular, it is possible to have chiral
and anti-chiral edge states on the same
edge. This is not possible for fermions
with atomic spin $F=\frac{1}{2}$.

%%%%%%%%%%%%%%%%%%%%%%%%%%%%%%

\begin{acknowledgments}
This work was supported in part by the ``Topological Material Science'' (No. JP15H05855) KAKENHI on Innovative Areas from JSPS of Japan, a Grant-in-Aid for Challenging Exploratory Research (No. JP15K13498), and a Grant-in-Aid for
Scientific Research B (No. JP17H02922). The research of Y.A is supported by grant 400/12 of the
Israeli Science Foundation.
\end{acknowledgments}

\newpage
\appendix

%%%%%%%%%%\section{Justification of the Model}%%%%%%%%%%%%%
\section{Justification of the model}
\label{Sec:Justification}
In this (relatively long) appendix we address, in some details,
 the question of whether the model analyzed above is
realizable in a system of cold atoms. First, in subsection \ref{Subsec:Design}
the design of the pertinent optical potential is explained. Then, in subsection \ref{Subsec:TB},
the tight-binding picture is developed ab-initio. It follows by the introduction of spin-orbit
interaction in subsection \ref{sec-tight-bind-subsec-spin-orbit}, and concludes in
subsection \ref{sec-tight-bind-subsec-long-wave} that
describes the long wave approximation.
\subsection{Designing the Optical Lattice}
The key question of generating SOC in cold atom systems
is addressed in this subsection. The basic idea is to subject the
2D fermion gas to a specific pattern of polarized laser beams.
While completing this task we have learnt of a recent papers
addressing this problem \cite{SO-opt-latt-PRL04,%
Demler-17,SOC-opt-latt-PRL05,Juzeliuas2010,%
Campbell2011,Lin2011,Wang2012,%
Dalibard,Cheuk2012,Huang2016,SOC-PRL05b,%
SOC-NewJPhys03,SOC-PRL10a,SOC-JPhB13,SOC-PRA14,Wu2016}.
\label{Subsec:Design}
%---------------- Sp0 vs MF ---------------------------
\begin{figure}[htb]
%H=6.28, L=14.65
\centering
  \includegraphics[width=40 mm,angle=0]
   {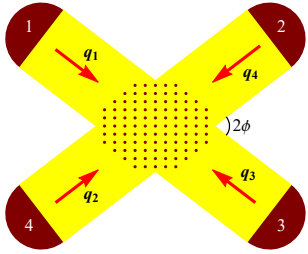}
 \caption{\footnotesize
   {\color{blue}(color online)}
   Two pairs of counter propagating laser beams.
   The lasers are denoted as $j$ ($j=1,2,3,4$).
   Corresponding wave vectors are $\mbfq_j$.
   The angle between crossing beans is $2\phi$.
   The red dots denote trapped atoms.}
 \label{Fig-beams-squarre}
\end{figure}

Consider alkali atoms interacting with a configuration of laser
beams producing an amplitude modulated
electric field $\mbfE(\mbfr,t)$, as displayed in
Fig. \ref{Fig-beams-squarre},
\begin{eqnarray}
  \mbfE(\mbfr,t) &=&
  \sum_{j=1}^{4}
  \mbfE_{j}(\mbfr,t)~
  e^{i \omega t}.
  \label{EF-tot-def}
\end{eqnarray}
Here
\begin{eqnarray}
  \mbfE_{j}(\mbfr,t) &=&
  \Big\{
      E_0+
      E_1
      e^{i \omega_1 t}
  \Big\}~
  e^{i \mbfq_j \mbfr}
  \times \nonumber \\ &\times&
  \Big\{
      \sqrt{1-\beta^2}~
      \mbfe_z+
      \beta
      \big[
          \mbfe_z
          \times
          \mbfe_j
      \big]
  \Big\},
  \label{EF-Ej-def}
\end{eqnarray}
where $\beta$, $E_0$ and $E_1$ are a real parameters.
Hereafter we assume that
$0 < \beta \ll \beta_0=1/\sqrt{2}$ and $E_0 \gg E_1 > 0$.
All the lasers produce light with the same wavelength
$\lambda_0$ and wave number $q_0=2 \pi/\lambda_0$.
The wave vectors $\mbfq_j$ ($j=1,2,3,4$) are,
\begin{eqnarray*}
  \mbfq_j ~=~
  q_0~\mbfe_j,
\end{eqnarray*}
where the unit vectors $\mbfe_j$ are,
\begin{eqnarray*}
  &&
  \mbfe_1 ~=~
  -\mbfe_3 ~=~
  \mbfe_x~
  \cos\phi-
  \mbfe_y~
  \sin\phi,
  \\
  &&
  \mbfe_2 ~=~
  -\mbfe_4 ~=~
  \mbfe_x~
  \cos\phi+
  \mbfe_y~
  \sin\phi,
\end{eqnarray*}
where $\mbfe_x$, $\mbfe_y$ and $\mbfe_x$ are unit vectors parallel
to the $x$-, $y$- and $z$-axes, the angle $\phi$ is close to
$\pi/4$.

We assume that detuning of the light frequency from
the resonant frequency $\omega_e$ of the $^{2}$S$_{1/2}$
and $^{2}$P$_{3/2}$ is much smaller than fine splitting
between the levels $^{2}$P$_{3/2}$ and $^{2}$P$_{1/2}$.
Therefore hereafter we are interested just in the states
$^{2}$S$_{1/2}$ and $^{2}$P$_{3/2}$. Corresponding
wave functions are $|g,\sigma\rangle$ and $|e,\mu\rangle$,
where $\sigma=\ua,\da$ and $\mu=\pm\frac{1}{2},\pm\frac{3}{2}$.
From the other hand, the detuning of the light frequency
$\omega$ from the resonance frequency $\omega_e$ is
large with respect to the radiative width of the excited states,
therefore spontaneous emission is suppressed and we can
adiabatically eliminate the excited states by writing an effective
Hamiltonian which involves only the ground states
\cite{SO-opt-latt-PRL04},
\begin{eqnarray*}
  H_{\mathrm{eff}} &=&
  \sum_{\sigma,\sigma'}
  V_{\sigma\sigma'}(\mbfr,t)~
  X^{\sigma\sigma'},
  \label{H-eff-def}
\end{eqnarray*}
where $X^{\sigma\sigma'}=|{g,\sigma}\rangle\langle{g,\sigma'}|$
is Hubbard operator.
The effective interaction $V_{\sigma\sigma'}(\mbfr)$ is,
\begin{eqnarray*}
  V_{\sigma\sigma'}(\mbfr,t) &=&
  \sum_{\mu}
  \frac{1}{\hbar(\omega-\omega_e)}~
  \big\langle
      g,\sigma
  \big|
      \mbfE(\mbfr,t)
      \cdot
      {\mathbf{d}}
  \big|
      e,\mu
  \big\rangle
  \times \nonumber \\ && \times
  \big\langle
      e,\mu
  \big|
      \mbfE^{*}(\mbfr,t)
      \cdot
      {\mathbf{d}}
  \big|
      g,\sigma'
  \big\rangle,
\end{eqnarray*}
where $\mbfE(\mbfr,t)$ is the electric field (\ref{EF-tot-def})
and ${\mathbf{d}}$ is the dipole moment operator of
the atom.

The total electronic orbital moment of atom in
the ground state is $J=\frac{1}{2}$.
Then the effective interaction $V_{\sigma\sigma'}(\mbfr,t)$
can be written in terms of a artificial magnetic field
${\mathbf{B}}(r)$ coupled to the total electronic orbital
momentum operator $\hat{\mathbf{J}}$,
\begin{eqnarray}
  V_{\sigma\sigma'}(\mbfr,t) &=&
  V(\mbfr,t)~
  \delta_{\sigma\sigma'}+
  {\bm{B}}(\mbfr,t)
  \cdot
  {\bm{J}}_{\sigma\sigma'}.
  \label{V-potential-vector}
\end{eqnarray}
Here the scalar potential $V(\mbfr,t)$ is proportional
to the local light intensity, while the vectorial field
${\bm{B}}(\mbfr,t)$ is proportional to the local
electromagnetic spin:
\begin{eqnarray}
  V(\mbfr,t) &=&
  -\alpha_0
  \mbfE^{*}(\mbfr,t)
  \cdot
  \mbfE(\mbfr,t),
  \label{opt-potential}
  \\
  {\bm{B}}(\mbfr,t) &=&
  -i \alpha_1
  \mbfE^{*}(\mbfr,t)
  \times
  \mbfE(\mbfr,t),
  \label{opt-magnetic-field}
\end{eqnarray}
where $\alpha_0$ and $\alpha_1$ are scalar
and vector dynamical polarizabilities of
the atoms.

Taking into account eqs. (\ref{EF-tot-def}) and (\ref{EF-Ej-def}),
we get the scalar potential (\ref{opt-potential}) in the form
\begin{eqnarray}
  V(\mbfr,t) &=&
  -\Big\{
      V_0+
      V_1
      \cos(\omega_1 t)
  \Big\}~
  {\mathcal{V}}(\mbfr),
  \nonumber
  \\
  {\mathcal{V}}(\mbfr)
  &=&
  16~
  \Bigg\{
       \Big[
           1-
           2
           \beta^2
       \Big]~
       \cos^{2}
       \bigg(
            \frac{\pi x}{a_x}
       \bigg)~
       \cos^{2}
       \bigg(
            \frac{\pi y}{a_y}
       \bigg)+
  \nonumber \\ && ~ +
       \beta^2~
       \sin^2\phi~
       \cos^2
       \bigg(
            \frac{\pi x}{a_x}
       \bigg)+
  \nonumber \\ && ~ +
       \beta^2~
       \cos^2\phi~
       \cos^2
       \bigg(
            \frac{\pi y}{a_y}
       \bigg)
  \Bigg\},
  \label{V-pot-res}
\end{eqnarray}
where
\begin{eqnarray}
  &&
  V_0 =
  \alpha_0~
  \big(
  E_{0}^{2}+
  E_{1}^{2}
  \big),
  \ \ \
  V_1 =
  2
  \alpha_0
  E_0
  E_1,
  \nonumber
  \\
  &&
  a_x =
  \frac{\lambda_0}{2\cos\phi},
  \ \ \ \ \
  a_y =
  \frac{\lambda_0}{2\sin\phi}.
  \label{V0-a0-def}
\end{eqnarray}

The potential $V(\mbfr)$ satisfies the periodic conditions,
\begin{eqnarray}
  V(\mbfr,t) ~=~
  V(\mbfr+\mbfa_x,t) ~=~
  V(\mbfr+\mbfa_y,t),
  \label{V-periodic}
\end{eqnarray}
where the lattice vectors $\mbfa_x$ and $\mbfa_y$ are,
\begin{eqnarray}
  \mbfa_x ~=~
  a_x~
  \mbfe_x,
  \ \ \ \ \
  \mbfa_y ~=~
  a_y~
  \mbfe_y,
  \label{a1-a2-vec-def}
\end{eqnarray}
$a_x=|\mbfa_x|$ and $a_y=|\mbfa_y|$ are given by eq.
(\ref{V0-a0-def}).

The artificial magnetic field (\ref{opt-magnetic-field}) lies in
the $x$-$y$ plane,
$$
  {\bm{B}}(\mbfr) ~=~
  B_x(\mbfr)~
  \mbfe_x+
  B_y(\mbfr)~
  \mbfe_y.
$$
Explicitly $B_x(\mbfr)$ and $B_y(\mbfr)$ are
\begin{subequations}
\begin{eqnarray}
  B_x(\mbfr) =
  B_0~
  \cos\phi~
  \sin
  \bigg(
      \frac{2 \pi x}{a_x}
  \bigg)~
  \cos^2
  \bigg(
       \frac{\pi y}{a_y}
  \bigg),
  \label{Bx-vec-res}
  \\
  B_y(\mbfr) =
  B_0~
  \sin\phi~
  \sin
  \bigg(
       \frac{2 \pi y}{a_y}
  \bigg)~
  \cos^2
  \bigg(
       \frac{\pi x}{a_x}
  \bigg),
  \
  \label{By-vec-res}
\end{eqnarray}
  \label{subeqs-B-vec-res}
\end{subequations}
where
\begin{eqnarray}
  B_0 &=&
  16~
  \alpha_1~
  \big\{
      E_{0}^{2}+
      E_{1}^{2}
  \big\}~
  \beta
  \sqrt{1-\beta^2}.
  \label{B0-def}
\end{eqnarray}
We assume that $\beta\ll1$ and $E_1 \ll E_0$,
so that the time-dependent part of the artificial
magnetic field which is proportional to
$\alpha_1 E_0 E_1 \beta$ is neglected hereafter.

The artificial magnetic field ${\bm{B}}(\mbfr)$ satisfies
the periodic conditions similar to eq. (\ref{V-periodic}),
\begin{eqnarray}
  {\bm{B}}(\mbfr) ~=~
  {\bm{B}}(\mbfr+\mbfa_x) ~=~
  {\bm{B}}(\mbfr+\mbfa_y),
  \label{B-periodic}
\end{eqnarray}
where the lattice vectors $\mbfa_x$ and $\mbfa_y$ are
given by Eq. (\ref{a1-a2-vec-def}).

%---------------- V and B ---------------------------
\begin{figure}[htb]
%H=6.28, L=14.65
\centering
  \subfigure[]
  {\includegraphics[width=50 mm,angle=0]
   {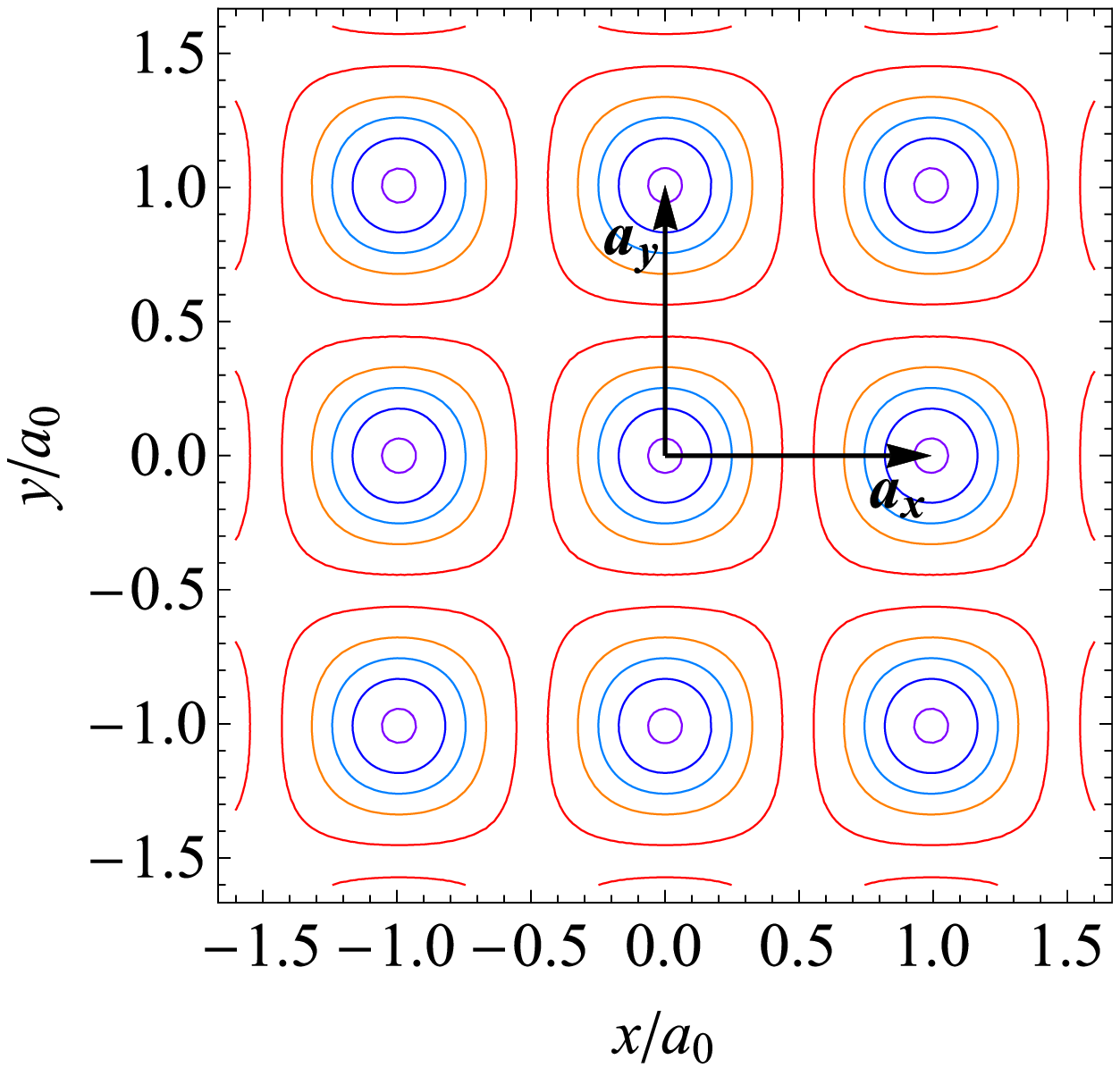}
   \label{Fig-V-square}}
  \subfigure[]
  {\includegraphics[width=50 mm,angle=0]
   {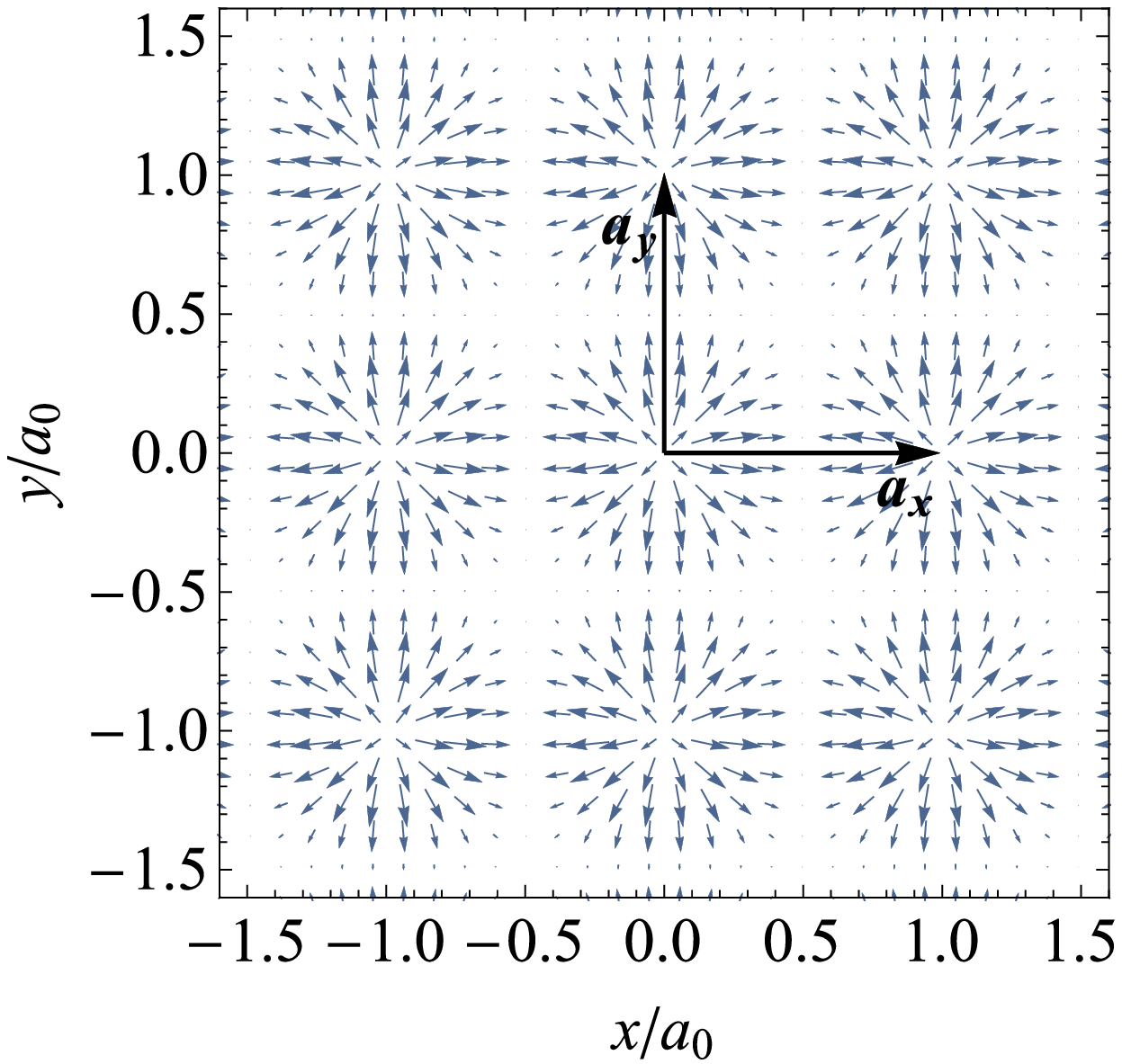}
   \label{Fig-B-square}}
 \caption{\footnotesize
   {\color{blue}(color online)}
   {\textbf{Panel (a)}}: The potential (\ref{V-pot-res}) for
   $\beta=0.1$ and $\phi=7\pi/30$.
   The purple, blue, sky blue, orange and red contours correspond
   to $V=0.96V_{\mathrm{min}}$, $0.73V_{\mathrm{min}}$,
   $0.5V_{\mathrm{min}}$, $0.27V_{\mathrm{min}}$ and
   $0.04V_{\mathrm{min}}$, respectively.
   {\textbf{Panel (b)}}: Artificial magnetic field
   (\ref{subeqs-B-vec-res}) for $\beta=0.1$ and $\phi=7\pi/30$.
   For both panels, $\mbfa_x$ and $\mbfa_y$ are given by
   eq. (\ref{a1-a2-vec-def}).}
 \label{Fig-V-B-square}
\end{figure}

Optical potential (\ref{V-pot-res}) and artificial magnetic field
(\ref{subeqs-B-vec-res}) are displayed in Fig. \ref{Fig-V-B-square}
for $\beta=0.1$ and $\phi=7\pi/30$.
It is seen that $V(\mbfr)$ has minima
$V_{\mathrm{min}}=-16V_0(1-\beta^2)$
at $\mbfr=\mbfn$ given by
\begin{eqnarray}
  \mbfn &=&
  n_x \mbfa_x+
  n_y \mbfa_y,
  \label{nA-nB-def}
\end{eqnarray}
$n_x$ and $n_y$ are integers.

First Brillouin zone of the rectangular lattice is given by
the conditions,
\begin{eqnarray*}
  \big|
      k_x
  \big|
  ~\leq~
  \frac{Q_x}{2},
  \ \ \ \ \
  \big|
      k_y
  \big|
  ~\leq~
  \frac{Q_y}{2},
\end{eqnarray*}
where
$$
  Q_x ~=~
  \frac{2 \pi}{a_x},
  \ \ \ \ \
  Q_y ~=~
  \frac{2 \pi}{a_y}.
$$

%%%%%%%%%%%%%%%%%%%%%%%%%%%%%%
\subsection{Tight Binding Approximation}
  \label{Subsec:TB}
The model considered here has two parameters of energy, the depth
$W_0$ of the potential wells and the recoiling energy
${\mathcal{E}}_{Q}$,
\begin{eqnarray}
  W_0 &=&
  -16 V_0
  \big(1-\beta^2\big),
  \label{V-depth}
  \\
  {\mathcal{E}}_{Q} &=&
  \frac{\hbar^2 (Q_{x}^{2}+Q_{y}^{2})}{2M}.
  \label{EQ-def}
\end{eqnarray}
When $W_0\gg{\mathcal{E}}_{Q}$, the atoms are localized near
the places of stable equilibrium and just hope from place to
place. In this case, we can describe the gas in framework of
the tight binding approximation.

In order to construct a tight binding Hamiltonian, we derive
wave functions and energy levels of atom localized near a minimum
position of the optical potential (\ref{V-pot-res}). Then we
estimate tunneling rates.

%%%%%%%%%%%%%%%%%%%%
\subsubsection{Quantum States of Localized Atoms}
  \label{sec-tight-bind-subsec-states}

When $0<\beta<\min(\beta_{c,x},\beta_{c,y})$ (which is assumed
hereafter), the optical potential (\ref{V-pot-res}) has minima at
$\mbfr=\mbfn$, where $\mbfn$ is given by eq. (\ref{nA-nB-def}),
$\beta_{c,x},\beta_{c,y}$ are given by eq. (\ref{subeqs-betac})
below.
For $\mbfr$ close to one of the minimum points [say, $\mbfr=(0,0)$],
$V(\mbfr)$ can be approximated as,
\begin{eqnarray}
  V(\mbfr) =
  -W_0+
  \frac{1}{2}~
  \Big\{
      K_x~
      x^2+
      K_y~
      y^2
  \Big\}+
  O\big(r^4\big).
  \label{V-pot-harmonic}
\end{eqnarray}
Here
\begin{eqnarray}
  K_x =
  K_0~
  \cos^2\phi~
  \bigg\{
       1-
       \frac{3\beta^2}{2}-
       \frac{\beta^2}{2}~
       \cos\big(2\phi\big)
  \bigg\},
  \label{Kx-harmonic-def}
  \\
  K_y =
  K_0~
  \sin^2\phi~
  \bigg\{
       1-
       \frac{3\beta^2}{2}+
       \frac{\beta^2}{2}~
       \cos\big(2\phi\big)
  \bigg\},
  \label{Ky-harmonic-def}
\end{eqnarray}
where
\begin{eqnarray*}
  K_0 &=&
  16 V_0 q_{0}^{2}.
\end{eqnarray*}

Then the wave functions of atom localized near the point
$\mbfr=(0,0)$ are,
\begin{eqnarray}
  \Phi_{\nu_x,\nu_y}^{\mathrm{(2D)}}(\mbfr) &=&
  \Phi_{\nu_x}^{(x)}(x)~
  \Phi_{\nu_x}^{(y)}(y),
  \label{WF-2D-harmonic}
\end{eqnarray}
where
\begin{eqnarray}
  \Phi_{\nu_{\alpha}}^{(\alpha)}(x_{\alpha}) &=&
  \frac{{\mathcal{N}}_{\alpha}}
       {\sqrt{2^{\nu_\alpha}~\nu_\alpha!}}~
  H_{\nu_\alpha}\big(\xi_\alpha\big)~
  e^{-\xi_{\alpha}^{2}/2}.
  \label{WF-1D-harmonic}
\end{eqnarray}
Here $\alpha$ is a Cartesian index, $\nu_{\alpha}=0,1,2,\ldots$,
$$
  {\mathcal{N}}_{j} ~=~
  \bigg(
       \frac{K_{\alpha} M}{\pi^2 \hbar^2}
  \bigg)^{1/8},
  \ \ \ \ \
  \xi_{\alpha} ~=~
  x_{\alpha}~
  \bigg(
       \frac{K_{\alpha}M}{\hbar^2}
  \bigg)^{1/4}.
$$
Corresponding energy levels (measured from $-W_0$) are,
\begin{eqnarray}
  {\mathcal{E}}_{\nu_x,\nu_y} &=&
  {\mathcal{E}}_{x,\nu_x}+
  {\mathcal{E}}_{y,n_y},
  \nonumber
  \\
  {\mathcal{E}}_{\alpha,n_{\alpha}}
  &=&
  \hbar
  \sqrt{\frac{K_{\alpha}}{M}}~
  \bigg(
       \nu_{\alpha}+
       \frac{1}{2}
  \bigg).
  \label{Energy-harmonic}
\end{eqnarray}

%---------------- energies harmonic ---------------------------
\begin{figure}[htb]
%H=6.28, L=14.65
\centering
  \includegraphics[width=50 mm,angle=0]
   {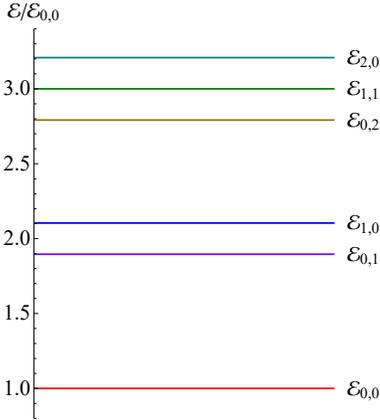}
 \caption{\footnotesize
   {\color{blue}(color online)}
   Energy levels (\ref{Energy-harmonic}) for
   $\beta=0.1$ and $\phi=7\pi/30$.}
 \label{Fig-energy-harmonic}
\end{figure}

Energy levels (\ref{Energy-harmonic}) are displayed in Fig.
\ref{Fig-energy-harmonic} for $\beta=0.1$ and $\phi=7\pi/30$.
We are interested in the close lying energy levels
${\mathcal{E}}_{1,0}$ and ${\mathcal{E}}_{0,1}$. Indeed,
the difference ${\mathcal{E}}_{1,0}-{\mathcal{E}}_{0,1}$ is,
\begin{eqnarray*}
  {\mathcal{E}}_{1,0}-
  {\mathcal{E}}_{0,1}
  &=&
  \frac{\hbar}{\sqrt{M}}~
  \Big\{
      \sqrt{K_x}-
      \sqrt{K_y}
  \Big\}.
\end{eqnarray*}
When $\phi=\pi/4$, then
$$
  K_{x}^{(0)} ~=~
  K_{y}^{(0)} ~=~
  \frac{K_0}{2}~
  \bigg\{
       1-
       \frac{3\beta^2}{2}
  \bigg\},
$$
and
$$
  {\mathcal{E}}_{1,0}^{(0)}
  ~=~
  {\mathcal{E}}_{0,1}^{(0)}
  ~=~
  2\hbar~
  \sqrt{\frac{K_0}{2M}~
        \bigg(
             1-
             \frac{3\beta^2}{2}
        \bigg)}.
$$
For $\phi=\frac{\pi}{4}-\delta$
with $|\delta|\ll1$, we can write,
\begin{eqnarray*}
  K_x &=&
  \frac{K_0}{2}~
  \bigg\{
       1-
       \frac{3\beta^2}{2}
  \bigg\}+
  K_0~
  \delta~
  \Big\{
      1-2\beta^2
  \Big\},
  \\
  K_y &=&
  \frac{K_0}{2}~
  \bigg\{
       1-
       \frac{3\beta^2}{2}
  \bigg\}-
  K_0~
  \delta~
  \Big\{
      1-2\beta^2
  \Big\}.
\end{eqnarray*}
Therefore ${\mathcal{E}}_{1,0}$ and ${\mathcal{E}}_{0,1}$ are
\begin{eqnarray*}
  {\mathcal{E}}_{1,0}
  ~=~
  {\mathcal{E}}_{1,0}^{(0)}-
  \veps_0,
  \ \ \ \ \
  {\mathcal{E}}_{0,1}
  ~=~
  {\mathcal{E}}_{1,0}^{(0)}+
  \veps_0,
\end{eqnarray*}
where
\begin{eqnarray}
  \veps_0 &=&
  \frac{\hbar\sqrt{K_0}}{\sqrt{M}}~
  \frac{\delta\big(1-2\beta^2\big)}
       {\sqrt{2-3\beta^2}}.
  \label{epsilon0-def}
\end{eqnarray}

%%%%%%%%%%%%%%%%%%%%
\subsubsection{Energy dispersion beyond
  the harmonic approximation}
  \label{sec-tight-bind-subsec-beyond-harm}

Let us consider bands with high band
number $\nu_x$ and $\nu_y=0$ or $1$,
such that the harmonic approximation
is still good for description of the motion
of the atom in the $y$-direction, and
falls for description of the motion
in the $x$-direction. The optical
potential can be approximated as,
\begin{eqnarray}
  V(\mbfr) &\approx&
  V_{\mathrm{1D}}(x)+
  V_{\mathrm{harm}}(y),
  \label{V=V1D+Vharm}
\end{eqnarray}
where
\begin{eqnarray}
  V_{\mathrm{1D}}(x)
  &=&
  -16 V_0~
  \bigg\{
       \big[
           1-
           \beta^2
           \big(
           1+
           \cos^2\phi
           \big)
       \big]
  \times \nonumber \\ && \times
       \cos^2
       \bigg(
            \frac{\pi x}{a_x}
       \bigg)+
       \beta^2
       \cos^2\phi
  \bigg\},
  \label{A-1D-approx}
  \\
  V_{\mathrm{harm}}(y)
  &=&
  \frac{K_y y^2}{a_y^2}.
  \label{V-1D-approx}
\end{eqnarray}
We assume here that $V_1=0$.
The effect of $V_1$ on the energy
dispersion is calculated below.

The 1D potential $V_{\mathrm{1D}}(x)$
is periodic with period $a_x$, therefore
the motion in the $x$-direction is described
by two quantum numbers, the band number
$\nu_x$ and the wave number $k_x$
belonging to the first Brillouin zone,
$$
  |k_x| ~<~ \frac{\pi}{a_x}.
$$
Then the wave function is a product of
the two wave functions, depending on
$x$ and $y$,
$$
  \Phi_{\nu_x,\nu_y,k_x}^{\mathrm{(2D)}}(\mbfr) =
  \Psi_{\nu_x,k_x}(x)~
  \Phi_{\nu_y}^{(y)}(y).
$$
The wave function $\Phi_{\nu_y}^{(y)}(y)$
($\nu_y=0,1$) is given by
Eq. (\ref{WF-1D-harmonic}),
whereas the wave function
$\Psi_{\nu_x}(x)$ are found from
the equation,
\begin{eqnarray}
  -\frac{\hbar^2}{2 M}~
  \Psi''_{\nu_x,k_x}(x)+
  V_{\mathrm{1D}}(x)
  \Psi_{\nu_x,k_x}(x)
  = \nonumber \\ =
  {\mathcal{E}}_{x,\nu_x,k_x}~
  \Psi_{\nu_x,k_x}(x).
  \label{eq-Schrodinger-periodic}
\end{eqnarray}

%---------------- energies periodic ---------------------------
\begin{figure}[htb]
%H=6.28, L=14.65
\centering
  \includegraphics[width=65 mm,angle=0]
   {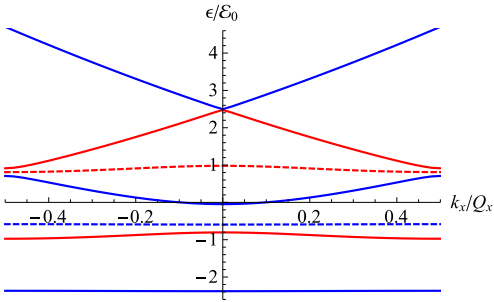}
 \caption{\footnotesize
   {\color{blue}(color online)}
   Energy levels calculated numerically from
   Eq. (\ref{eq-Schrodinger-periodic}) for
   $V_0=0.2{\mathcal{E}}_{0}$,
   $\beta=0.1$ and $\phi=7\pi/30$.
   The solid blue and red lines correspond
   to even and odd $\nu_x$.
   The dashed blue and red lines describe
   ${\mathcal{E}}_{x,0}+\hbar\omega_1$ and
   ${\mathcal{E}}_{x,1}+\hbar\omega_1$.}
 \label{Fig-energy-periodic}
\end{figure}

Energy dispersion calculated numerically
from Eq. (\ref{eq-Schrodinger-periodic}) is
shown in Fig. \ref{Fig-energy-periodic} for
$V_0=0.2{\mathcal{E}}_{0}$, $\beta=0.1$
and $\phi=7\pi/30$. It is seen that
${\mathcal{E}}_{x,\nu_x,k_x}$ with
$\nu_x=0,1$ does not depend on $k_x$
and can be approximated by
Eq. (\ref{Energy-harmonic}).

In similar way, we can consider motion of the atoms with
$\nu_x=0,1$ and any $\nu_y$ and find dispersion
${\mathcal{E}}_{y,\nu_y,k_y}$. The Schr\"odinger equation for
${\mathcal{E}}_{y,\nu_y,k_y}$ can be obtained from
Eq. (\ref{eq-Schrodinger-periodic}) by replacing $V_{1D}(x)$ by
$V_{1D}(y)$ and $V_{\mathrm{harm}}(y)$ by $V_{\mathrm{harm}}(x)$,
where
\begin{eqnarray*}
  V_{\mathrm{1D}}(y)
  &=&
  -16 V_0~
  \bigg\{
       \big[
           1-
           \beta^2
           \big(
           1+
           \sin^2\phi
           \big)
       \big]
  \times \nonumber \\ && \times
       \cos^2
       \bigg(
            \frac{\pi y}{a_y}
       \bigg)+
       \beta^2
       \sin^2\phi
  \bigg\},
  \\
  V_{\mathrm{harm}}(x)
  &=&
  \frac{K_x x^2}{a_x^2}.
\end{eqnarray*}
A 2D energy $\veps_{\boldsymbol\nu,\mbfk}^{(0)}$ of the trapped
atoms can be written as,
\begin{equation}
  \veps_{\boldsymbol\nu,\mbfk}^{(0)}=
  {\mathcal{E}}_{x,\nu_x,k_x}^{(0)}+
  {\mathcal{E}}_{y,\nu_y,k_y}^{(0)}.
  \label{energy2D=energy-x-y}
\end{equation}

%%%%%%%%%%%%%%%%%%%%
\subsubsection{Tunneling due to the time dependent potential}
  \label{sec-tight-bind-subsec-tunnel}

We now calculate corrections from the time-dependent potential
on the energy dispersion relation $\veps_{\boldsymbol\nu,\mbfk}^{(0)}$
with $\boldsymbol\nu=(0,1)$ and $(1,0)$. We assume here that
$|V_1|$ is very small with respect to the excitation energies
$|\veps_{\boldsymbol\nu,\mbfk}^{(0)}+\hbar\omega_1-%
\veps_{\boldsymbol\nu',\mbfk}^{(0)}|$,
where $\veps_{\boldsymbol\nu',\mbfk}^{(0)}$ are excited
energy levels. In this case the quantum transitions between
the energy levels are forbidden. In order to calculate correction
to the energies due to ellastic scattering of the light,
we apply time-dependent perturbation theory. Within this theory,
the wave functions of the state $\boldsymbol\nu=(\nu_x,\nu_y)$
are,
\begin{eqnarray}
  \Psi_{\boldsymbol\nu,\mbfk}(\mbfr,t) &=&
  \sum_{\mbfn'}
  A_{\boldsymbol\nu,\boldsymbol\nu';\mbfk}(t)
  \Phi_{\boldsymbol\nu',\mbfk}^{\mathrm{(2D)}}(\mbfr,t),
  \label{WF-time-dependent}
\end{eqnarray}
where
$$
  \Phi_{\boldsymbol\nu,\mbfk}^{\mathrm{(2D)}}(\mbfr,t)
  =
  \Phi_{\boldsymbol\nu}^{\mathrm{(2D)}}(\mbfr)
  e^{i \mbfk \mbfr-\frac{i}{\hbar}\veps_{\boldsymbol\nu,\mbfk}^{(0)} t},
$$
and $\Phi_{\boldsymbol\nu}^{\mathrm{(2D)}}(\mbfr)$
is given by Eq. (\ref{WF-2D-harmonic}).

The functions $A_{\boldsymbol\nu,\boldsymbol\nu';\mbfk}(t)$
satisfy the equation,
\begin{eqnarray}
  i \hbar
  \dot{A}_{\boldsymbol\nu,\boldsymbol\nu';\mbfk}(t)
  &=&
  V_1 \cos(\omega_1 t)
  \sum_{\boldsymbol\nu''}
  {\mathcal{V}}_{\boldsymbol\nu',\boldsymbol\nu''}~
  A_{\boldsymbol\nu,\boldsymbol\nu'';\mbfk}(t)
  \times \nonumber \\ && \times
  \exp
  \bigg[
       \frac{i t}{\hbar}~
       \Big(
           \veps_{\boldsymbol\nu',\mbfk}^{(0)}-
           \veps_{\boldsymbol\nu'',\mbfk}^{(0)}
       \Big)
  \bigg],
  \label{eq-for-A}
\end{eqnarray}
where $\boldsymbol\nu=(\nu_x,\nu_y)$,
$\mbfk=(k_x,k_y)$,
\begin{eqnarray}
  {\mathcal{V}}_{\boldsymbol\nu',\boldsymbol\nu''}
  =
  \int
  \Big(
      \Phi_{\boldsymbol\nu'}^{\mathrm{(2D)}}(\mbfr)
  \Big)^{*}
  {\mathcal{V}}(\mbfr)
  \Phi_{\boldsymbol\nu''}^{\mathrm{(2D)}}(\mbfr)~
  d^2 \mbfr.
  \label{V1-nn-ME}
\end{eqnarray}

In the zeroth approximation, the coefficients
$A_{\boldsymbol\nu,\boldsymbol\nu'}^{(0)}$
are
$$
  A_{\boldsymbol\nu,\boldsymbol\nu'}^{(0)}~=~
  \delta_{\boldsymbol\nu,\boldsymbol\nu'},
$$
where $\delta_{\boldsymbol\nu,\boldsymbol\nu'}$
is the Kronecker delta.
Equation for the coefficients
$A_{\boldsymbol\nu,\boldsymbol\nu';\mbfk}^{(1)}$
in the first approximation can be get from Eq. (\ref{eq-for-A})
by substituting in the right hand side
$A_{\boldsymbol\nu,\boldsymbol\nu'}^{(0)}$ instead of
$A_{\boldsymbol\nu,\boldsymbol\nu'}(t)$,
\begin{eqnarray}
  i \hbar
  \dot{A}_{\boldsymbol\nu,\boldsymbol\nu';\mbfk}^{(1)}(t)
  =
  V_1 \cos(\omega_1 t)
  {\mathcal{V}}_{\boldsymbol\nu',\boldsymbol\nu}
  e^{-i \Omega_{\boldsymbol\nu,\boldsymbol\nu';\mbfk} t},
  \label{eq-for-A-perturb}
\end{eqnarray}
where the resonant frequencies are
$$
  \Omega_{\boldsymbol\nu,\boldsymbol\nu';\mbfk}
  =
  \frac{1}{\hbar}~
  \Big(
      \veps_{\boldsymbol\nu,\mbfk}^{(0)}-
      \veps_{\boldsymbol\nu',\mbfk}^{(0)}
  \Big).
$$

Solving eq. (\ref{eq-for-A-perturb}),
\begin{eqnarray}
  A_{\boldsymbol\nu,\boldsymbol\nu';\mbfk}^{(1)}(t)
  &=&
  \frac{V_1
        {\mathcal{V}}_{\boldsymbol\nu',\boldsymbol\nu}
        e^{-i (\Omega_{\boldsymbol\nu,\boldsymbol\nu';\mbfk}+\omega_1) t}}
       {2
        \Big(
            \veps_{\boldsymbol\nu,\mbfk}^{(0)}-
            \veps_{\boldsymbol\nu',\mbfk}^{(0)}+
            \hbar\omega_1
        \Big)}+
  \nonumber \\ &+&
  \frac{V_1
        {\mathcal{V}}_{\boldsymbol\nu',\boldsymbol\nu}
        e^{-i (\Omega_{\boldsymbol\nu,\boldsymbol\nu';\mbfk}-\omega_1) t}}
       {2
        \Big(
            \veps_{\boldsymbol\nu,\mbfk}^{(0)}-
            \veps_{\boldsymbol\nu',\mbfk}^{(0)}-
            \hbar\omega_1
        \Big)}.
  \label{A-res}
\end{eqnarray}

The energy dispersion relation of the trapped atoms
calculated in the second order perturbation theory
with $V_1$ is
\begin{eqnarray}
  \veps_{\boldsymbol\nu,\mbfk}
  &=&
  \veps_{\boldsymbol\nu,\mbfk}^{(0)}+
  \frac{V_{1}^{2}}{2\hbar}
  \sum_{\boldsymbol\nu' \neq \nu}
  \frac{\Omega_{\boldsymbol\nu,\boldsymbol\nu';\mbfk}~
        \big|
            {\mathcal{V}}_{\boldsymbol\nu',\boldsymbol\nu}
        \big|^{2}}
       {\Omega_{\boldsymbol\nu,\boldsymbol\nu';\mbfk}^{2}-
            \omega_{1}^{2}}.
  \label{disp-due-to-V1}
\end{eqnarray}

We are interested here in the energy bands $\boldsymbol\nu=(0,1)$
and $(1,0)$.
We perform the numerical calculations for $V_0=0.2{\mathcal{E}}_{0}$,
$V_1=0.028{\mathcal{E}}_{0}$ and $\hbar\omega_1=1.79{\mathcal{E}}_{0}$.
Non-perturbed energies ${\mathcal{E}}_{\alpha,\nu_{\alpha},k_{\alpha}}^{(0)}$
[$\alpha=x,y$] (\ref{energy2D=energy-x-y}) are shown in
Fig. \ref{Fig-energy-periodic}, solid curves. The dashed curves show
${\mathcal{E}}_{\alpha,\nu_{\alpha},k_{\alpha}}^{(0)}+\hbar\omega_1$.
It is seen the $|V_1|$ is small with respect to the excitation energy
$|\hbar\omega_1+{\mathcal{E}}_{\alpha,\nu'_{\alpha},k_{\alpha}}^{(0)}-%
{\mathcal{E}}_{\alpha,\nu_{\alpha},k_{\alpha}}^{(0)}|$
(where ${\mathcal{E}}_{\alpha,\nu'_{\alpha},k_{\alpha}}^{(0)}$
are the energies of excited states), and therefore the perturbation
theory is correct.
Moreover, we can write the energies (\ref{disp-due-to-V1}) as
\begin{subequations}
\begin{eqnarray}
  \veps_{(0,1),\mbfk} &=&
  {\mathcal{E}}_{x,0,k_x}+
  {\mathcal{E}}_{y,1,k_y},
  \label{E01-time-vs-veps}
  \\
  \veps_{(1,0),\mbfk} &=&
  {\mathcal{E}}_{x,1,k_x}+
  {\mathcal{E}}_{y,0,k_y},
  \label{E10-time-vs-veps}
\end{eqnarray}
where
\begin{eqnarray}
  {\mathcal{E}}_{x,0,k_x} &=&
  {\mathcal{E}}_{x,0,k_x}^{(0)}+
  \frac{V_{1}^{2}}{4}
  \times \nonumber \\ &\times&
  \frac{\big|
            {\mathcal{V}}_{(2,1),(0,1)}
        \big|^{2}}
       {{\mathcal{E}}_{x,0,k_x}^{(0)}-
        {\mathcal{E}}_{x,2,k_x}^{(0)}+
        \hbar\omega_1},
  \label{veps-0x}
  \\
  {\mathcal{E}}_{y,0,k_y} &=&
  {\mathcal{E}}_{y,0,k_y}^{(0)}+
  \frac{V_{1}^{2}}{4}
  \times \nonumber \\ &\times&
  \frac{\big|
            {\mathcal{V}}_{(1,2),(1,0)}
        \big|^{2}}
       {{\mathcal{E}}_{y,0,k_y}^{(0)}-
        {\mathcal{E}}_{y,2,k_y}^{(0)}+
        \hbar\omega_1},
  \label{veps-0y}
\end{eqnarray}
  %%%%%
\begin{eqnarray}
  {\mathcal{E}}_{x,1,k_x} &=&
  {\mathcal{E}}_{x,1,k_x}^{(0)}+
  \frac{V_{1}^{2}}{4}
  \times \nonumber \\ &\times&
  \sum_{\nu_x=2,3}
  \frac{\big|
            {\mathcal{V}}_{(\nu_x,0),(0,1)}
        \big|^{2}}
       {{\mathcal{E}}_{x,1,k_x}^{(0)}-
        {\mathcal{E}}_{x,\nu_x,k_x}^{(0)}+
        \hbar\omega_1},
  \label{veps-1x}
  \\
  {\mathcal{E}}_{y,1,k_y} &=&
  {\mathcal{E}}_{y,1,k_y}^{(0)}+
  \frac{V_{1}^{2}}{4}
  \times \nonumber \\ &\times&
  \sum_{\nu_y=2,3}
  \frac{\big|
            {\mathcal{V}}_{(0,\nu_y),(1,0)}
        \big|^{2}}
       {{\mathcal{E}}_{y,1,k_y}^{(0)}-
        {\mathcal{E}}_{y,\nu_y,k_y}^{(0)}+
        \hbar\omega_1}.
  \label{vwps-1y}
\end{eqnarray}
  \label{subeqs-e01-e10-time}
\end{subequations}

It is important that ${\mathcal{E}}_{x,0,k_x}$ and
${\mathcal{E}}_{x,1,k_x}$ depend on $k_x$ and
do not depend on $k_y$. From the other side,
${\mathcal{E}}_{y,0,k_y}$ and
${\mathcal{E}}_{y,1,k_y}$ depend on $k_y$
and do not depend on $k_x$.

\begin{figure}[htb]
%H=6.28, L=14.65
\centering
  \includegraphics[width=65 mm,angle=0]
   {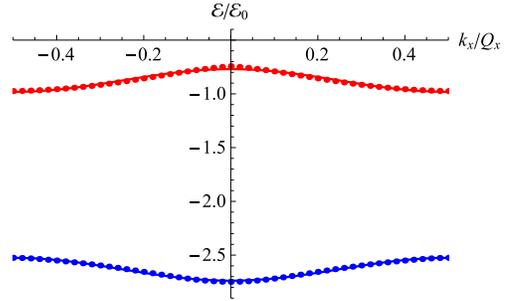}
 \caption{\footnotesize
   {\color{blue}(color online)}
   Energy levels calculated numerically from
   Eq. (\ref{eq-Schrodinger-periodic}) for
   $V_0=0.2{\mathcal{E}}_{0}$,
   $V_1=0.028{\mathcal{E}}_{0}$,
   $\beta=0.1$ and $\phi=7\pi/30$.
   The solid blue and red lines correspond
   to even and odd $\nu_x$.}
 \label{Fig-energy-fit}
\end{figure}

We plot ${\mathcal{E}}_{x,\nu_x,k_x}$ ($\nu_x=0,1$) for
$V_0=0.2{\mathcal{E}}_{0}$,
$V_1=0.028{\mathcal{E}}_{0}$ and
$\hbar\omega_1=1.79{\mathcal{E}}_{0}$ in Fig. \ref{Fig-energy-fit},
dotted curve.
It can be shown that the plot for ${\mathcal{E}}_{y,\nu_y,k_y}$
is very similar but slightly different from Fig. \ref{Fig-energy-fit} and
is not shown here.
Note that Eq. (\ref{subeqs-e01-e10-time}) has a cumbersome form.
However, it is clear that when the atoms are strongly localized at
the bottom of the wells of the optical lattice potential, the probability
of an atom to jump into adjacent sites is relatively low, whereas
the probability to jump into a next adjacent site is negligible.
This means that a tight-binding description is a good approximation.
By the other worlds, the energies
${\mathcal{E}}_{\alpha,\nu_{\alpha},k_{\alpha}}$ can be
fitted  by a simple equation as,
\begin{eqnarray}
  {\mathcal{E}}_{\alpha,\nu_{\alpha},k_{\alpha}}
  &\approx&
  \veps_{\nu_{\alpha}}+
  \Lambda_{\nu_{\alpha}}
  \cos(k_{\alpha} a_{\alpha}),
  \label{veps-nu-fit}
\end{eqnarray}
and find the parameters $\veps_{\nu}$ and $\Lambda_{\nu}$
from the condition of minimal standard deviation,
\begin{eqnarray*}
  \sum_{k}
  \Big(
      {\mathcal{E}}_{\alpha,\nu_{\alpha},k_{\alpha}}^{\mathrm{(num)}}-
      {\mathcal{E}}_{\alpha,\nu_{\alpha},k_{\alpha}}^{\mathrm{(fit)}}
  \Big)^{2},
\end{eqnarray*}
where ${\mathcal{E}}_{\alpha,\nu_{\alpha},k_{\alpha}}^{\mathrm{(num)}}$
are the energies calculated numerically from Eq. (\ref{subeqs-e01-e10-time}),
and ${\mathcal{E}}_{\alpha,\nu_{\alpha},k_{\alpha}}^{\mathrm{(fit)}}$ are
calculated from Eq. (\ref{veps-nu-fit}).
The fitting parameters calculated numerically are
\begin{eqnarray*}
  &&
  \veps_0 =
  -2.6315~
  {\mathcal{E}}_{0},
  \ \ \ \ \
  \Lambda_0 =
  -0.1061~
  {\mathcal{E}}_{0},
  \\
  &&
  \veps_1 =
  -0.8758~
  {\mathcal{E}}_{0},
  \ \ \ \ \
  \Lambda_1 =
  0.1056~
  {\mathcal{E}}_{0}.
\end{eqnarray*}

%---------------- energies fit ---------------------------

Tight binding energies ${\mathcal{E}}_{0,\alpha,k_{\alpha}}$ and
${\mathcal{E}}_{1,\alpha,k_{\alpha}}$ (\ref{veps-nu-fit}) are shown
in Fig. \ref{Fig-energy-fit}, solid curves. The solid and dotted
curves are close one to another, so that the approximation
(\ref{veps-nu-fit}) is good.

From the other side, the dispersion
(\ref{veps-nu-fit}) can be obtained
from the tight-binding model, where
$\Lambda_0$ and $\Lambda_1$
are the tunneling rates of the atoms
at the level $\nu=0$ and $1$.
Taking into account that
$\Lambda_0\approx\Lambda_1$,
we conclude that the tight-binding
model is described by a single parameter,
the tunneling rate $\Lambda \approx \Lambda_0\approx\Lambda_1$.
%The tunneling rates can be increased using
%the photon assisted tunneling technique
%\cite{Demler-17,SOC-PRL10a,SOC-JPhB13,SOC-PRA14}.

The wave functions $\Phi_{1,0}^{\mathrm{(2D)}}(\mbfr)$ and
$\Phi_{0,1}^{\mathrm{(2D)}}(\mbfr)$ satisfy the following
properties,
\begin{subequations}
\begin{eqnarray}
  \Phi_{1,0}^{\mathrm{(2D)}}\big(-x,y\big)
  &=&
  -\Phi_{1,0}^{\mathrm{(2D)}}\big(x,y\big),
  \label{parity-10-x}
  \\
  \Phi_{0,1}^{\mathrm{(2D)}}\big(-x,y\big)
  &=&
  \Phi_{0,1}^{\mathrm{(2D)}}\big(x,y\big),
  \label{parity-01-x}
\end{eqnarray}
\begin{eqnarray}
  \Phi_{1,0}^{\mathrm{(2D)}}\big(x,-y\big)
  &=&
  \Phi_{1,0}^{\mathrm{(2D)}}\big(x,y\big),
  \label{parity-10-y}
  \\
  \Phi_{0,1}^{\mathrm{(2D)}}\big(x,-y\big)
  &=&
  -\Phi_{0,1}^{\mathrm{(2D)}}\big(x,y\big).
  \label{parity-01-y}
\end{eqnarray}
  \label{subeqs-parity}
\end{subequations}
When an atom tunnels from the site $\mbfn$ to the site
$\mbfn+\mbfa_x$, the parity with respect to $y\to-y$
inversion is a good quantum number. Similarly, when
an atom tunnels from the site $\mbfn$ to the site
$\mbfn+\mbfa_y$, the parity with respect to $x\to-x$
inversion is a good quantum number.
Because of this, tunneling of atoms from the quantum state
$\Phi_{1,0}^{\mathrm{(2D)}}(\mbfr-\mbfn)$ in the site $\mbfn$ to
the quantum state
$\Phi_{0,1}^{\mathrm{(2D)}}(\mbfr-\mbfn-\mbfa_{\alpha})$ in
the site $\mbfn+\mbfa_{\alpha}$ is forbidden. Similarly,
tunneling of atoms from the quantum state
$\Phi_{0,1}^{\mathrm{(2D)}}(\mbfr-\mbfn)$ to the quantum state
$\Phi_{1,0}^{\mathrm{(2D)}}(\mbfr-\mbfn-\mbfa_{\alpha})$ is
forbidden.

%%%%%%%%%%%%%%%%%%%%
\subsubsection{Tight Binding Hamiltonian}
  \label{sec-tight-bind-subsec-tight-bind}

Summarising the results of subsections
\ref{sec-tight-bind-subsec-states},
\ref{sec-tight-bind-subsec-beyond-harm} and
\ref{sec-tight-bind-subsec-tunnel}, we are ready to construct
a tight binding Hamiltonian $H_0$. Let $\psi_{{\mathrm{c}},f}(\mbfn)$ and
$\psi_{{\mathrm{c}},f}^{\dag}(\mbfn)$ be annihilation and creation operators
of atom in the quantum state
$\Phi_{1,0}^{\mathrm{(2D)}}(\mbfr-\mbfn)$ in the site $\mbfn$
with magnetic quantum number $f$. Similarly, $\psi_{{\mathrm{v}},f}(\mbfn)$
and $\psi_{{\mathrm{v}},f}^{\dag}(\mbfn)$ be annihilation and creation
operators of atom in the quantum state
$\Phi_{0,1}^{\mathrm{(2D)}}(\mbfr-\mbfn)$. Here the index ${\mathrm{c}}$ and
${\mathrm{v}}$ means ``conduction'' and ``valence'' band. Then the tight
binding Hamiltonian is
\begin{eqnarray}
  H_0 &=&
  \veps_0
  \sum_{f,\mbfn}
  \Big\{
      \psi_{{\mathrm{c}},f}^{\dag}(\mbfn)
      \psi_{{\mathrm{c}},f}(\mbfn)-
      \psi_{{\mathrm{v}},f}^{\dag}(\mbfn)
      \psi_{{\mathrm{v}},f}(\mbfn)
  \Big\}+
  \nonumber \\ && +
  \sum_{f,\xi}
  \sum_{\langle\mbfn,\mbfn'\rangle}
  \Lambda_{\xi}\big(\mbfa\big)
  \psi_{\xi,f}^{\dag}(\mbfn)
  \psi_{\xi,f}(\mbfn').
  \label{H0-tight-bind-def}
\end{eqnarray}
Here $\langle\mbfn,\mbfn'\rangle$ denotes neighboring sites,
$\mbfa=\mbfn'-\mbfn$ and
\begin{eqnarray*}
  &&
  \Lambda_{v}(\pm\mbfa_x) ~=~ \Lambda,
  \ \ \ \ \
  \ \ \
  \Lambda_{v}(\pm\mbfa_y) ~=~ -\Lambda,
  \\
  &&
  \Lambda_{c}(\pm\mbfa_x) ~=~ -\Lambda,
  \ \ \ \ \
  \Lambda_{c}(\pm\mbfa_y) ~=~ \Lambda.
\end{eqnarray*}
$\veps_0$ is given by eq. (\ref{epsilon0-def}).
We chose the Fermi energy to be zero.

In order to diagonalize $H_0$, we apply Fourier transformations,
\begin{eqnarray}
  \psi_{\xi,f}(\mbfn) &=&
  \frac{1}{\sqrt{\mathcal{N}}}~
  \sum_{\mbfk}
  c_{\xi,\mbfk,f}~
  e^{i \mbfk \mbfn - i \pi n_y},
  \label{Fourier-def}
\end{eqnarray}
where ${\mathcal{N}}$ is the number of sites of the optical
lattice. Then $H_0$ takes the form,
\begin{eqnarray}
  H_0 &=&
  \sum_{\mbfk,f}
  \veps_{\mbfk}~
  \Big\{
      c_{{\mathrm{c}},\mbfk,f}^{\dag}
      c_{{\mathrm{c}},\mbfk,f}-
      c_{{\mathrm{v}},\mbfk,f}^{\dag}
      c_{{\mathrm{v}},\mbfk,f}
  \Big\},
  \label{H0-diagonal}
\end{eqnarray}
where
\begin{eqnarray}
  \veps_{\mbfk} =
  \veps_0-
  2\Lambda
  \Big\{
      \cos\big(k_x a_x)+
      \cos\big(k_y a_y\big)
  \Big\}.
  \label{En0-k}
\end{eqnarray}

It should be noted the following:
The lowest energy of the ``conduction'' band is
$\Delta_0$, whereas the highest energy of the ``valence'' band
is $-\Delta_0$, where
\begin{eqnarray}
  \Delta_0 ~=~
  \veps_0-
  4\Lambda.
  \label{Delta0-def}
\end{eqnarray}
Overlapping of the conduction and valence bands depends either
$\Delta_0$ is positive or negative.
\begin{itemize}
\item When $\Delta_0$ is positive, there is a gap $2\Delta_0$
      separating the conduction and valence bands and the optical
      lattice is an insulator. In this case we need to apply
      external magnetic field to get topological edge states.

\item When $\Delta_0$ is negative, there is overlapping of
      the conduction and valence bands. Spin-orbital interaction
      (considered in subsection
      \ref{sec-tight-bind-subsec-spin-orbit} below) opens a gap in
      the bulk of the optical lattice and keep the edge states
      gapless. In this case we get edge states even without
      magnetic field.
\end{itemize}

%---------------- V and B ---------------------------
\begin{figure}[htb]
%H=6.28, L=14.65
\centering
  \subfigure[]
  {\includegraphics[width=40 mm,angle=0]
   {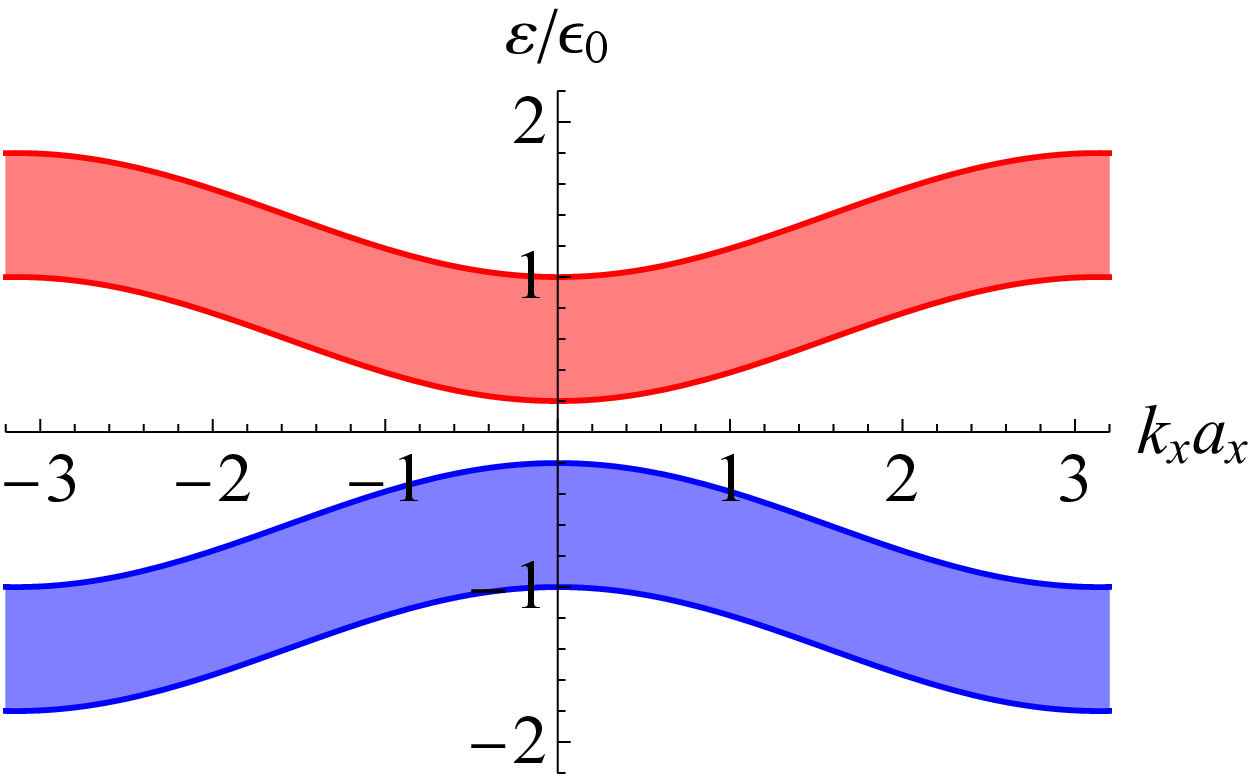}
   \label{Fig-Sp0-low}}
  \subfigure[]
  {\includegraphics[width=40 mm,angle=0]
   {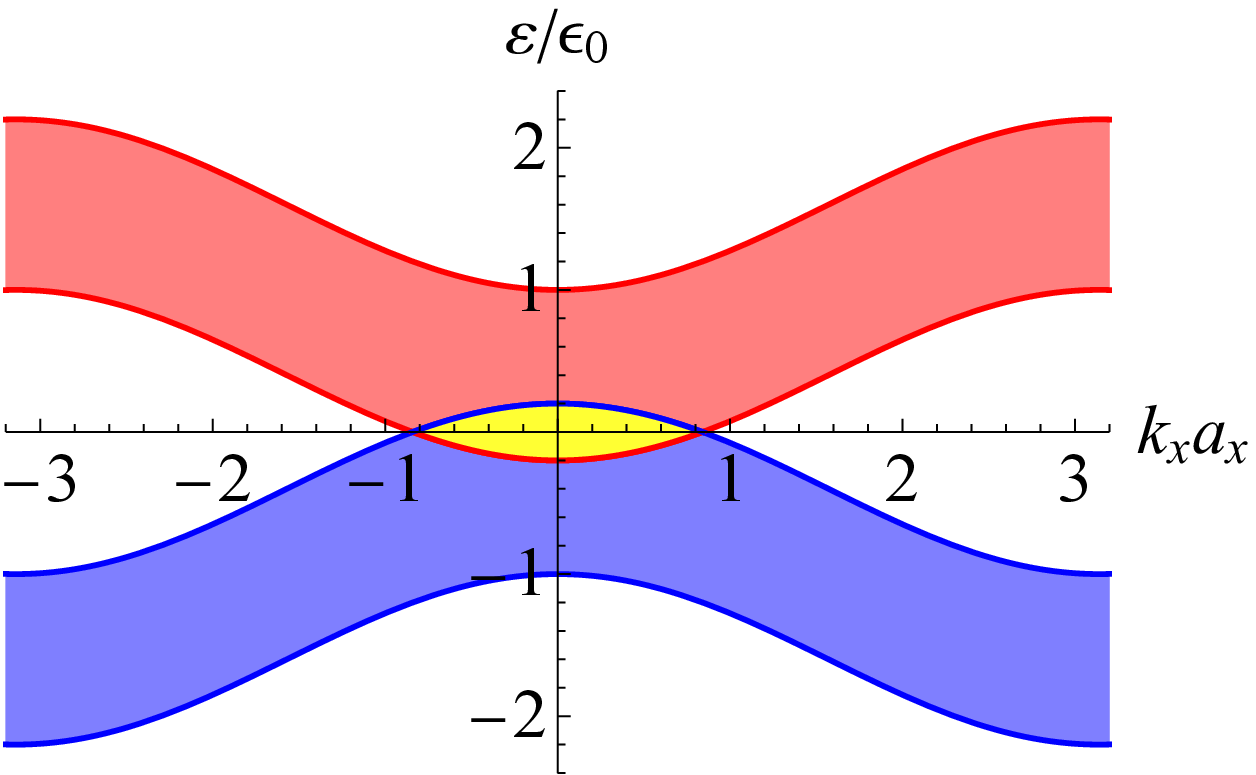}
   \label{Fig-Sp0-high}}
 \caption{\footnotesize
   {\color{blue}(color online)}
   Energy spectrum (\ref{En0-k}) for $\Lambda_0=0.2\veps_0$
   [panel (a)] and $\Lambda=0.3\veps_0$ [panel (b)].
   The red and blue filled areas denote the conduction and
   valence bands, the yellow filled area in the panel (b) denotes
   overlapping of the bands.}
 \label{Fig-Sp0-low-high}
\end{figure}

The cases $\Delta_0>0$ and $\Delta_0<0$ are illustrated in Fig.
\ref{Fig-Sp0-low-high}.

%%%%%%%%%%%%%%%%%%%%
\subsection{Spin-Orbit Interaction}
  \label{sec-tight-bind-subsec-spin-orbit}

Spin orbit interaction Hamiltonian is
\begin{eqnarray}
  H_{\mathrm{SO}} &=&
  \frac{1}{2}~
  \sum_{\xi,\xi'}
  \sum_{f,f'}
  \sum_{\langle\mbfn,\mbfn'\rangle}
  {\mathcal{S}}_{\xi,\xi';f,f'}(\mbfa)
  \times \nonumber \\ && \times
  \psi_{\xi,f}^{\dag}(\mbfn)
  \psi_{\xi',f'}(\mbfn'),
  \label{HSO-tight-bind-def}
\end{eqnarray}
where $\langle\mbfn,\mbfn'\rangle$ denotes
neighboring sites, $\mbfa=\mbfn'-\mbfn$,
\begin{eqnarray}
  {\mathcal{S}}_{\xi,\xi';f,f'}(\mbfa) &=&
  \frac{1}{2I+1}
  \int{d^2\mbfr}~
  \Phi_{\xi}(\mbfr)
  \Phi_{\xi'}(\mbfr-\mbfa)
  \times \nonumber \\ && \times
  \Big(
      {\bm{B}}(\mbfr)
      \cdot
      {\bm{F}}_{f,f'}
  \Big),
  \label{SO-matrix-element}
\end{eqnarray}
where
${\bm{B}}(\mbfr)$ is given by
eq. (\ref{subeqs-B-vec-res}). Here we use the expression
$$
  \sum_{f,f'}
  \big|
      f
  \big\rangle
  \big\langle
      f
  \big|
      \hat{\bm{J}}
  \big|
      f'
  \big\rangle
  \big\langle
      f'
  \big|
  ~=~
  \frac{\hat{\bm{F}}}{2I+1},
$$
[which is true for the case $F=I\pm\frac{1}{2}$,
where $J=\frac{1}{2}$ is the electronic orbital moment and
$I$ is the nuclear spin].
Here we use the notations $\Phi_{{\mathrm{c}}}(\mbfr)$ and $\Phi_{{\mathrm{v}}}(\mbfr)$
for $\Phi_{1,0}^{\mathrm{(2D)}}(\mbfr)$ and
$\Phi_{0,1}^{\mathrm{(2D)}}(\mbfr)$.

Let us consider in details tunneling of atoms between the sites
$\mbfn=0$ and $\mbfn'=\mbfa_x$. The matrix element
(\ref{SO-matrix-element}) can be written as
\begin{eqnarray*}
  {\mathcal{S}}_{\xi,\xi';f,f'}(\mbfa_x) =
  {\mathcal{S}}_{\xi,\xi'}^{x}(\mbfa_x)~
  F_{f,f'}^{x}+
  {\mathcal{S}}_{\xi,\xi'}^{y}(\mbfa_x)~
  F_{f,f'}^{y},
\end{eqnarray*}
where
\begin{eqnarray*}
  {\mathcal{S}}_{\xi,\xi'}^{x}(\mbfa_x) &=&
  \frac{1}{2I+1}
  \int{d^2\mbfr}~
  \Phi_{\xi}(\mbfr)
  \Phi_{\xi'}(\mbfr-\mbfa_x)
  B^{x}(\mbfr),
  \\
  {\mathcal{S}}_{\xi,\xi'}^{y}(\mbfa_x) &=&
  \frac{1}{2I+1}
  \int{d^2\mbfr}~
  \Phi_{\xi}(\mbfr)
  \Phi_{\xi'}(\mbfr-\mbfa_x)
  B^{y}(\mbfr).
\end{eqnarray*}
The integration is restricted by the interval
$$
  |y| ~\lesssim~ y_0,
  \ \ \
  x_0 ~<~ x ~<~ a_x-x_0,
$$
where $x_0$ and $y_0$ are classical turning points for atom
at the energy level ${\mathcal{E}}_{1,0}$ or
${\mathcal{E}}_{0,1}$.
Consider ${\mathcal{S}}_{\xi,\xi'}^{x}(\mbfa_x)$.
It is important that $B_x(\mbfr)$ is even with respect to
the inversion $y\to-y$ and odd with respect to
the glide inversion $x\to{a}_{x}-x$. The functions
$\Phi_{\mathrm{c}}(\mbfr)\Phi_{\mathrm{v}}(\mbfr-\mbfa_x)$ and
$\Phi_{\mathrm{v}}(\mbfr)\Phi_{\mathrm{c}}(\mbfr-\mbfa_x)$ are odd with respect to
the $y\to-y$ inversion. Therefore
$$
  {\mathcal{S}}_{{\mathrm{c,v}}}^{x}(\mbfa_x)
  ~=~
  {\mathcal{S}}_{\mathrm{v,c}}^{x}(\mbfa_x)
  ~=~ 0.
$$
From the other side, the functions
$\Phi_{\xi}(\mbfr)\Phi_{\xi}(\mbfr-\mbfa_x)$
($\xi={\mathrm{c,v}}$) are even with respect to
the $x\to{a}_{x}-x$ glide inversion. Therefore
$$
  {\mathcal{S}}_{\mathrm{c,c}}^{x}(\mbfa_x)
  ~=~
  {\mathcal{S}}_{\mathrm{v,v}}^{x}(\mbfa_x)
  ~=~ 0.
$$

Consider ${\mathcal{S}}_{\xi,\xi'}^{y}(\mbfa_x)$.
$B_y(\mbfr)$ is odd with respect to
the inversion $y\to-y$ and even with respect to
the glide inversion $x\to{a}_{x}-x$. The functions
$\Phi_{\xi}(\mbfr)\Phi_{\xi}(\mbfr-\mbfa_x)$ ($\xi={\mathrm{c,v}}$) are
even with respect to the $y\to-y$ inversion. Therefore
$$
  {\mathcal{S}}_{\mathrm{c,c}}^{y}(\mbfa_x)
  ~=~
  {\mathcal{S}}_{\mathrm{v,v}}^{y}(\mbfa_x)
  ~=~ 0.
$$
The functions
$\Phi_{\mathrm{c}}(\mbfr)\Phi_{\mathrm{v}}(\mbfr-\mbfa_x)$ and
$\Phi_{\mathrm{v}}(\mbfr)\Phi_{\mathrm{c}}(\mbfr-\mbfa_x)$ are odd with respect to
the $y\to-y$ inversion. Therefore
$$
  {\mathcal{S}}_{\mathrm{c,v}}^{y}(\mbfa_x)
  ~\neq~ 0,
  \ \ \ \ \
  {\mathcal{S}}_{\mathrm{v,c}}^{y}(\mbfa_x)
  ~\neq~ 0.
$$
Using eq. (\ref{subeqs-B-vec-res}), we can write
\begin{eqnarray}
  {\mathcal{S}}_{\mathrm{c,v}}^{y}(\mbfa_x)
  &=&
  \frac{B_0 \sin\phi}{2I+1}
  \int\limits_{-y_0}^{y_0}
  \Phi_{0}(y)
  \Phi_{1}(y)
  \sin
  \bigg(
       \frac{2 \pi y}{a_x}
  \bigg)
  \times \nonumber \\ &\times&
  \int\limits_{x_0}^{a_x-x_0}
  \Phi_{1}(x)
  \Phi_{0}(x-a_x)
  \cos^2
  \bigg(
       \frac{\pi x}{a_x}
  \bigg).
  \label{S-cv-ax-res}
\end{eqnarray}
Similarly, we can derive ${\mathcal{S}}_{\mathrm{v,c}}^{y}(\mbfa_x)$
\begin{eqnarray}
  {\mathcal{S}}_{\mathrm{v,c}}^{y}(\mbfa_x)
  &=&
  \frac{B_0 \sin\phi}{2I+1}
  \int\limits_{-y_0}^{y_0}
  \Phi_{1}(y)
  \Phi_{0}(y)
  \sin
  \bigg(
       \frac{2 \pi y}{a_x}
  \bigg)
  \times \nonumber \\ &\times&
  \int\limits_{x_0}^{a_x-x_0}
  \Phi_{0}(x)
  \Phi_{1}(x-a_x)
  \cos^2
  \bigg(
       \frac{\pi x}{a_x}
  \bigg).
  \label{S-vc-ax-res}
\end{eqnarray}
Note that $\Phi_{0}(x)$ is positive for both $x>0$ and $x<0$,
whereas $\Phi_{1}(x)$ is positive for $x>0$ and negative for
$x<0$. The function $\Phi_1(x)$ in the right hand side of
eq. (\ref{S-cv-ax-res}) is positive, whereas the function
$\Phi_1(x-a_x)$ in the right hand side of eq. (\ref{S-vc-ax-res})
is negative. Then we conclude that
\begin{eqnarray*}
  {\mathcal{S}}_{\mathrm{c,v}}^{y}(\mbfa_x)
  ~=~
  -{\mathcal{S}}_{\mathrm{v,c}}^{y}(\mbfa_x).
\end{eqnarray*}

Similarly, we can see that for the tunneling between the sites
$\mbfn=0$ and $\mbfn'=\mbfa_y$, only
${\mathcal{S}}_{\mathrm{c,v}}^{x}(\mbfa_y)$ and
${\mathcal{S}}_{\mathrm{c,v}}^{x}(\mbfa_y)$ are nontrivial,
\begin{eqnarray*}
  {\mathcal{S}}_{\mathrm{c,v}}^{x}(\mbfa_y)
  &=&
  \frac{B_0 \cos\phi}{2I+1}
  \int\limits_{-x_0}^{x_0}
  \Phi_{0}(x)
  \Phi_{1}(x)
  \sin
  \bigg(
       \frac{2 \pi x}{a_x}
  \bigg)
  \times \nonumber \\ &\times&
  \int\limits_{y_0}^{a_y-y_0}
  \Phi_{1}(y)
  \Phi_{0}(y-a_y)
  \cos^2
  \bigg(
       \frac{\pi y}{a_y}
  \bigg),
  \\
  {\mathcal{S}}_{\mathrm{v,c}}^{x}(\mbfa_y)
  &=&
  \frac{B_0 \cos\phi}{2I+1}
  \int\limits_{-x_0}^{x_0}
  \Phi_{1}(x)
  \Phi_{0}(x)
  \sin
  \bigg(
       \frac{2 \pi x}{a_x}
  \bigg)
  \times \nonumber \\ &\times&
  \int\limits_{y_0}^{a_y-y_0}
  \Phi_{0}(y)
  \Phi_{1}(y-a_y)
  \cos^2
  \bigg(
       \frac{\pi y}{a_y}
  \bigg).
\end{eqnarray*}
Taking into account that $\Phi_1(y)$ is odd with respect to
the inversion $y\to-y$, we can see that
\begin{eqnarray*}
  {\mathcal{S}}_{\mathrm{c,v}}^{x}(\mbfa_y)
  ~=~
  -{\mathcal{S}}_{\mathrm{v,c}}^{x}(\mbfa_y).
\end{eqnarray*}
When the angle $\phi$ is close to $\pi/4$, then
\begin{eqnarray*}
  {\mathcal{S}}_{\mathrm{c,v}}^{x}(\mbfa_y)
  ~\approx~
  {\mathcal{S}}_{\mathrm{c,v}}^{y}(\mbfa_x)
  ~\equiv~
  \lambda_{\mathrm{SO}}.
\end{eqnarray*}
Here
\begin{eqnarray}
  \lambda_{\mathrm{SO}} \approx
  \frac{\pi^3 B_0}{2\sqrt{2}~(2I+1)}~
  \frac{x_{0}^{2}}{a_{0}^{2}}~
  \exp
  \bigg(
      -\frac{a_{0}^{2}}{4x_{0}^{2}}
  \bigg),
  \label{lambda-SO-res}
\end{eqnarray}
where ${a}_{x}\approx{a}_{y}\approx{a}_{0}$,
$$
  x_0 ~=~
  \Bigg(
       \frac{2\sqrt{2}~\hbar^2}
            {K_0 M \big(2-3\beta^2\big)}
  \Bigg)^{1/4}.
$$
The spin-orbit coupling rate can be increased using
the photon assisted tunneling technique
\cite{Demler-17,SOC-PRL10a,SOC-JPhB13,SOC-PRA14}.

Then the Hamiltonian (\ref{HSO-tight-bind-def}) can be written as
\begin{eqnarray}
  H_{\mathrm{SO}} &=&
  \lambda_{\mathrm{SO}}
  \sum_{f,f'}
  \sum_{\langle\mbfn,\mbfn'\rangle}
  \frac{1}{|\mbfa|}~
  \Big(
      \big[
          \mbfa
          \times
          \mbfe_z
      \big]
      \cdot
      {\mathbf{F}}_{f,f'}
  \Big)
  \times \nonumber \\ && \times
  \psi_{{\mathrm{c}},f}^{\dag}(\mbfn)
  \psi_{{\mathrm{v}},f'}(\mbfn'),
  \label{HSO-tight-bind-approx}
\end{eqnarray}
where $\langle\mbfn,\mbfn'\rangle$ denotes the neighboring sites
and
$$
  \mbfa  ~=~
  \mbfn'-\mbfn.
$$

Applying Fourier transformations (\ref{Fourier-def}), we can write
$H_{\mathrm{SO}}$ in the form,
\begin{eqnarray*}
  H_{\mathrm{SO}} =
  \sum_{\mbfk,f,f'}
  \big(
      \tilde{\boldsymbol\lambda}_{\mbfk}
      \cdot
      {\mathbf{F}}_{f,f'}
  \big)
  \Big\{
      i
      c_{{\mathrm{c}},\mbfk,f}^{\dag}
      c_{{\mathrm{v}},\mbfk,f'}+
      {\mathrm{h.c.}}
  \Big\},
\end{eqnarray*}
where
\begin{eqnarray*}
  \tilde{\boldsymbol\lambda}_{\mbfk} &=&
  2 \lambda_{\mathrm{SO}}~
  \Big\{
      \sin(k_y a_y)~
      \mbfe_x-
      \sin(k_x a_x)~
      \mbfe_y
  \Big\}.
\end{eqnarray*}
Finally, we can make the unitary transformations,
$$
  c_{{\mathrm{v}},\mbfk,f} ~\to~
  i c_{{\mathrm{v}},\mbfk,f},
$$
and applying proper unitary transformations of the matrices
$\hat{F}^{x,y,z}$, we get a standard form for the Dresselhaus
spin-orbit interaction,
\begin{eqnarray}
  H_{\mathrm{SO}} =
  \sum_{\mbfk,f,f'}
  \big(
      \boldsymbol\lambda_{\mbfk}
      \cdot
      {\mathbf{F}}_{f,f'}
  \big)
  \Big\{
      c_{{\mathrm{c}},\mbfk,f}^{\dag}
      c_{{\mathrm{v}},\mbfk,f'}+
      {\mathrm{h.c.}}
  \Big\},
  \label{HSO-k-space}
\end{eqnarray}
where
\begin{eqnarray*}
  \boldsymbol\lambda_{\mbfk} &=&
  2 \lambda_{\mathrm{SO}}~
  \Big\{
      \sin(k_x a_x)~
      \mbfe_x+
      \sin(k_y a_y)~
      \mbfe_y
  \Big\}.
\end{eqnarray*}

%%%%%%%%%%%%%%%%%%%%
\subsection{Long Wave Approximation}
  \label{sec-tight-bind-subsec-long-wave}

Let us consider quantum states with $ka_0\ll1$ [hereafter we
assume that ${a_x}\approx{a_y}\approx{a_0}$ and use $a_0$ instead
of $a_x$ and $a_y$] and apply the long wave approximation.
In this approximation, we can write
\begin{eqnarray*}
  &&
  \veps_{\mbfk} ~\approx~
  \Delta_0+
  \frac{\hbar^2 k^2}{2 M_0},
  \ \ \ \ \
  \lambda_{\mbfk} ~\approx~
  \hbar v~
  \big(
      \mbfk
      \cdot
      \hat{\bm{S}}
  \big),
\end{eqnarray*}
where $\Delta_0$ is given by eq. (\ref{Delta0-def}),
\begin{eqnarray}
  M_0 =
  \frac{\hbar^2}{2 \Lambda_0 a_{0}^{2}},
  \ \ \ \ \
  v =
  \frac{2}{\hbar}~
  \lambda_{\mathrm{SO}} a_0.
  \label{M0-A-def}
\end{eqnarray}
Then the Hamiltonian $H=H_0+H_{\mathrm{SO}}$ of the atoms in
the optical lattice is written as,
\begin{subequations}
\begin{eqnarray}
  H_0 ~=~
  \sum_{\xi,\mbfk,s}
  \xi
  \bigg(
       \Delta_0+
       \frac{\hbar^2 k^2}{2 M_0}
  \bigg)~
  c_{\xi,\mbfk,s}^{\dag}~
  c_{\xi,\mbfk,s},
  \label{H0-long-wave}
  \\
  H_{\mathrm{SO}} ~=~
  \hbar v
  \sum_{\xi,\mbfk,s,s'}
  \Big(
      \mbfk
      \cdot
      {\mathbf{S}}_{s,s'}
  \Big)~
  c_{\xi,\mbfk,s}^{\dag}~
  c_{\bar\xi,\mbfk,s'},
  \label{HSO-long-wave}
\end{eqnarray}
  \label{subeqs-H-long-wave}
\end{subequations}
where $\xi={\mathrm{c,v}}$ or $\xi=\pm1$ for the ``conduction'' or ``valence'' bands.
%It can be shown that when $\Delta_0M_0<0$, the Hamiltonian
%(\ref{subeqs-H-long-wave}) describes helical edge states. When
%$\Delta_0M_0>0$, there are no edge states. In this case, in order
%to get edge states, we need to put external magnetic field.
%Since the magnetic field violates the time reversal symmetry,
%we get chiral edge states.
%Here we consider chiral edge states for $\Delta_0M_0>0$ and
%external magnetic field, whereas helical edge states will be
%considered in another paper.

%%%%%%%%%%%%%%%%%%%%%%%%%%%%%%
\section{Numerical Calculations}
  \label{append-num-calc}

Here we apply numerical calculations to solve eqs.
(\ref{eq-for-kappa}) and (\ref{eq-for-energy}) for
$\Delta_0=0$, $B=2\epsilon_0$ and $k_x=0.1k_0$.
We consider the cases $\eta=\pm1$ in turn.

%%%%%%%%%%%%%%%%%%%%
\subsection{Numerical Calculations for $\eta=1$}
  \label{append-num-calc-subsec-p1}
Substituting $\Delta_0=0$, $B=2\epsilon_0$ and $k_x=0.1k_0$ to
eq. (\ref{eq-for-kappa}) and solving it, we get $\kappa_1$,
$\kappa_2$, $\kappa_3$ and $\kappa_4$  as functions of
$\veps_{1}(k_x)$.
We substitute $\kappa_{n}$ ($n=1,2,3,4$) into
eq. (\ref{subeqs-M-3x3}) and get the matrices which depends just
on energy $\veps_{1}(k_x)$. Then with eqs. (\ref{D-s-n=det}),
and (\ref{eq-for-energy}), we get
an equation for $\veps_{1}(k_x)$.
%---------------- kappa vs energy ---------------------------
\begin{figure}[htb]
%H=6.28, L=14.65
\centering
  \includegraphics[width=60 mm,angle=0]
   {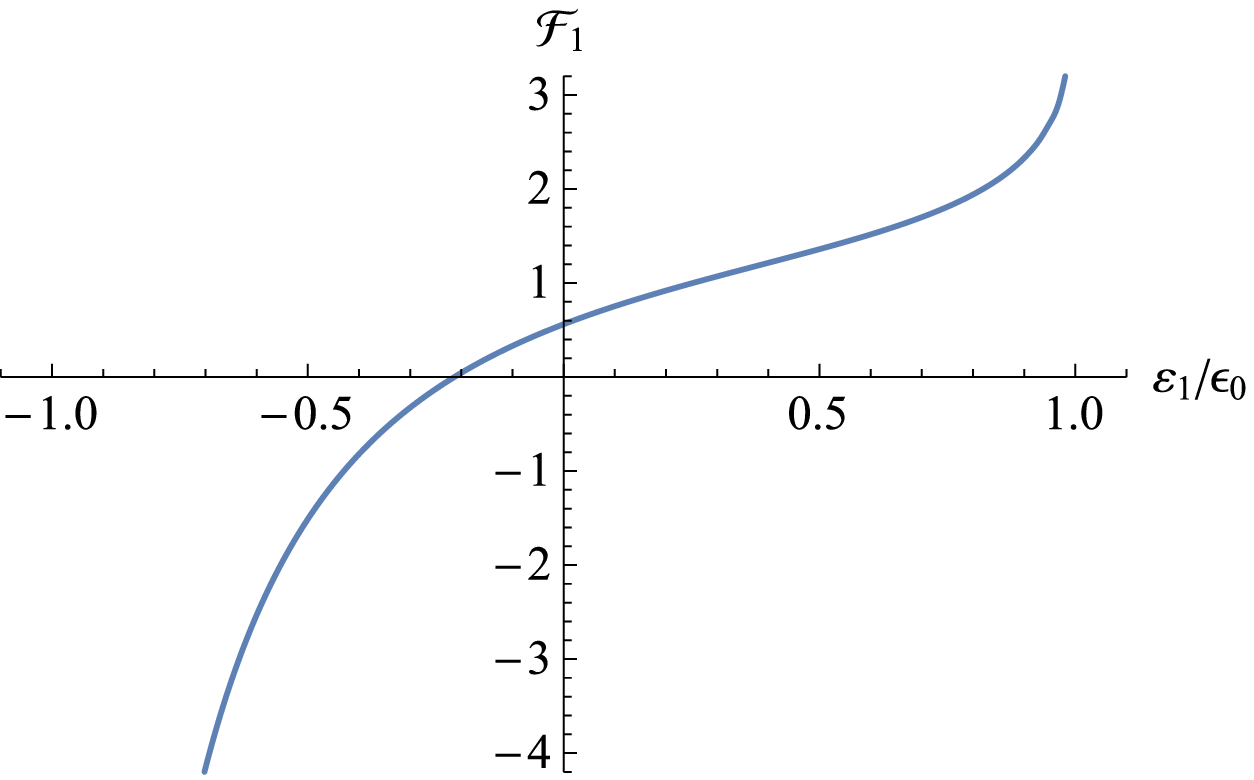}
 \caption{\footnotesize
   {\color{blue}(color online)}
  Function ${\mathcal{F}}_{1}(\veps_1)$,
  eq. (\ref{eq-for-energy}), for $\Delta_0=0$,
  $B=2\epsilon_0$, $k_x=0.1k_0$ and $\eta=1$.
  $\epsilon_0$ and $k_0$ are given by eq. (\ref{epsilon0-k0-def}).}
 \label{Fig-F-vs-energy-p1}
\end{figure}
The function ${\mathcal{F}}_{1}(\veps_1)$ is displayed in
Fig. \ref{Fig-F-vs-energy-p1}.
Here $\veps_1(k_x)$ is restricted by the condition,
$$
  \big|
      \veps_1(k_x)
  \big|
  ~<~
  \veps_{{\mathrm{c}},\frac{1}{2},1}(k_x)
  ~=~
  1.00016\epsilon_0.
$$
Solving eq. (\ref{eq-for-energy})
numerically, we get
\begin{eqnarray*}
  \veps_1 &=&
 -0.213735\epsilon_0.
\end{eqnarray*}

%%%%%%%%%%%%%%%%%%%%
\subsection{Numerical Calculations for $\eta=-1$}
  \label{append-num-calc-subsec-m1}

Consider now topological edge states for
$\Delta_0=0$, $B=2\epsilon_0$ and $k_x=0.1k_0$ and $\eta=-1$.
Solving eq. (\ref{eq-for-kappa}), we get $\kappa$'s
as functions of $\veps_{\bar{1}}(k_x)$.
It should be noted that $\kappa_1$ and $\kappa_2$ are complex.

%---------------- kappa vs energy ---------------------------
\begin{figure}[htb]
%H=6.28, L=14.65
\centering
  \includegraphics[width=60 mm,angle=0]
   {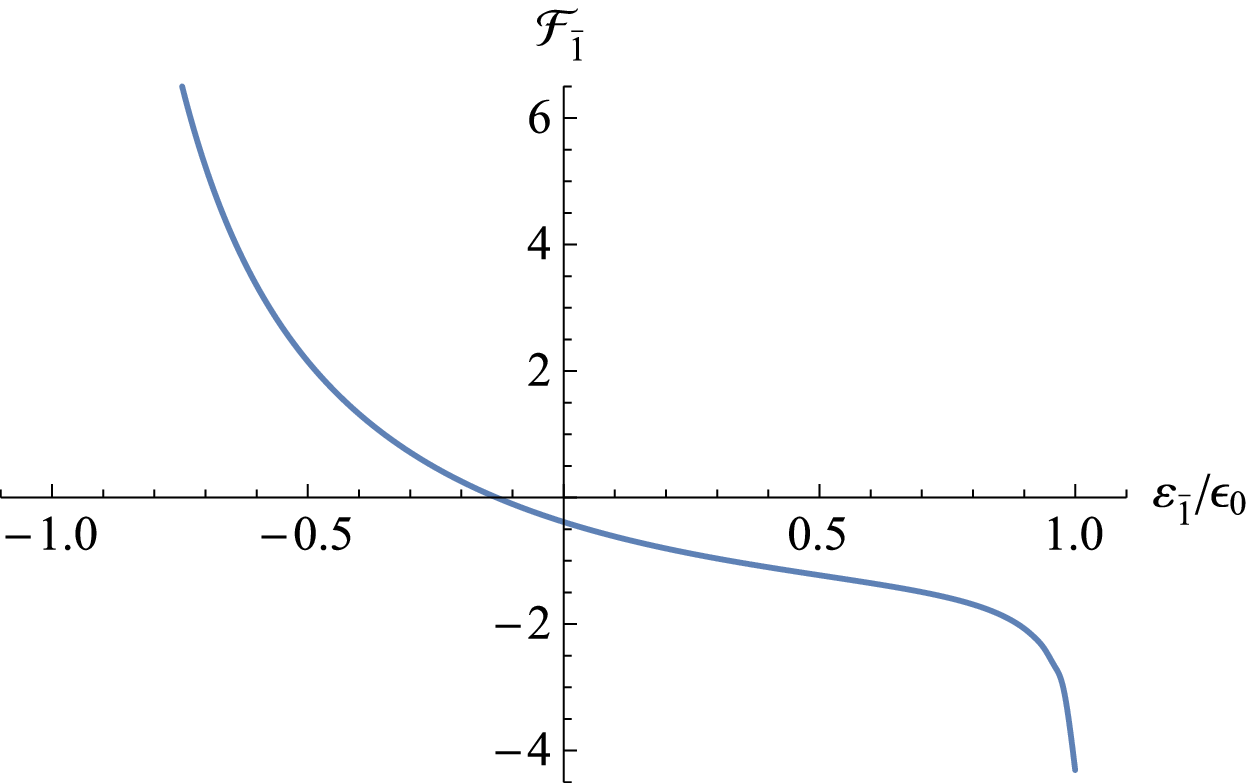}
 \caption{\footnotesize
   {\color{blue}(color online)}
   Function ${\mathcal{F}}_{\bar{1}}(\veps_{\bar{1}})$,
   eq. (\ref{eq-for-energy}),
   for $\Delta_0=0$, $B=2\epsilon_0$,
   $k_x=0.1k_0$ and $\eta=-1$.
   $\epsilon_0$ and $k_0$ are given by eq. (\ref{epsilon0-k0-def}).}
 \label{Fig-F-vs-energy-m1}
\end{figure}
We substitute $\kappa_{n}$ ($n=1,2,3,4$) into
eq. (\ref{subeqs-M-3x3}) and get the matrices which depends just
on energy $\veps_{\bar{1}}(k_x)$. Then with eqs. (\ref{D-s-n=det}),
(\ref{eq-for-energy}) and (\ref{eq-for-energy}), we get
an equation for $\veps_{\bar{1}}(k_x)$.
The function ${\mathcal{F}}_{\bar{1}}(\veps_{\bar{1}})$ is displayed in
Fig. \ref{Fig-F-vs-energy-m1}.
Here $\veps_{\bar{1}}(k_x)$ is restricted by the condition,
$$
  \big|
      \veps_{\bar{1}}(k_x)
  \big|
  ~<~
  \veps_{{\mathrm{c}},\frac{1}{2},\bar{1}}(k_x)
  ~=~
  1.00982\epsilon_0.
$$

Solving eq. (\ref{eq-for-energy})
numerically, we get
\begin{eqnarray*}
  \veps_{\bar{1}} &=&
  -0.131167\epsilon_0.
\end{eqnarray*}

%%%%%%%%%%%%%%%%%%%%%%%%%%%%%%
\section{Minima and Maxima Points of $V(\mbfr)$}
  \label{append-min-max}

We consider properties of the optical potential (\ref{V-pot-res})
for different values of $\beta$ within the interval
$$
  0 ~<~ \beta ~<~ 1.
$$
The potential $V(\mbfr)$ satisfies the periodic conditions
(\ref{V-periodic}).

When an atoms is placed in external potential $V(\mbfr)$,
the force ${\mathbf{F}}(\mbfr)$ acting on the atom is,
\begin{eqnarray}
  {\mathbf{F}}(\mbfr) &=&
  -\nabla V(\mbfr).
  \label{force-def}
\end{eqnarray}
At equilibrium points, the force acting at the atom vanishes.
Therefore we have a set of equations,
\begin{subequations}
\begin{eqnarray}
  \sin
  \bigg(
       \frac{2 \pi x}{a_x}
  \bigg)~
  \bigg\{
       1-
       \beta^2~
       \Big[
           1+
           \cos\big(2\phi\big)
       \Big]+
  \ \ \ \
  \nonumber \\ +
       \Big[
           1-2\beta^2
       \Big]
       \cos
       \bigg(
            \frac{2 \pi y}{a_y}
       \bigg)
  \bigg\}
  ~=~ 0,
  \label{eq1-equilibrium}
  \\
  \sin
  \bigg(
       \frac{2 \pi y}{a_y}
  \bigg)~
  \bigg\{
       1-
       \beta^2
       \Big[
           1-
           \cos\big(2\phi\big)
       \Big]+
  \ \ \ \
  \nonumber \\ \times
       \Big[
           1-2\beta^2
       \Big]
       \cos
       \bigg(
           \frac{2 \pi x}{a_x}
       \bigg)
  \bigg\}
  ~=~ 0,
  \label{eq2-equilibrium}
\end{eqnarray}
  \label{subeqs-for-equilibrium}
\end{subequations}
where $a_x$ and $a_y$ are given by eq. (\ref{V0-a0-def}).

Note that there are values $\beta_{c,x}$ and
$\beta_{c,y}$ of $\beta$
\begin{subequations}
\begin{eqnarray}
  \beta_{c,x} &=&
  \sqrt{\frac{2}{3+\cos(2\phi)}},
  \label{beta-cx}
  \\
  \beta_{c,y} &=&
  \sqrt{\frac{2}{3-\cos(2\phi)}},
  \label{beta-cy}
\end{eqnarray}
  \label{subeqs-betac}
\end{subequations}
\begin{subequations}
such that for $\beta<\beta_{c,x}$, the equation
\begin{eqnarray}
  1-
  \beta^2
  \Big[
      1+
      \cos\big(2\phi\big)
  \Big]+
  \ \ \ \ \
  \nonumber \\ +
  \big(
      1-2\beta^2
  \big)~
  \cos
  \bigg(
       \frac{2 \pi x}{a_x}
  \bigg)
  ~=~ 0,
  \label{eq-x-equilibrium-additional}
\end{eqnarray}
has no solutions. Similarly, for $\beta<\beta_{c,y}$, the equation
\begin{eqnarray}
  1-
  \beta^2
  \Big[
      1-
      \cos\big(2\phi\big)
  \Big]+
  \ \ \ \ \
  \nonumber \\ +
  \big(
      1-2\beta^2
  \big)~
  \cos
  \bigg(
       \frac{2 \pi y}{a_y}
  \bigg)
  ~=~ 0,
  \label{eq-y-equilibrium-additional}
\end{eqnarray}
  \label{subeqs-for-equilibrium-additional}
\end{subequations}
has no solutions. In this case the equilibrium positions are given
by the equations
\begin{subequations}
\begin{eqnarray}
  \sin
  \bigg(
       \frac{2 \pi x}{a_x}
  \bigg)
  ~=~ 0,
  \label{eq-x-equilibrium-main}
  \\
  \sin
  \bigg(
       \frac{2 \pi y}{a_y}
  \bigg)
  ~=~ 0.
  \label{eq-y-equilibrium-main}
\end{eqnarray}
  \label{subeqs-for-equilibrium-main}
\end{subequations}
For $\beta>\beta_{c,x}$ and $\beta>\beta_{c,y}$, eqs.
(\ref{subeqs-for-equilibrium-additional}) and
(\ref{subeqs-for-equilibrium-main}) have nontrivial solutions.

In what following, we assume that $0<\phi<\pi/4$, and therefore
$$
  \frac{1}{\sqrt{2}}
  ~<~
  \beta_{c,x}
  ~\leq~
  \beta_{c,y}
  ~<~ 1.
$$

In order to answer the question is the equilibrium point
$\mbfr_0=(x_0,y_0)$ stable, unstable or saddle, we investigate
the following matrix,
\begin{eqnarray}
  \hat{\mathcal{M}}(\mbfr_0) &=&
  \frac{1}{16V_0 q_{0}^{2}}~
  \left(
    \begin{array}{cc}
      V_{x,x}(\mbfr_0)
      &
      V_{x,y}(\mbfr_0)
      \\
      V_{y,x}(\mbfr_0)
      &
      V_{y,y}(\mbfr_0)
    \end{array}
  \right),
  \label{M-matrix-def}
\end{eqnarray}
where
\begin{eqnarray*}
  &&
  V_{x,x}(\mbfr_0) ~=~
  \frac{\partial^{2} V(\mbfr_0)}
       {\partial x_{0}^{2}},
  \ \ \
  V_{y,y}(\mbfr_0) ~=~
  \frac{\partial^{2} V(\mbfr_0)}
       {\partial y_{0}^{2}},
  \\
  &&
  V_{x,y}(\mbfr_0) ~=~
  V_{y,x}(\mbfr_0) ~=~
  \frac{\partial^{2} V(\mbfr_0)}
       {\partial x_0 \partial y_0}.
\end{eqnarray*}
Explicitly, they are
\begin{eqnarray*}
  \frac{V_{x,x}(\mbfr)}{16 V_0 q_{0}^{2}}
  &=&
  \cos
  \bigg(
       \frac{2 \pi x}{a_x}
  \bigg)~
  \bigg\{
       1-
       \beta^2
       \big[
           1+
           \cos(2\phi)
       \big]+
   \nonumber \\ && +
       \big[
           1-
           2
           \beta^2
       \big]
       \cos
       \bigg(
            \frac{2 \pi y}{a_y}
       \bigg)
  \bigg\}~
  \cos^2\phi,
  \\
  \frac{V_{y,y}(\mbfr)}{16 V_0 q_{0}^{2}} &=&
  \cos
  \bigg(
       \frac{2 \pi y}{a_y}
  \bigg)~
  \bigg\{
       1-
       \beta^2
       \big[
           1-
           \cos(2\phi)
       \big]+
   \nonumber \\ && +
       \big[
           1-
           2
           \beta^2
       \big]
       \cos
       \bigg(
            \frac{2 \pi x}{a_x}
       \bigg)
  \bigg\}~
  \sin^2\phi,
  \\
  \frac{V_{x,y}(\mbfr)}{16 V_0 q_{0}^{2}}
  &=&
  -\big[1-2\beta^2\big]~
  \sin
  \bigg(
       \frac{2 \pi x}{a_x}
  \bigg)~
  \sin
  \bigg(
       \frac{2 \pi y}{a_y}
  \bigg)
  \times \nonumber \\ && \times
  \cos\phi~
  \sin\phi.
\end{eqnarray*}

There are two eigenvalues of $\hat{\mathcal{M}}(\mbfr_0)$,
\begin{eqnarray}
  {\mathcal{M}}_{\pm}(\mbfr_0) &=&
  {\mathcal{A}}(\mbfr_0)
  \pm
  \sqrt{{\mathcal{B}}^{2}(\mbfr_0)+
        {\mathcal{C}}^{2}(\mbfr_0)},
  \label{M-eigen-values}
\end{eqnarray}
where
\begin{eqnarray*}
  {\mathcal{A}}(\mbfr_0) &=&
  \frac{V_{x,x}(\mbfr_0)+
        V_{y,y}(\mbfr_0)}
       {32 V_0 q_{0}^{2}},
  \\
  {\mathcal{B}}(\mbfr_0) &=&
  \frac{V_{x,x}(\mbfr_0)-
        V_{y,y}(\mbfr_0)}
       {32 V_0 q_{0}^{2}},
  \\
  {\mathcal{C}}(\mbfr_0) &=&
  \frac{V_{x,y}(\mbfr_0)}
       {16 V_0 q_{0}^{2}}.
\end{eqnarray*}
There are three cases,
\begin{itemize}
\item When ${\mathcal{M}}_{+}(\mbfr_0)>0$ and
      ${\mathcal{M}}_{-}(\mbfr_0)>0$, the equilibrium point
      $\mbfr_0$ is stable.

\item When ${\mathcal{M}}_{+}(\mbfr_0)<0$ and
      ${\mathcal{M}}_{-}(\mbfr_0)<0$, the equilibrium point
      $\mbfr_0$ is unstable.

\item When ${\mathcal{M}}_{+}(\mbfr_0)>0$ and
      ${\mathcal{M}}_{-}(\mbfr_0)<0$, the equilibrium point
      $\mbfr_0$ is saddle.
\end{itemize}

We investigate the equilibrium points for two
intervals, $\beta<\beta_{c,x}$ and $\beta>\beta_{c,y}$,
in turn [where $\beta_{c,x}$ and $\beta_{c,y}$ are given
by eq. (\ref{subeqs-betac})].

%%%%%%%%%%%%%%%%%%%%
\subsection{The Case $0<\beta<\beta_{c,x}$}
  \label{subsec-beta<beta_c}

When $\beta<\beta_{c,x}$, the equilibrium points are given by
eq. (\ref{subeqs-for-equilibrium-main}).
Explicitly, they are
\begin{eqnarray}
  \boldsymbol\alpha_{n_x,n_y} &=&
  \frac{n_x \mbfa_x}{2}+
  \frac{n_y \mbfa_y}{2},
  \label{equilib-b<bc}
\end{eqnarray}
where $\mbfa_x$ and $\mbfa_y$ are given by eq. (\ref{a1-a2-vec-def}),
$n_x$ and $n_x$ are integers.

The sign of ${\mathcal{M}}_{\pm}(\boldsymbol\alpha_{n_x,n_y})$,
eq. (\ref{M-eigen-values}), depend either $n_x$ and $n_y$ are
even or odd. There are four cases:
\begin{itemize}
\item When $n_x$ and $n_y$ are even,
      $$
        n_x=2m_x,
        \ \ \
        n_y=2m_y,
      $$
      then
      \begin{eqnarray*}
        {\mathcal{M}}_{+}
        \big(
            \boldsymbol\alpha_{2m_x,2m_y}
        \big)
        &=&
        2 {\mathfrak{c}}^{2}(\phi,\beta)~
        \bigg(
             1-
             \frac{\beta^2}{\beta_{c,x}^{2}}
        \bigg)
        ~>~ 0,
        \\
        {\mathcal{M}}_{-}
        \big(
            \boldsymbol\alpha_{2m_x,2m_y}
        \big)
        &=&
        2 {\mathfrak{s}}^{2}(\phi,\beta)~
        \bigg(
             1-
             \frac{\beta^2}{\beta_{c,y}^{2}}
        \bigg)
        ~>~ 0,
      \end{eqnarray*}
      where
      \begin{eqnarray*}
        {\mathfrak{c}}(\phi,\beta) ~=~
        \left\{
          \begin{array}{ccc}
            \cos\phi
            &
            {\text{for}}
            &
            0 ~<~ \beta ~<~ \frac{1}{\sqrt{2}},
            \\
            \sin\phi
            &
            {\text{for}}
            &
            \frac{1}{\sqrt{2}} ~<~ \beta ~<~ \beta_{c,x},
          \end{array}
        \right.
        \\
        {\mathfrak{s}}(\phi,\beta) ~=~
        \left\{
          \begin{array}{ccc}
            \sin\phi
            &
            {\text{for}}
            &
            0 ~<~ \beta ~<~ \frac{1}{\sqrt{2}},
            \\
            \cos\phi
            &
            {\text{for}}
            &
            \frac{1}{\sqrt{2}} ~<~ \beta ~<~ \beta_{c,x}.
          \end{array}
        \right.
      \end{eqnarray*}
      Therefore, $\boldsymbol\alpha_{2m_x,2m_y}$ are
      {\textit{stable}} equilibrium points.
      The potential energy (\ref{V-pot-res}) at
      $\mbfr=\boldsymbol\alpha_{2m_1,2m_2}$ is,
      \begin{eqnarray*}
        V\big(\boldsymbol\alpha_{2m_x,2m_y}\big) &=&
        -16~
        \big(1-\beta^2\big)~
        V_0.
      \end{eqnarray*}

\item When $n_x$ and $n_y$ are odd,
      $$
        n_x=2m_x+1,
        \ \ \ \ \
        n_y=2m_y+1,
      $$
      then
      \begin{eqnarray*}
        {\mathcal{M}}_{+}
        \big(
            \boldsymbol\alpha_{2m_x+1,2m_y+1}
        \big)
        &=&
        -\frac{\beta^2}{2}
        \sin(2\phi)
        ~<~ 0,
        \\
        {\mathcal{M}}_{-}
        \big(
            \boldsymbol\alpha_{2m_x+1,2m_y+1}
        \big)
        &=&
        -\frac{\beta^2}{2}
        \sin(2\phi)
        ~<~ 0.
      \end{eqnarray*}
      Therefore, $\boldsymbol\alpha_{2m_x+1,2m_y+1}$ are
      {\textit{unstable}} equilibrium points.
      The potential energy (\ref{V-pot-res}) at
      $\mbfr=\boldsymbol\alpha_{2m_x+1,2m_y+1}$ is,
      \begin{eqnarray*}
        V\big(\boldsymbol\alpha_{2m_x+1,2m_y+1}\big) &=& 0.
      \end{eqnarray*}

\item When $n_x$ is even and $n_y$ is odd,
      $$
        n_x=2m_x,
        \ \ \ \ \
        n_y=2m_y+1,
      $$
      then
      \begin{eqnarray*}
        {\mathcal{M}}_{+}
        \big(
            \boldsymbol\alpha_{2m_x,2m_y+1}
        \big)
        &=&
        \frac{\beta^2}{2}~
        \sin^{2}(2\phi)
        ~>~ 0,
        \\
        {\mathcal{M}}_{-}
        \big(
            \boldsymbol\alpha_{2m_x,2m_y+1}
        \big)
        &=&
        -2\sin^2\phi~
        \bigg(
             1-
             \frac{\beta^2}{\beta_{c,y}^{2}}
        \bigg)
        ~<~ 0.
      \end{eqnarray*}
      Therefore, $\boldsymbol\alpha_{2m_x,2m_y+1}$ are
      {\textit{saddle}} equilibrium points.
      The potential energy (\ref{V-pot-res}) at
      $\mbfr=\boldsymbol\alpha_{2m_x,2m_y+1}$ is,
      \begin{eqnarray*}
        V\big(\boldsymbol\alpha_{2m_x,2m_y+1}\big) &=&
        -16 \beta^2 V_0 \sin^2\phi.
      \end{eqnarray*}

\item When $n_x$ is odd and $n_y$ is even,
      $$
        n_x=2m_x+1,
        \ \ \ \ \
        n_y=2m_y,
      $$
      then
      \begin{eqnarray*}
        {\mathcal{M}}_{+}
        \big(
            \boldsymbol\alpha_{2m_x+1,2m_y}
        \big)
        &=&
        \frac{\beta^2}{2}~
        \sin^2\big(2\phi\big)
        ~>~ 0,
        \\
        {\mathcal{M}}_{-}
        \big(
            \boldsymbol\alpha_{2m_1+1,2m_2}
        \big)
        &=&
        -2\cos^2\phi~
        \bigg(
             1-
             \frac{\beta^2}{\beta_{c,x}^{2}}
        \bigg)
        ~<~ 0.
      \end{eqnarray*}
      Therefore, $\boldsymbol\alpha_{2m_1+1,2m_2}$ are saddle
      equilibrium points.
      The potential energy (\ref{V-pot-res}) at
      $\mbfr=\boldsymbol\alpha_{2m_1+1,2m_2}$ is,
      \begin{eqnarray*}
        V\big(\boldsymbol\alpha_{2m_x+1,2m_y}\big) &=&
        -16 \beta^2 V_0 \cos^2\phi.
      \end{eqnarray*}
\end{itemize}

%%%%%%%%%%%%%%%%%%%%
\begin{figure}[htb]
\centering
  \subfigure[]
  {\includegraphics[width=50 mm,angle=0]
  {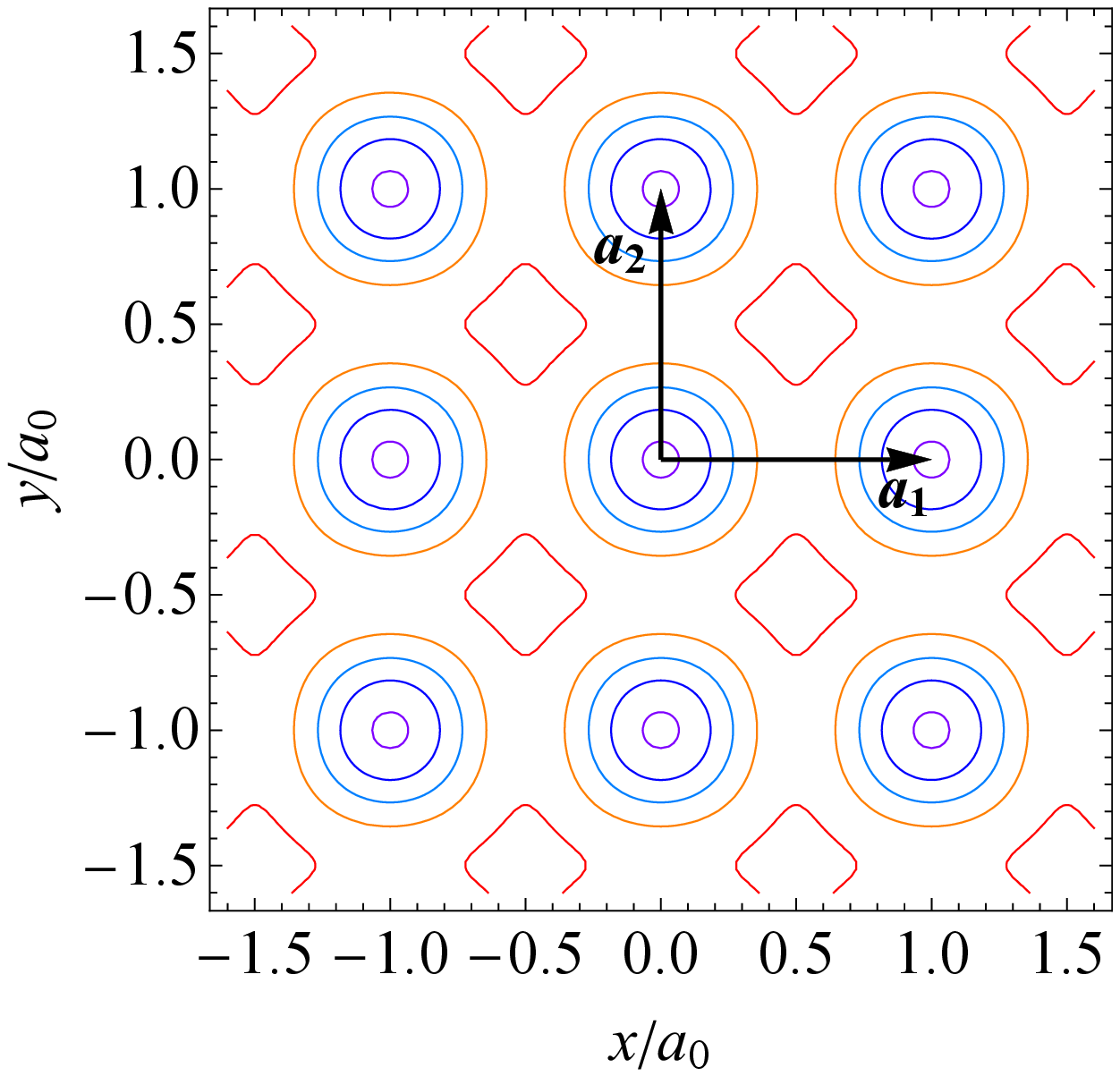}
   \label{Fig-min-max-b<bc}}
  \subfigure[]
  {\includegraphics[width=50 mm,angle=0]
   {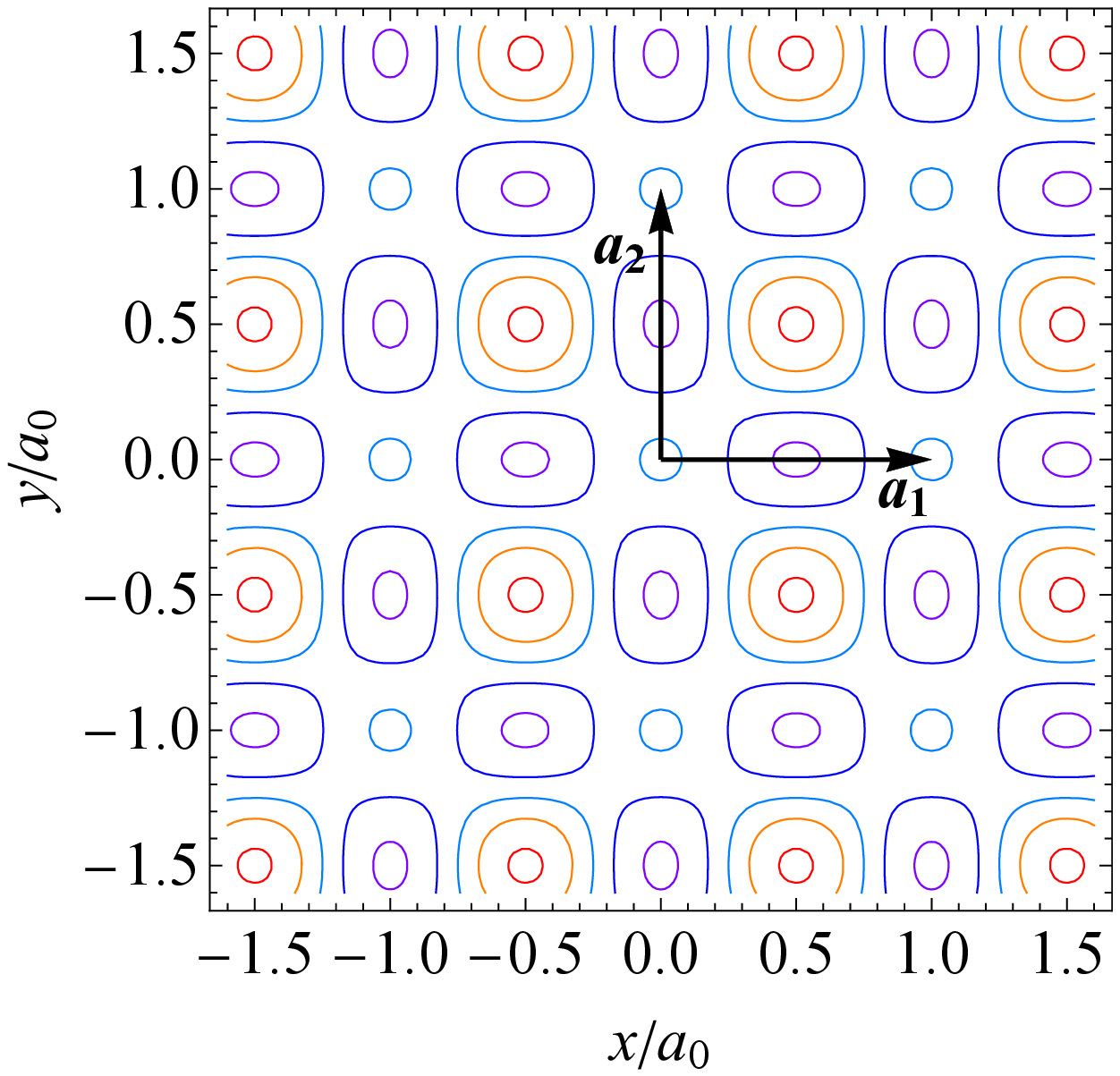}
   \label{Fig-min-max-b>bc}}
 \caption{\footnotesize
   ({\color{blue}color online})
   The potential energy (\ref{V-pot-res}) for $\beta=0.4$ [panel (a)]
   and $\beta=0.9$ [panel (b)]. For both panels,
   $\phi=\frac{99\pi}{400}$.
   The purple, blue, sky blue, orange
   and red curves denote $V(\mbfr)=0.04V_{\mathrm{min}}$,
   $0.27V_{\mathrm{min}}$, $0.5V_{\mathrm{min}}$,
   $0.73V_{\mathrm{min}}$ and $0.96V_{\mathrm{min}}$, where
   $V_{\mathrm{min}}$ is the minimal value of $V(\mbfr)$ for given
   $\beta$.}
 \label{Fig-min-max}
\end{figure}

The potential energy (\ref{V-pot-res}) is displayed in Fig.
\ref{Fig-min-max-b<bc} for $\phi=\frac{99\pi}{400}$ and
$\beta=0.4$ (which lies within the interval
$0<\beta<\beta_{c,x}$).
The minima points $\boldsymbol\alpha_{2m_x,2m_y}$ and the maxima
points $\boldsymbol\alpha_{2m_x+1,2m_y+1}$ are well seen.

%%%%%%%%%%%%%%%%%%%%
\subsection{The Case $\beta>\beta_{c,y}$}
  \label{subsec-beta>beta_c}

When $\beta_{c,y}<\beta<1$, the equilibrium points are given by
eqs. (\ref{subeqs-for-equilibrium-additional}) and
(\ref{subeqs-for-equilibrium-main}).
Explicitly, they are
\begin{subequations}
\begin{eqnarray}
  &&
  \boldsymbol\alpha_{n_x,n_y}
  =
  \frac{n_x \mbfa_x}{2}+
  \frac{n_y \mbfa_y}{2},
  \label{equilib-alpha-b>bc}
  \\
  &&
  \boldsymbol\beta_{\zeta_x,n_x;\zeta_y,n_y}
  =
  \sum_{j=x,y}
  \zeta_j
  \bigg(
       \frac{\varphi_{\beta,j}}{\pi}+
       n_j
  \bigg)~
  \mbfa_j.
  \label{equilib-beta-b>bc}
\end{eqnarray}
  \label{subeqs-equilib-b>bc}
\end{subequations}
Here $n_x$ and $n_y$ are integers, $\eta_j=\pm1$ ($j=x,y$),
and
\begin{eqnarray*}
  \varphi_{\beta,x} =
  \arccos
  \Bigg(
       \frac{1-
             \beta^2
             \big[
                 1+
                 \cos(2\phi)
             \big]}
            {2\beta^2-1}
  \Bigg),
  \\
  \varphi_{\beta,y} =
  \arccos
  \Bigg(
       \frac{1-
             \beta^2
             \big[
                 1-
                 \cos(2\phi)
             \big]}
            {2\beta^2-1}
  \Bigg).
\end{eqnarray*}
Note that for any $\beta$ within the interval
$\beta_{c,y}\leq\beta\leq1$,
$$
  0 ~\leq~
  \frac{1-
        \beta^2
        \big[
            1\pm
            \cos(2\phi)
        \big]}
       {2\beta^2-1}
  ~\leq~ 1,
$$
and therefore
$$
  0 ~\leq~
  \varphi_{\beta,j}
  ~\leq~ \frac{\pi}{2},
  \ \ \ \ \
  j=x,y.
$$
We investigate now the sign of ${\mathcal{M}}_{\pm}(\mbfr)$,
eq. (\ref{M-eigen-values}), for all equilibrium points
(\ref{subeqs-equilib-b>bc}).

First, we consider sign of
${\mathcal{M}}_{\pm}(\boldsymbol\alpha_{n_1,n_2})$, where
$\boldsymbol\alpha_{n_1,n_2}$ is given by
eq. (\ref{equilib-alpha-b>bc}).
There are four cases, when $n_1$ and $n_2$ are even or odd.
We consider all these cases in turn.
\begin{itemize}
\item When $n_x$ and $n_y$ are even,
      $$
        n_x=2m_x,
        \ \ \ \ \
        n_y=2m_y,
      $$
      then
      \begin{eqnarray*}
        {\mathcal{M}}_{+}
        \big(
            \boldsymbol\alpha_{2m_x,2m_y}
        \big)
        &=&
        -2\sin^2\phi~
        \bigg(
             \frac{\beta^2}
                  {\beta_{c,y}^{2}}-
             1
        \bigg)
        ~<~ 0,
        \\
        {\mathcal{M}}_{-}
        \big(
            \boldsymbol\alpha_{2m_1,2m_2}
        \big)
        &=&
        -2\cos^2\phi~
        \bigg(
             \frac{\beta^2}
                  {\beta_{c,x}^{2}}-
             1
        \bigg)
        ~<~ 0.
      \end{eqnarray*}
      Therefore, $\boldsymbol\alpha_{2m_x,2m_y}$ are unstable
      equilibrium points.
      The potential energy (\ref{V-pot-res}) at
      $\mbfr=\boldsymbol\alpha_{2m_x,2m_y}$ is,
      \begin{eqnarray*}
        V
        \big(
            \boldsymbol\alpha_{2m_x,2m_y}
        \big)
        &=&
        -16~
        \big(1-\beta^2\big)~
        V_0.
      \end{eqnarray*}

\item When $n_x$ and $n_y$ are odd,
      $$
        n_x=2m_x+1,
        \ \ \ \ \
        n_y=2m_y+1,
      $$
      then
      \begin{eqnarray*}
        {\mathcal{M}}_{+}
        \big(
            \boldsymbol\alpha_{2m_x+1,2m_y+1}
        \big)
        &=&
        -\frac{\beta^2}{2}~
        \sin^2\big(2\phi\big)
        ~<~ 0,
        \\
        {\mathcal{M}}_{-}
        \big(
            \boldsymbol\alpha_{2m_x+1,2m_y+1}
        \big)
        &=&
        -\frac{\beta^2}{2}~
        \sin^2\big(2\phi\big)
        ~<~ 0.
      \end{eqnarray*}
      Therefore, $\boldsymbol\alpha_{2m_x+1,2m_y+1}$
      are unstable equilibrium points.
      The potential energy (\ref{V-pot-res}) at
      $\mbfr=\boldsymbol\alpha_{2m_x+1,2m_y+1}$ is,
      \begin{eqnarray*}
        V
        \big(
            \boldsymbol\alpha_{2m_x+1,2m_y+1}
        \big)
        &=& 0.
      \end{eqnarray*}

\item When $n_x$ is even and $n_y$ is odd,
      $$
        n_x=2m_x,
        \ \ \ \ \
        n_y=2m_y+1,
      $$
      then
      \begin{eqnarray*}
        {\mathcal{M}}_{+}
        \big(
            \boldsymbol\alpha_{2m_x,2m_y+1}
        \big)
        &=&
        \frac{\beta^2}{2}~
        \sin^2\big(2\phi\big)
        ~>~ 0,
        \\
        {\mathcal{M}}_{-}
        \big(
            \boldsymbol\alpha_{2m_x,2m_y+1}
        \big)
        &=&
        2 \sin^2\phi~
        \bigg(
             \frac{\beta^2}{\beta_{c,y}^{2}}-
             1
        \bigg)
        ~>~ 0.
      \end{eqnarray*}
      Therefore, $\boldsymbol\alpha_{2m_1,2m_2+1}$ are stable
      equilibrium points.
      The potential energy (\ref{V-pot-res}) at
      $\mbfr=\boldsymbol\alpha_{2m_1,2m_2+1}$ is,
      \begin{eqnarray*}
        V
        \big(
            \boldsymbol\alpha_{2m_1,2m_2+1}
        \big)
        &=&
        -16 \beta^2 V_0 \sin^2\phi.
      \end{eqnarray*}

\item When $n_x$ is odd and $n_y$ is even,
      $$
        n_x=2m_x+1,
        \ \ \ \ \
        n_y=2m_y,
      $$
      then
      \begin{eqnarray*}
        {\mathcal{M}}_{+}
        \big(
            \boldsymbol\alpha_{2m_x+1,2m_y}
        \big)
        &=&
        \frac{\beta^2}{2}~
        \sin^2\big(2\phi\big)
        ~>~ 0,
        \\
        {\mathcal{M}}_{-}
        \big(
            \boldsymbol\alpha_{2m_x+1,2m_y}
        \big)
        &=&
        2 \cos^2\phi~
        \bigg(
             \frac{\beta^2}{\beta_{c,x}^{2}}-
             1
        \bigg)
        ~>~ 0.
      \end{eqnarray*}
      Therefore, $\boldsymbol\alpha_{2m_x+1,2m_y}$ are stable
      equilibrium points.
      The potential energy (\ref{V-pot-res}) at
      $\mbfr=\boldsymbol\alpha_{2m_x+1,2m_y}$ is,
      \begin{eqnarray*}
        V
        \big(
            \boldsymbol\alpha_{2m_x+1,2m_y}
        \big)
        &=&
        -16 \beta^2 V_0 \cos^2\phi.
      \end{eqnarray*}
\end{itemize}

{\textbf{Next step}}, we consider sign of
${\mathcal{M}}_{\pm}(\boldsymbol\beta_{\zeta_x,n_x;\zeta_y,n_y})$,
where $\boldsymbol\beta_{\zeta_x,n_x;\zeta_y,n_y}$ is given by
eq. (\ref{equilib-beta-b>bc}). Explicitly, we get
\begin{eqnarray*}
  {\mathcal{M}}_{+}
  \big(
      \boldsymbol\beta_{\zeta_x,n_x;\zeta_y,n_y}
  \big)
  =
  \frac{\beta^2}{2\big(2\beta^2-1\big)}
  \times
  \ \ \ \ \
  \\ \times
  \sqrt{\frac{\big(
            \beta^2-
            \beta_{c,x}^{2}
        \big)~
        \big(
            \beta^2-
            \beta_{c,y}^{2}
        \big)}
       {9-\cos^2(2\phi)}}
  ~>~ 0,
  \\
  {\mathcal{M}}_{-}
  \big(
      \boldsymbol\beta_{\zeta_x,n_x;\zeta_y,n_y}
  \big)
  =
  -\frac{\beta^2}{2\big(2\beta^2-1\big)}
  \times
  \ \ \ \ \
  \\ \times
  \sqrt{\frac{\big(
            \beta^2-
            \beta_{c,x}^{2}
        \big)~
        \big(
            \beta^2-
            \beta_{c,y}^{2}
        \big)}
       {9-\cos^2(2\phi)}}
  ~<~ 0.
\end{eqnarray*}
Therefore, $\boldsymbol\beta_{\zeta_x,n_x;\zeta_y,n_y}$ are saddle
equilibrium points.
The potential energy (\ref{V-pot-res}) at
$\mbfr=\boldsymbol\beta_{\zeta_x,n_x;\zeta_y,n_y}$ is,
\begin{eqnarray*}
  V
  \big(
      \boldsymbol\beta_{\zeta_x,n_x;\zeta_y,n_y}
  \big)
  &=&
  -\frac{4\beta^4 V_0 \sin^2\big(2\phi\big)}
        {2\beta^2-1}.
\end{eqnarray*}

The potential energy (\ref{V-pot-res}) is displayed in Fig.
\ref{Fig-min-max-b>bc} for $\phi=\frac{99\pi}{400}$ and
$\beta=0.9$ (which lies within the interval
$\beta_{c,y}<\beta<1$).
The minima points $\boldsymbol\alpha_{2m_x+1,2m_y}$ and
$\boldsymbol\alpha_{2m_x,2m_y+1}$, as well as the maxima
points $\boldsymbol\alpha_{2m_x,2m_y}$ and
$\boldsymbol\alpha_{2m_x+1,2m_y+1}$ are well seen.
Note that for $\beta>\beta_{c,y}$, $V(\mbfr)$ has two minima
points per unit cell.
%%%%%%%%%%%Design%%%%%%%%%%%%%%

%%%%%%%%%%%%%%%%%%%%%%%%%%%%%%
\section{Mirror Chern Numbers}
  \label{append-Chern}

Eigenfunctions $|\psi_{\xi,s,\eta}(\mbfk)\rangle$ of
the Hamiltonian (\ref{H-4D-H-4D}) are,
\begin{eqnarray}
  \big|
      \psi_{\xi,s,\eta}(\mbfk)
  \big\rangle
  &=&
  \sum_{f}
  \chi_{\xi,s,f,\eta}(k)~
  e^{-i f \phi}
  |f\rangle.
  \label{WF-vs-s}
\end{eqnarray}
Here
\begin{eqnarray*}
  \chi_{\xi,s,f,\eta}(k) &=&
  \big(-1\big)^{\frac{3}{2}-f}
  {\mathcal{N}}_{\xi,s,\eta}(k)~
  {\mathrm{det}}
  \Big(
      \hat{\mathcal{M}}_{\xi,s,f,\eta}(k)
  \Big),
\end{eqnarray*}
where ${\mathcal{N}}_{\xi,s,\eta}(k)$ is a normalization constant,
and the  matrices $\hat{\mathcal{M}}_{\xi,s,f,\eta}(k)$ are,
\begin{eqnarray*}
  \hat{\mathcal{M}}_{\xi,s,\frac{3}{2},\eta}(k) &=&
  \left(
    \begin{array}{ccc}
      \sqrt{3}~d_k &
      0 &
      0
      \\
      g_{\xi,s,\frac{1}{2},\eta}(k) &
      2 d_k &
      0
      \\
      2 d_k &
      g_{\xi,s,\frac{\bar{1}}{2},\eta}(k) &
      \sqrt{3}~d_k
    \end{array}
  \right),
  \\
  \hat{\mathcal{M}}_{\xi,s,\frac{1}{2},\eta}(k) &=&
  \left(
    \begin{array}{ccc}
      g_{\xi,s,\frac{3}{2},\eta}(k) &
      0 &
      0
      \\
      \sqrt{3}~d_k &
      2~d_k &
      0
      \\
      0 &
      g_{\xi,s,\frac{\bar{1}}{2},\eta}(k) &
      \sqrt{3}~d_k
    \end{array}
  \right),
  \\
  \hat{\mathcal{M}}_{\xi,s,\frac{\bar{1}}{2},\eta}(k) &=&
  \left(
    \begin{array}{ccc}
      g_{\xi,s,\frac{3}{2},\eta}(k) &
      \sqrt{3}~d_k &
      0
      \\
      \sqrt{3}~d_k &
      g_{\xi,s,\frac{1}{2},\eta}(k) &
      0
      \\
      0 &
      2~d_k &
      \sqrt{3}~d_k
    \end{array}
  \right),
  \\
  \hat{\mathcal{M}}_{\xi,s,\frac{\bar{3}}{2},\eta}(k) &=&
  \left(
    \begin{array}{ccc}
      g_{\xi,s,\frac{3}{2},\eta}(k) &
      \sqrt{3}~d_k &
      0
      \\
      \sqrt{3}~d_k &
      g_{\xi,s,\frac{1}{2},\eta}(k) &
      2~d_k
      \\
      0 &
      2~d_k &
      g_{\xi,s,\frac{\bar{1}}{2},\eta}(k)
    \end{array}
  \right),
\end{eqnarray*}
where
\begin{eqnarray*}
  &&
  g_{\xi,s,f,\eta}(k) =
  h_{\eta,f}(k)-
  \veps_{\xi,s,\eta}(k),
  \\
  &&
  h_{\eta,f}(k) =
  \big(-1\big)^{\frac{3}{2}-f}
  \eta \Delta_{\mbfk}+
  f B.
\end{eqnarray*}
Here $\Delta_{\mbfk}$ is given by
Eq. (\ref{Mk-def}) and $d_k\equiv|\mbfd_{\mbfk}|=\hbar v k$.

Wave functions (\ref{WF-vs-s}) are described by the following
quantum numbers:
\begin{itemize}
\item Iso-spin quantum number $\xi={\mathrm{c,v}}$ for
      the conductance or valence band.

\item Sub-band quantum number $s=\frac{1}{2}$ and
      $\frac{3}{2}$.

\item Block quantum number $\eta=\pm1$.
\end{itemize}
Note that the sub-band quantum number $s$ takes just
positive values. This is because of the following reason:
\begin{itemize}
\item When $\eta=1$, then the modes with $f=\frac{3}{2}$ and
      $f=-\frac{1}{2}$ belong to the conduction band, whereas
      the modes with $f=-\frac{3}{2}$ and $f=\frac{1}{2}$ belong
      to the valence band.

\item When $\eta=-1$, then the modes with $f=-\frac{3}{2}$ and
      $f=\frac{1}{2}$ belong to the conduction band, whereas
      the modes with $f=\frac{3}{2}$ and $f=-\frac{1}{2}$ belong
      to the valence band.
\end{itemize}
Therefore we describe the modes of the conduction or valence band
by the positive $s=|f|$ and additional quantum number $\xi$.

Existence of the topological edge states can be checked
from the wave functions (\ref{WF-vs-s}). For this purpose, we
consider mirror Chern numbers $C_{\eta}$, which is given by eq. (\ref{Chern-number-def}).

It is convenient to use polar coordinates,
\begin{eqnarray*}
  k_x ~=~
  k\cos\phi,
  \ \ \ \ \
  k_y ~=~
  k\sin\phi.
\end{eqnarray*}
The vector
$|\vec\psi_{\eta,\sigma,\xi}(\mbfk)\rangle$,
Eq. (\ref{vec-psi-def}), in the polar
coordinates is,
\begin{eqnarray}
  \big|
      \vec\psi_{\xi,s,\eta}(\mbfk)
  \big\rangle
  =
  \mbfe_k~
  \frac{\partial |\psi_{\xi,s,\eta}(\mbfk)\rangle}
       {\partial k}+
  \frac{\mbfe_{\phi}}{k}~
  \frac{\partial |\psi_{\xi,s,\eta}(\mbfk)\rangle}
       {\partial \phi}.
  \label{vec-psi-polar}
\end{eqnarray}
Then the Berry curvature (\ref{Berry-curv-def}) takes the form,
\begin{eqnarray}
  F_{\xi,s,\eta}(\mbfk) =
  i
  \bigg(
       \partial_{k} A_{\xi,s,\eta;\phi}(\mbfk)-
       \partial_{\phi} A_{\xi,s,\eta;k}(\mbfk)
  \bigg),
  \label{Berry-polar}
\end{eqnarray}
where
\begin{subequations}
\begin{eqnarray}
  A_{\xi,s,\eta;k}(\mbfk) &=&
  \big\langle
      \psi_{\xi,s,\eta}(\mbfk)
  \big|
      \partial_k
  \big|
      \psi_{\xi,s,\eta}(\mbfk)
  \big\rangle,
  \label{A-k-def}
  \\
  A_{\xi,s,\eta;\phi}(\mbfk) &=&
  \big\langle
      \psi_{\xi,s,\eta}(\mbfk)
  \big|
      \partial_{\phi}
  \big|
      \psi_{\xi,s,\eta}(\mbfk)
  \big\rangle.
  \label{A-phi-def}
\end{eqnarray}
  \label{subeqs-A-k-A-phi-def}
\end{subequations}

Taking into account eq. (\ref{WF-vs-s}), we can write
\begin{eqnarray}
  F_{\xi,s,\eta}(\mbfk) =
  \frac{1}{k}
  \sum_{f}
  f~
  \partial_{k}
  \Big(
      \chi_{\xi,s,f,\eta}^{2}(k)
  \Big).
  \label{Berry-res}
\end{eqnarray}
Note that the Berry curvature (\ref{Berry-res}) depends just on
$k$ but not on $\phi$.

Expression (\ref{Berry-res}) allows us to derive the following
expression the mirror Chern numbers (\ref{Chern-number-def}),
\begin{eqnarray}
  C_{\eta} &=&
  \sum_{s}
  \Big\{
      \lim_{k\to\infty}
      \big\langle
          S^{z}_{{\mathrm{v}},s,\eta}(k)
      \big\rangle-
  \nonumber \\ && -
      \lim_{k\to0}
      \big\langle
          S^{z}_{{\mathrm{v}},s,\eta}(k)
      \big\rangle
  \Big\},
  \label{Chern-average}
\end{eqnarray}
where
\begin{eqnarray}
  \big\langle
      S^{z}_{\xi,s,\eta}(k)
  \big\rangle
  &=&
  \sum_{f}
  f~
  \chi_{\xi,s,f,\eta}^{2}(k).
  \label{Sz-average-def}
\end{eqnarray}
Eq. (\ref{Chern-average}) shows that in order to find the mirror Chern
numbers, we need just wave functions (\ref{WF-vs-s}) for $k=0$
and $k\to\infty$.

Consider $\langle{S}_{{\mathrm{v}},s,\eta}(k)\rangle$ for
$k\to\infty$. Taking into account that the spin-orbital
interaction is linear with $k$, and the kinetic energy is
quadratic with $k$, we can neglect the spin-orbital interaction
and get diagonal Hamiltonian. Then we can write
\begin{eqnarray}
  \big\langle
      S_{{\mathrm{v}},\frac{1}{2},\eta}(\infty)
  \big\rangle
  =
  \frac{\eta}{2},
  \ \ \ \ \
  \big\langle
      S_{{\mathrm{v}},\frac{3}{2},\eta}(\infty)
  \big\rangle
  =
  -\frac{3\eta}{2}.
  \label{Sz-average-infty}
\end{eqnarray}
The values $\langle{S}_{{\mathrm{v}},s,\eta}(k)\rangle$ for
$k\to0$ depends on the values of $\Delta_0$ and $B$. There are
five relevant intervals in the half-plane $\Delta_0$-$B$ ($B>0$)
displayed in Fig. \ref{Fig-intervals-MF}. We derive
$\langle{S}_{{\mathrm{v}},s,\eta}(0)\rangle$ and the mirror Chern
numbers (\ref{Chern-average}) for each of the intervals.

{\underline{\textbf{Interval (1)}}}:
When $\frac{2}{3}\Delta_0>B>0$, then
\begin{eqnarray*}
  \big\langle
      S_{{\mathrm{v}},\frac{1}{2},1}(0)
  \big\rangle
  =
  \frac{1}{2},
  \ \ \ \ \
  \big\langle
      S_{{\mathrm{v}},\frac{3}{2},1}(0)
  \big\rangle
  =
  -\frac{3}{2},
  \\
  \big\langle
      S_{{\mathrm{v}},\frac{1}{2},\bar{1}}(0)
  \big\rangle
  =
  -\frac{1}{2},
  \ \ \ \ \
  \big\langle
      S_{{\mathrm{v}},\frac{3}{2},\bar{1}}(0)
  \big\rangle
  =
  \frac{3}{2}.
\end{eqnarray*}
Taking into account eq. (\ref{Sz-average-infty}), we get
$$
  C_{1}=C_{\bar{1}}=0,
$$
and we have no topological edge states.

{\underline{\textbf{Interval (2)}}}:
When $2\Delta_0>B>\frac{2}{3}\Delta_0>0$, then
\begin{eqnarray*}
  &&
  \big\langle
      S_{{\mathrm{v}},\frac{1}{2},1}(0)
  \big\rangle
  =
  \frac{1}{2},
  \ \ \ \ \
  \big\langle
      S_{{\mathrm{v}},\frac{3}{2},1}(0)
  \big\rangle
  =
  -\frac{3}{2},
  \\
  &&
  \big\langle
      S_{{\mathrm{v}},\frac{1}{2},\bar{1}}(0)
  \big\rangle
  =
  -\frac{3}{2},
  \ \ \
  \big\langle
      S_{{\mathrm{v}},\frac{3}{2},\bar{1}}(0)
  \big\rangle
  =
  -\frac{1}{2}.
\end{eqnarray*}
Taking into account eq. (\ref{Sz-average-infty}), we get
$$
  C_{1}=0,
  \ \ \ \ \
  C_{\bar{1}}=3,
$$
and we have no topological edge states with $\eta=1$, and there
are three chiral topological modes with $\eta=-1$.

{\underline{\textbf{Interval (3)}}}: When $B>2|\Delta_0>0|$, then
\begin{eqnarray*}
  &&
  \big\langle
      S_{{\mathrm{v}},\frac{1}{2},1}(0)
  \big\rangle
  =
  -\frac{1}{2},
  \ \ \ \ \
  \big\langle
      S_{{\mathrm{v}},\frac{3}{2},1}(0)
  \big\rangle
  =
  -\frac{3}{2},
  \\
  &&
  \big\langle
      S_{{\mathrm{v}},\frac{1}{2},\bar{1}}(0)
  \big\rangle
  =
  -\frac{3}{2},
  \ \ \ \ \
  \big\langle
      S_{{\mathrm{v}},\frac{3}{2},\bar{1}}(0)
  \big\rangle
  =
  -\frac{1}{2}.
\end{eqnarray*}
Taking into account eq. (\ref{Sz-average-infty}), we get
$$
  C_{1}=1,
  \ \ \ \ \
  C_{\bar{1}}=3,
$$
and we have a chiral mode with $\eta=1$, and three chiral modes with
$\eta=-1$.

{\underline{\textbf{Interval (4)}}}:
When $-2\Delta_0>B>-\frac{2}{3}\Delta_0>0$, then
\begin{eqnarray*}
  &&
  \big\langle
      S_{{\mathrm{v}},\frac{1}{2},1}(0)
  \big\rangle
  =
  -\frac{3}{2},
  \ \ \
  \big\langle
      S_{{\mathrm{v}},\frac{3}{2},1}(0)
  \big\rangle
  =
  -\frac{1}{2},
  \\
  &&
  \big\langle
      S_{{\mathrm{v}},\frac{1}{2},\bar{1}}(0)
  \big\rangle
  =
  \frac{1}{2},
  \ \ \ \ \
  \big\langle
      S_{{\mathrm{v}},\frac{3}{2},\bar{1}}(0)
  \big\rangle
  =
  -\frac{3}{2}.
\end{eqnarray*}
Taking into account eq. (\ref{Sz-average-infty}), we get
$$
  C_{1}=1,
  \ \ \ \ \
  C_{\bar{1}}=2,
$$
and we have a chiral mode with $\eta=1$, and
two chiral modes with $\eta=-1$.

{\underline{\textbf{Interval (5)}}}:
When $-\frac{2}{3}\Delta_0>B>0$, then
\begin{eqnarray*}
  &&
  \big\langle
      S_{{\mathrm{v}},\frac{1}{2},1}(0)
  \big\rangle
  =
  -\frac{1}{2},
  \ \ \
  \big\langle
      S_{{\mathrm{v}},\frac{3}{2},1}(0)
  \big\rangle
  =
  \frac{3}{2},
  \\
  &&
  \big\langle
      S_{{\mathrm{v}},\frac{1}{2},\bar{1}}(0)
  \big\rangle
  =
  \frac{1}{2},
  \ \ \ \ \
  \big\langle
      S_{{\mathrm{v}},\frac{3}{2},\bar{1}}(0)
  \big\rangle
  =
  -\frac{3}{2}.
\end{eqnarray*}
Taking into account eq. (\ref{Sz-average-infty}), we get
      $$
        C_{1}=-2,
        \ \ \ \ \
        C_{\bar{1}}=2,
      $$
      and we have two chiral modes with $\eta=1$, and
      two chiral modes with $\eta=-1$.

\bibliography{reference}

\end{document}